\documentclass[aps,prapplied,twocolumn,footinbib,floatfix,superscriptaddress]{revtex4-1}
\usepackage{graphicx}
\usepackage{indentfirst}
\usepackage{braket}
\usepackage{float}
\usepackage{amsmath}
\usepackage{physics}
\usepackage{CJK}
\usepackage{esint}
\usepackage{color}
\usepackage[T1]{fontenc}
\usepackage{subfigure}
\usepackage{amsfonts}
\usepackage{footmisc}
\usepackage{scrextend}
\usepackage{multirow}
\usepackage[outdir=./]{epstopdf}

\usepackage[english]{babel}
\usepackage{url}
\usepackage{comment}
\usepackage{array}
\usepackage{subfiles}
\usepackage{tikz}
\usetikzlibrary{shapes,arrows}

\usepackage[hyperfootnotes=false]{hyperref}
\usepackage{cleveref}
\definecolor{darkblue}{rgb}{0,0,0.5}
\hypersetup{
colorlinks=true,
linkcolor=black,
filecolor=blue,
citecolor=darkblue,  
urlcolor=black,
}

\def\be{\begin{equation}}
\def\ee{\end{equation}}
\def\ba{\begin{eqnarray}}
\def\ea{\end{eqnarray}}
\def\bal{\begin{aligned}}
\def\eal{\end{aligned}}
\def\o{\overline}

\usepackage{bm}

\def\bp{\begin{pmatrix}}
\def\ep{\end{pmatrix}}

\begin{document}

\title{Practical route to entanglement-assisted communication over noisy bosonic channels}

\author{Haowei Shi}
\affiliation{
James C. Wyant College of Optical Sciences, University of Arizona, Tucson, Arizona 85721, USA
}

\author{Zheshen Zhang}
\affiliation{
Department of Materials Science and Engineering, University of Arizona, Tucson, Arizona 85721, USA
}
\affiliation{
James C. Wyant College of Optical Sciences, University of Arizona, Tucson, Arizona 85721, USA
}

\author{Quntao Zhuang}
\email{zhuangquntao@email.arizona.edu}
\affiliation{
Department of Electrical and Computer Engineering, University of Arizona, Tucson, Arizona 85721, USA
}
\affiliation{
James C. Wyant College of Optical Sciences, University of Arizona, Tucson, Arizona 85721, USA
}

\begin{abstract} 
Entanglement offers substantial advantages in quantum information processing, but loss and noise hinder its applications in practical scenarios. Although it has been well known for decades that the classical communication capacity over lossy and noisy bosonic channels can be significantly enhanced by entanglement, no practical encoding and decoding schemes are available to realize any entanglement-enabled advantage. Here, we report structured encoding and decoding schemes for such an entanglement-assisted communication scenario. Specifically, we show that phase encoding on the entangled two-mode squeezed vacuum state saturates the entanglement-assisted classical communication capacity over a very noisy channel and overcomes the fundamental limit of covert communication without entanglement assistance. We then construct receivers for optimum hypothesis testing protocols under discrete phase modulation and for optimum noisy phase estimation protocols under continuous phase modulation. Our results pave the way for entanglement-assisted communication and sensing in the radio-frequency and microwave spectral ranges.
\end{abstract} 

\maketitle

\section{Introduction}
Entanglement's benefit for quantum information processing has been revealed by pioneer works in communication~\cite{bennett2002entanglement}, sensing~\cite{pirandola2018advances,giovannetti2004quantum,giovannetti2011advances}, and computation~\cite{shor1999polynomial}. Notably, the advantage enabled by the initial entanglement even survives loss and noise in certain entanglement-breaking scenarios, as predicted~\cite{Lloyd2008,tan2008quantum,zhuang2017,barzanjeh2015microwave} and experimentally demonstrated~\cite{zhang2013,zhang2015,barzanjeh2019experimental} in the entanglement-enhanced sensing protocol called quantum illumination.

It is also known, in theory, that pre-shared entanglement increases the classical communication capacity, i.e., the maximum rate of reliable communication of classical bits (cbits), over a quantum channel $\Phi$ (a completely-positive trace-preserving map). In an ideal case, the superdense-coding~\cite{bennett1992} protocol allows for sending two cbits on a single qubit, with the assistance of one entanglement bit (ebit). Formally, one characterizes the rate limit of such entanglement-assisted (EA) communication by the classical capacity with unlimited entanglement-assistance~\cite{bennett2002entanglement,bennett1999entanglement,holevo02,hsieh2008entanglement}, $C_E\left(\Phi\right)$~\footnote{The initial proof is for finite-dimensional systems, and Refs.~\cite{holevo2003entanglement,holevo2013classical} proved the EA capacity formula for infinite dimensional channel with more rigor.}. Compared with the classical capacity without entanglement-assistance, i.e., the Holevo-Schumacher-Westmoreland capacity, $C\left(\Phi\right)$~\cite{hausladen1996classical,schumacher1997sending,holevo1998capacity}, the improvement enabled by entanglement can be drastic even over a noisy channel $\Phi$. In particular, it is known~\cite{bennett2002entanglement} that the ratio of $C_E\left(\Phi\right)/C\left(\Phi\right)$ can diverge logarithmically with the inverse of signal power over a noisy and lossy bosonic channel~\footnote{Similar large improvement can also happen in large-dimension depolarizing channels~\cite{holevo02,holevo2013information}}. Such an EA scenario is widely applicable to radio-frequency (RF) communication, deep-space communication~\cite{banaszek2019approaching}, and covert communication~\cite{bash2015quantum,bullock2019fundamental}.

Despite the large advantage of EA capacity, a practical EA encoding and decoding scheme that achieves \emph{any} advantage over the classical capacity is unknown in the high noise regime. Previous experiments~\cite{prevedel_2011,chiuri_2013} focused on ideal scenarios with qubits; Although the EA capacity formula for bosonic Gaussian channel is well established~\cite{holevo01,de2017gaussian}, the achievability proof in Ref.~\cite{bennett2002entanglement} relies on approximating an infinite-dimensional channel as a channel with finite but large dimension; thus a structured encoding scheme is not given for bosonic channels. In fact, simple schemes like continuous-variable (CV) superdense coding~\cite{ban1999quantum,braunstein2000,ban2000quantum} do not beat the classical capacity in the noisy and weak signal regime~\cite{sohma2003}, making experimental demonstrations of the EA capacity advantage elusive~\cite{mizuno2005experimental,barzanjeh2013,li2002quantum}.
More recent encoding protocols in Refs.~\cite{wilde2012information,anshu2019building,qi2018applications,khabbazi2019union} use mode permutations or mode selections to encode classical information. Despite being convenient for theoretical analysis, these protocols require large quantum memories to store all quantum states and are thus difficult to implement with available technology.

The main contributions of this paper are 1) discovery of the optimum encoding scheme, 2) showing EA communication advantage in presence of lossy entanglement distribution, and 3) construction of practical quantum receivers for EA classical communication over lossy and noisy bosonic channels. We first prove that phase encoding on two-mode squeezed vacuum (TMSV) is asymptotically optimal as channel noise increases (Section~\ref{sec:encoding_capacity}). With phase encoding, we also show that the EA advantage can still be appreciable when entanglement distribution is lossy, thereby further reinforcing the robustness of EA communication (Section~\ref{sec:noisy_preshare}). Next, we show that such an EA communication protocol is in fact secure and allows one to break the square-root law of covert communication~\cite{bullock2019fundamental} by a logarithmic factor (Section~\ref{sec:encoding_covertness}). Then, we propose practical quantum receivers, based on prior results of Refs.~\cite{zhuang2017,Guha2009}, to offer a constant advantage over the classical capacity $C\left(\Phi\right)$ in the weak signal power regime (Section~\ref{sec:BPSK}). As a by-product, we show that our design in the context of continuous encoding also enables optimal phase estimation and asymptotically saturates the quantum Fisher information (QFI) upper bound~\cite{gagatsos2017bounding} (Section~\ref{sec:phase_estimation}), as the noise increases. Finally, we project the performance of a proof-of-concept experiment, based on the parameters reported in Ref.~\cite{zhang2015} (Section~\ref{sec:exp}).

We begin our paper by a brief overview.

\section{Overview}

In most of our discussions, we assume that an entangled signal-idler pair is pre-shared before the communication, potentially through a ground-satellite and/or fiber-based quantum network. The focus of this paper is on EA communication protocols assuming pre-shared entanglement is available, with the understanding that building a full-up quantum network is a challenging task. In a less ideal situation, we show that an EA advantage remains (see Section~\ref{sec:noisy_preshare}) in the absence of a full-scale quantum network that completely overcomes entanglement distribution loss.

\begin{figure}
\centering
\includegraphics[width=0.4\textwidth]{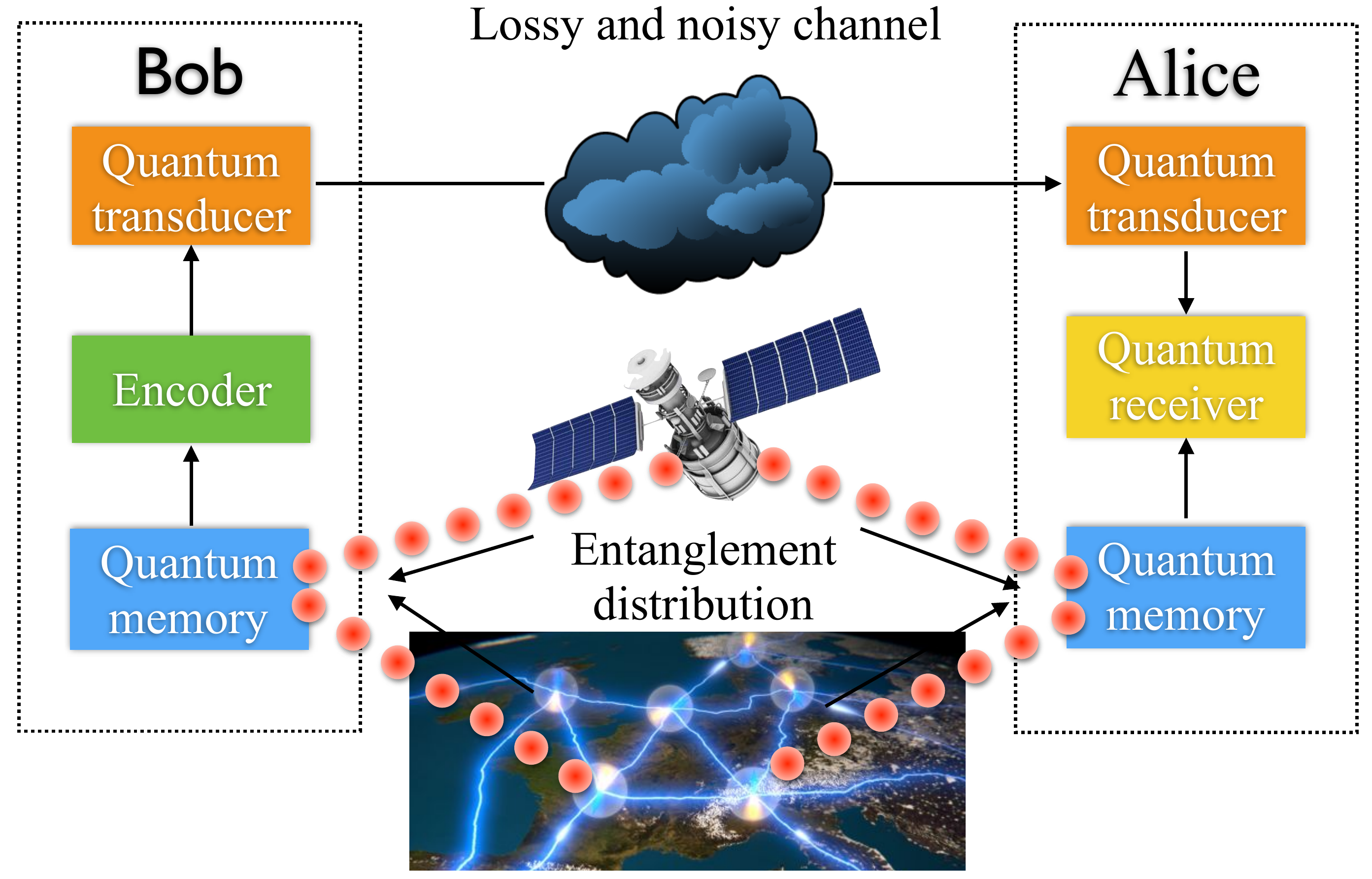}
\caption{Concept of entanglement-assisted communication. The entanglement is pre-shared through quantum internet based on satellites and/or ground-links.
\label{concept}
}
\end{figure}

Figure~\ref{concept} shows a general picture of EA communication: Bob encodes classical information on the signal mode of the pre-shared entanglement retrieved from a quantum memory. A quantum transducer is then employed to convert the wavelength of the signal mode to that of the information carrier, e.g., an RF field. The wavelength-converted signal is then sent to Alice through a lossy and noisy channel. After transducing the received signal from Bob, Alice jointly measures the signal mode and the entangled idler mode retrieved from a quantum memory to decode the classical information. We show that the EA communication scheme outperforms even the best classical scheme without entanglement assistance, with a significant advantage especially over a lossy and noisy communication channel in a weak-signal regime. Notably, such an EA communication advantage can be achieved with practical sources and receivers.

Before going into technical details, we elucidate the applicable scenarios for EA communication. Noisy communication channels are commonly used in the RF domain due to blackbody radiation. In optical communications, although the ambient noise is not naturally present, the classical communication traffic in, e.g., optical fibers of the Internet, may still be regarded as the effective noise. In addition, the communication channel can also be noisy in adversarial scenarios with active jamming. There also exist multiple communication settings limited to weak signal power. For example, in deep space communication, devices deployed on satellites or deep-space stations are likely to be power constrained. Besides, in a scenario where Alice and Bob wish to stay undetected, known as covert communication, the signal power is minimized and embedded in a bright noise background (see Section~\ref{sec:encoding_covertness} for details).

\section{Lossy and noisy bosonic channels: A compendium}
\label{sec:bosonic_capacity}

\begin{figure}
\centering
\includegraphics[width=0.4\textwidth]{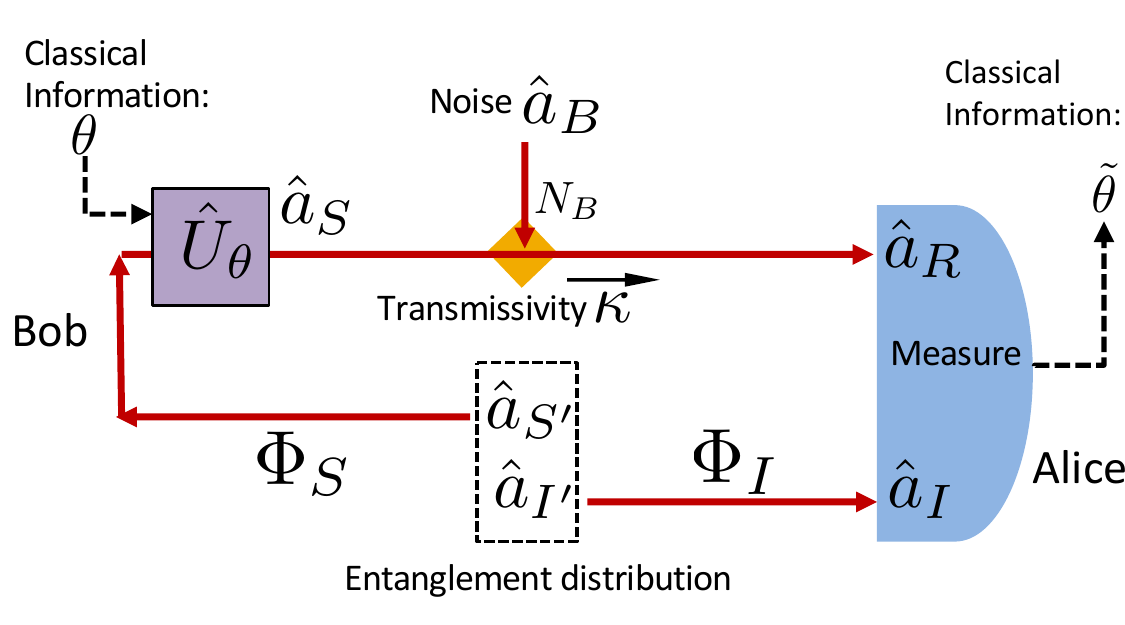}
\caption{Schematic of the entanglement-assisted classical communication protocol. The pre-shared entanglement is distributed through two channels $\Phi_S$ and $\Phi_I$. Classical information $\theta$ is encoded on the signal $\hat{a}_S$, which is sent over a noisy channel, represented by the beam splitter with transmissivity $\kappa$ and noise $N_B$, and then jointly measured with the entangled idler $\hat{a}_I$ to decode the classical information $\tilde{\theta}$.
\label{Schematic}
}
\end{figure}

Communications typically involve transmitting electromagnetic waves carrying classical information through optical fibers or free space, both of which can be modeled as a bosonic thermal lossy channel $\mathcal{L}^{\kappa,N_B}$ with the following input-output mode relation in the Heisenberg picture: 
\be 
\hat{a}_R=\sqrt{\kappa}\hat{a}_S+\sqrt{1-\kappa}\hat{a}_B,
\label{IO_relation}
\ee
as illustrated in Fig.~\ref{Schematic}. Here, the input mode is subject to an average energy constraint $\expval{\hat{a}_S^\dagger \hat{a}_S}=N_S$, and the noise mode $\hat{a}_B$ is in a thermal state with mean photon number $N_B/(1-\kappa)$, where $\kappa$ is the transmissivity of the bosonic channel.

Without EA, the classical capacity is known as~\cite{GiovannettiV2014}
\be 
C(\mathcal{L}^{\kappa,N_B})=g(\kappa N_S+N_B)-g(N_B),
\label{C_formula}
\ee 
obtained by maximizing the Holevo information~\cite{holevo1973bounds,holevo1998capacity,schumacher1997sending} over the ensemble of states. Here, $g(n)=(n+1)\log_2(n+1)-n \log_2 n$ is the entropy of a thermal state with mean photon number $n$. The capacity is achieved by an ensemble of Gaussian-modulated coherent states in conjunction with a joint-detection receiver, which are in general difficult to build. In some special situations, however, practical receivers are known to achieve the classical capacity~\cite{Shapiro2009ieee,banaszek2019approaching}. For example, in the limit of $\kappa N_S \gg 1$ and $N_B \ll 1$, the optical heterodyne receiver approaches the classical capacity. Moreover, in the large noise case of $N_B \gg 1$, the classical capacity $C(\mathcal{L}^{\kappa,N_B})=\kappa N_S/\ln(2)N_B+O(1/N_B^2)$ and is always saturated by a heterodyne or a homodyne receiver (see Appendix~\ref{App:classical} for details).



Classical communication can be enhanced by pre-shared entanglement. The bosonic EA classical communication operates in the following way (see Fig.~\ref{Schematic} for an example). One starts with entangled signal-idler pairs $\hat{a}_{S^\prime}, \hat{a}_{I^\prime}$, which are delivered to the sender Bob and the receiver Alice through channels $\Phi_S$ and $\Phi_I$. In this Section, both channels $\Phi_S$ and $\Phi_I$ are assumed lossless and noiseless, i.e., perfect unlimited pre-shared entanglement can be shared by Alice and Bob. Entanglement pre-shared through a common lossy channel will be considered in Section~\ref{sec:noisy_preshare}. The encoded signal $\hat{a}_S$, with mean photon number $N_S$, is sent through the noisy channel. A joint measurement on the received signal-idler pairs $\hat{a}_R, \hat{a}_I$ is performed to decode information. The EA classical capacity is~\cite{bennett2002entanglement}
\be
C_E(\mathcal{L}^{\kappa,N_B})=g(N_S)+g(N_S^\prime)-g(A_+)-g(A_-),
\label{CE_formula}
\ee 
where 
$A_\pm=(D-1\pm(N_S^\prime-N_S))/2$, $N_S^\prime=\kappa N_S+N_B$ and $D=\sqrt{(N_S+N_S^\prime+1)^2-4\kappa N_S(N_S+1)}$. 
Various aspects of EA communication have been explored, including extensions to limited pure entanglement~\cite{shor2004classical}, noisy entanglement~\cite{zhuang2017additive}, trade-off capacities~\cite{wilde2012quantum,wilde2012information}, and superaddivity issues~\cite{zhu2017,zhu2018superadditivity}. 

Comparing the capacity formulas with and without EA, one has
\be 
\lim_{N_B\to\infty}C_E/C=(1+N_S)\ln\left(1+1/N_S\right),
\ee
which diverges as $\ln(1/N_S)$ (see Fig.~\ref{Ce_over_C}). Thus, in the weak signal and strong noise regime, EA can offer a large capacity advantage over unassisted classical communications. Moreover, it is known that encoding on the
TMSV, 
\be
\ket{\psi^{N_S}}_{S^\prime I^\prime}=\sum_{n=0}^\infty \sqrt{N_S^n/(N_S+1)^{n+1}}\ket{n}_{S^\prime}\ket{n}_{I^\prime},
\label{wf_TMSV}
\ee 
achieves the EA classical capacity over a bosonic thermal lossy channel~\cite{wilde2012information,anshu2019building,qi2018applications}, but previously proposed encoding either needs large quantum memories or non-Gaussian operations without structured realizations, both of which are beyond the reach of current technology. Also, there is no known structured receiver that achieves the EA classical capacity.

It is worth mentioning that both the coherent states for classical communication and the TMSV for EA classical communication belong to the class of Gaussian states~\cite{Weedbrook_2012}, whose Wigner functions have a Gaussian shape. Gaussian states are important for quantum information processing, because they enable nonclassical resources such as squeezing and entanglement, and moreover they often allow analytical solution in various problems. An $n$-mode Gaussian state $\hat{\rho}$ comprising modes $\hat{a}_k, 1\le k \le n$, is fully characterized by the mean and the covariances of real quadrature field operators $\hat{q}_k=\hat{a}_k+\hat{a}_k^\dagger, \hat{p}_k=i\left(\hat{a}_k^\dagger-\hat{a}_k\right)$. Formally, we can define a real $2n$-dim vector of operators
${\hat{\bm x}}=\left(\hat{q}_1,\hat{p}_1,\cdots, \hat{q}_n,\hat{p}_n\right)$, then the mean $\bar{\bm x}=\expval{\hat{\bm x}}_{\hat{\rho}}$ and the elements of the $2n$-by-$2n$ covariance matrix are given by
\be
{\bm \Lambda}_{ij}=\frac{1}{2} \expval{\{\hat{x}_i-\bar{x}_i,\hat{x}_j-\bar{x}_j\}}_{\hat{\rho}},
\ee
where $\{,\}$ is the anticommutator and $\expval{\hat{A}}_{\hat{\rho}}={\rm Tr} \left(\hat{A}\hat{\rho}\right)$. As an example, from the wavefunction in Eq.~\ref{wf_TMSV} we can obtain the covariance matrix of a TMSV as
\begin{align}
& 
{\mathbf{{\mathbf{\Lambda}}}}_{\rm TMSV} =
\left(
\begin{array}{cccc}
(2N_S+1) {\mathbf I}&2C_0{\mathbf Z}\\
2C_0 {\mathbf Z}&(2N_S+1){\mathbf I}
\end{array} 
\right),
&
\label{cov_TMSV}
\end{align}
where ${\mathbf I}$, ${\mathbf Z}$ are two-by-two Pauli matrices, and $C_0=\sqrt{N_S\left(N_S+1\right)}$ is the amplitude of the phase-sensitive cross correlation.

\section{Optimal encoding---phase modulation}
\label{sec:encoding}
\subsection{Channel capacity with perfect pre-shared entanglement}
\label{sec:encoding_capacity}
In this section, we show that a set of states produced by phase modulation on an ideal TMSV is the asymptotic optimal encoding scheme, in that it achieves $C_E\left(\mathcal{L}^{\kappa,N_B}\right)$ for $N_B\gg1$.
Mathematically, phase modulation is described by the unitary $\hat{U}_\theta=\exp\left(i\theta \hat{a}^\dagger\hat{a}\right)$~\cite{Weedbrook_2012} that maps $\hat{a}\to e^{i\theta}\hat{a}$, where $\hat{a}$ is the annihilation operator of the incoming field. Under phase encoding (see Fig.~\ref{Schematic}), the joint state of the returned signal $\hat{a}_R$ and retained idler $\hat{a}_I$ at the receiver is 
\be 
\hat{\rho}^\theta_{RI}\equiv \mathcal{L}^{\kappa,N_B}\left[(\hat{U}_\theta \otimes \hat{I}) \hat{\psi}^{N_S}_{S^\prime I^\prime} (\hat{U}_\theta^\dagger \otimes \hat{I})\right],
\ee 
where $\hat{\psi}^{N_S}_{S^\prime I^\prime}$ is the density operator of the TMSV. From the input-output relation in Eq.~\ref{IO_relation} and the covariance matrix of TMSV in Eq.~\ref{cov_TMSV}, we can obtain the covariance matrix of the zero-mean Gaussian state $\hat{\rho}^\theta_{RI}$ as
\ba
&
{\mathbf{\Lambda}}_\theta =
\left(
\begin{array}{cccc}
(2\left(N_B+\kappa N_S\right)+1) {\mathbf I}&2C_p{\mathbf R}_\theta\\
2C_p{\mathbf R}_\theta&(2N_S+1){\mathbf I}
\end{array} 
\right),
\label{hk}
&
\ea
where ${\mathbf R}_\theta={\rm Re}\left[ \exp\left(i\theta\right) \left({\mathbf Z}-i{\mathbf X}\right)\right]$. The amplitude of the cross correlation in each mode pair is $C_p= \sqrt{\kappa}C_0$.

\begin{figure}
\centering
\includegraphics[width=0.3\textwidth]{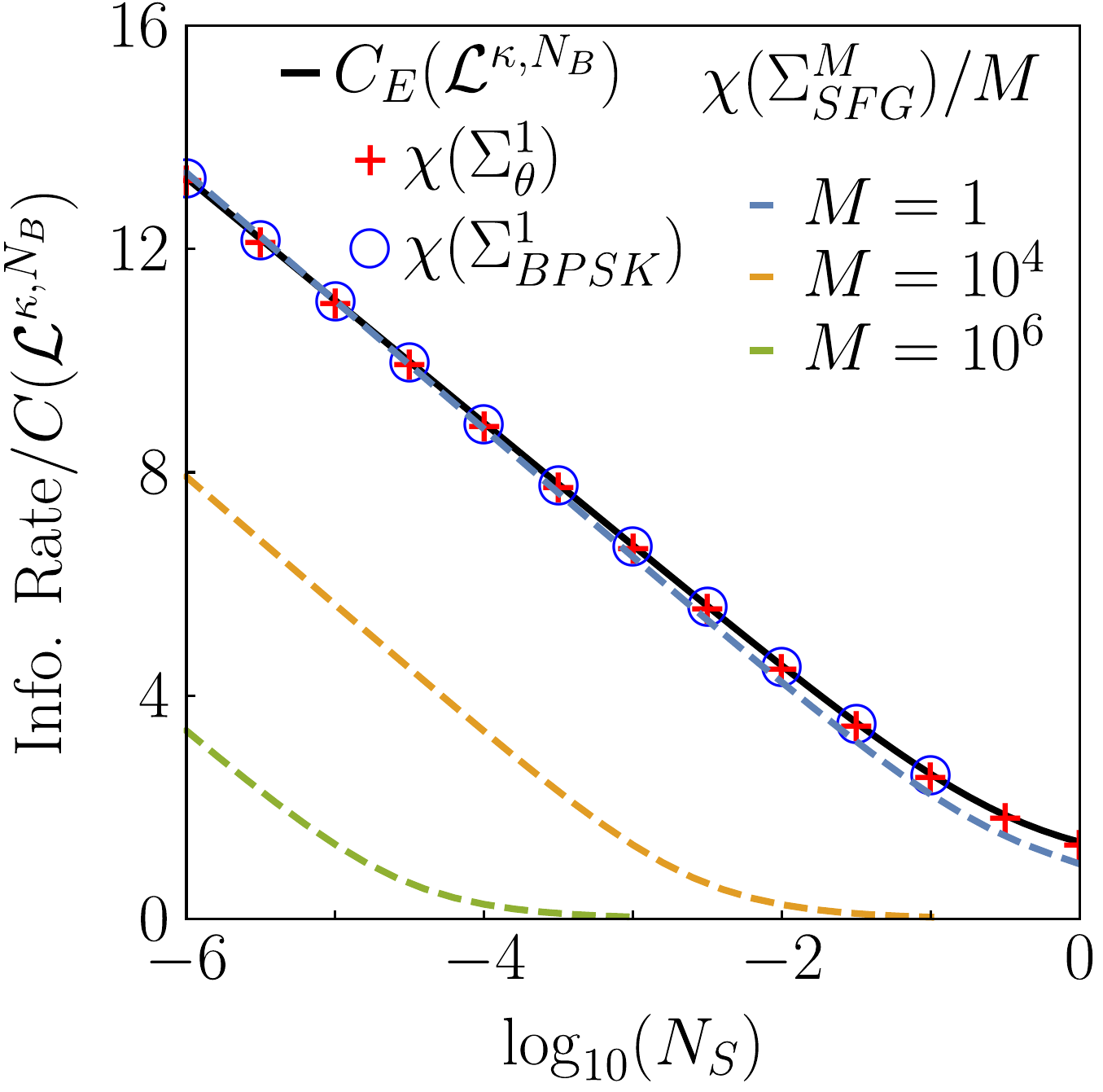}
\caption{The information (info.) rate divided by the unassisted classical capacity $C(\mathcal{L}^{\kappa,N_B})$ vs. the transmitted power $N_S$. Channel transmissivity $\kappa=0.1$ and noise $N_B=10$. The entanglement-assisted classical capacity $C_E(\mathcal{L}^{\kappa,N_B})$ (black solid line) has a large advantage when power is low, and the mode-wise phase encoding $\Sigma_{\theta}^1$ (red crosses) and $\Sigma_{BPSK}^1$ (blue circles) have Holevo information $\chi\left(\Sigma_{\theta}^1\right)$ and $\chi\left(\Sigma_{BPSK}^1\right)$ achieving the entanglement-assisted classical capacity (black solid). For phase modulation in $M$-mode blocks, the sum-frequency generation (SFG) process gives an estimation of the Holevo information per mode $\chi(\Sigma^M_{\rm SFG})/M$ for various repetition encoding block sizes $M$ (dashed lines). 
\label{Ce_over_C}
}
\end{figure}

Thus, the set of states at the receiver is given by 
$
\Sigma_{\theta}^1\equiv \Big\{ \hat{\rho}^\theta_{RI}, \theta\sim U[0,2\pi)\Big\},
$ 
where the phase $\theta$ is uniformly distributed. Under optimal decoding, the accessible information after the channel can be obtained by $\chi\left(\Sigma_{\theta}^1\right)$, where 
\be 
\chi\left(\{\hat{\rho}_x,p(x)\}\right)=S\left(\int_x p(x) \hat{\rho}_x\right)-\int_x p(x)S\left( \hat{\rho}_x\right)
\label{eq:Hol_info}
\ee
is the Holevo information and $S(\cdot)$ is the von Neumann entropy.

The conditional entropy $S(\hat{\rho}^\theta_{RI})$ can be straightforwardly calculated because the state is Gaussian~\cite{Weedbrook_2012}. The calculation for the unconditional entropy is nevertheless more involved, as detailed in Appendix~\ref{sec:uncond_entropy} and shown in Fig.~\ref{Ce_over_C} the numerical results in red crosses. By asymptotic expansion in the $N_B\gg1$ limit, we can show (details in Appendix~\ref{sec:uncond_entropy}) that
\ba 
&&\chi\left(\Sigma_{\theta}^1\right)=C_E(\mathcal{L}^{\kappa,N_B})+O(1/N_B^2)
\nonumber
\\
&&=\frac{1}{N_B}{\kappa N_S (1+N_S)\log_2 (1+\frac{1}{N_S})}+O(1/N_B^2).
\label{holevo_phase}
\ea 
Because phase encoding achieves the EA capacity $C_E(\mathcal{L}^{\kappa,N_B})$, it is the optimal encoding over a lossy and noisy bosonic channel in the asymptotic limit of $N_B\gg1$.

While continuous phase encoding is asymptotically optimal, encoding with a set of discrete phases are more practical in real-world operations.
As an example, Section \ref{sec:BPSK} demonstrates the binary phase-shift keying (BPSK) as a handy implementation that overcomes the classical capacity. In BPSK, the ensemble of the quantum states at the receiver is $\Sigma_{BPSK}^1=\{\hat{\rho}^\theta_{RI},\theta\sim U\{0,\pi\}$. Similarly, the Holevo information for $\Sigma_{BPSK}^1$ is calculated and depicted in Fig.~\ref{Ce_over_C}, showing the asymptotically optimalality for BPSK encoding in the large noise limit.

In many protocols such as quantum illumination and floodlight quantum key distribution~\cite{zhuang2016floodlight}, repetition coding of the same $\theta$ on $M$ signal-idler mode pairs, i.e., $\Sigma_{\theta}^M\equiv \{\otimes_{k=1}^M \hat{\rho}_{R_kI_k}^\theta, \theta\sim U[0,2\pi)\}$, is used to obtain sufficiently large mutual information per encoding so that efficient error correction codes can be employed. Let $M$ mode pairs be a phase modulation block, the derivation of the Holevo information per mode $\chi(\Sigma_{\theta}^M)/M$ is computationally challenging when $M\gg1$. However, we obtain a precise estimation of the information per mode $\chi(\Sigma^M_{\rm SFG})/M$ based on the results in Section~\ref{sec:SFG_est}, where a SFG process~\cite{zhuang2017} on the modes within each phase modulation block is devised. Figure~\ref{Ce_over_C} shows a good agreement between the estimation $\chi(\Sigma^M_{\rm SFG})/M$ (blue dashed) and the exact result $\chi(\Sigma_{\theta}^1)$ (red crosses) for $M=1$.

\subsection{Channel capacity with imperfect pre-shared entanglement distributed via a lossy channel}
\label{sec:noisy_preshare}
Previous analysis has assumed perfect pre-shared entanglement distributed through lossless and noiseless channels $\Phi_S$ and $\Phi_I$. It shows that phase encoding on perfect pre-shared TMSV states leads to a $\ln(1/N_S)$ capacity advantage in the weak signal and high noise regime. Although perfect pre-shared entanglement can be built up in a full-up quantum network in the future, current technology only allows for non-ideal distribution of entanglement. As such, imperfections in $\Phi_S$ and $\Phi_I$ need to be accounted for in a practical scenario.

Suppose the entangled TMSV signal-idler pairs are generated by Alice, so the idler distribution channel remains perfect, i.e., $\Phi_I = I$, while the signal is distributed to Bob through noiseless lossy channel $\Phi_S$, i.e., a pure-loss bosonic channel $\mathcal{L}^{\kappa_0,0}$. Since phase encoding $\hat{U}_\theta$ commutes with $\mathcal{L}^{\kappa_0,0}$, as verified by the input-output relation in Eq.~\ref{IO_relation} of a general thermal loss channel $\mathcal{L}^{\kappa,N_B}$ and by using the fact that thermal state is invariant under phase rotation, we may consider an equivalent protocol in which the initial signal modes $\hat{a}_{S^\prime}$'s are first phase encoded, subsequently go through the channels $\Phi_S=\mathcal{L}^{\kappa_0,0}$ and $\mathcal{L}^{\kappa,N_B}$ consecutively, and are finally received by Alice. In the equivalent protocol, the overall noisy channel 
\be 
\Phi_{\rm All}=\mathcal{L}^{\kappa,N_B}\circ \mathcal{L}^{\kappa_0,0}=\mathcal{L}^{\kappa_0\kappa,N_B}.
\ee 
To match the mean photon number going through the channel $\mathcal{L}^{\kappa,N_B}$ with the classical case, the mean photon number $N_{S^\prime}$ of $\hat{a}_{S^\prime}$ is constrained by $\kappa_0 N_{S^\prime}=N_S$.
Because the overall channel is again a lossy thermal channel, in the $\kappa\ll1, N_B\gg1$ limit, the accessible information at Alice's receiver reads
\ba 
&&\chi\left(\Sigma_{\theta,\kappa_0}^1\right)=C_E(\mathcal{L}^{\kappa_0\kappa, N_B}, N_S/\kappa_0)+O(1/N_B^2)
\nonumber
\\
&&=\frac{1}{N_B}\kappa N_S \left(1+\frac{N_S}{\kappa_0}\right)\log_2 \left(1+\frac{\kappa_0}{N_S}\right)+O(1/N_B^2),\nonumber\\
\ea 
where the input power $N_S/\kappa_0$ is made explicit.
Under $N_S/\kappa_0\ll1$, the accessible information $\chi\left(\Sigma_{\theta, \kappa_0}^1\right)$ remains a factor of $\ln(\kappa_0/N_S)$ larger than the unassisted capacity $C(\mathcal{L}^{\kappa,N_B}, N_S)$.

Note that prior to EA communication, pre-shared entanglement always needs to be built up via an entanglement-distribution channel. It thus behooves us to consider using the pure-loss entanglement distribution channel for classical communication without entanglement assistance, which is anticipated to outperform EA communication over a lossy and noisy channel. However, various scenarios preclude direct utilization of the entanglement-distribution channel for classical communication, as we enumerate two examples in the following. First, the performance of RF communications can be improved with pre-shared entanglement distributed via optical links. In this case, the RF-communication link is well modeled by a lossy thermal channel, while the optical entanglement distribution link is a pure-loss channel. Second, a pure-loss channel for entanglement establishment may not always be available. With long-lived quantum memories, entanglement can be distributed in the presence of an entanglement-distribution channel and is subsequently stored with high fidelity until retrieved on-demand for EA communication.

\subsection{Entanglement-assisted covert communication}
\label{sec:encoding_covertness}

\begin{figure}
\centering
\includegraphics[width=0.3\textwidth]{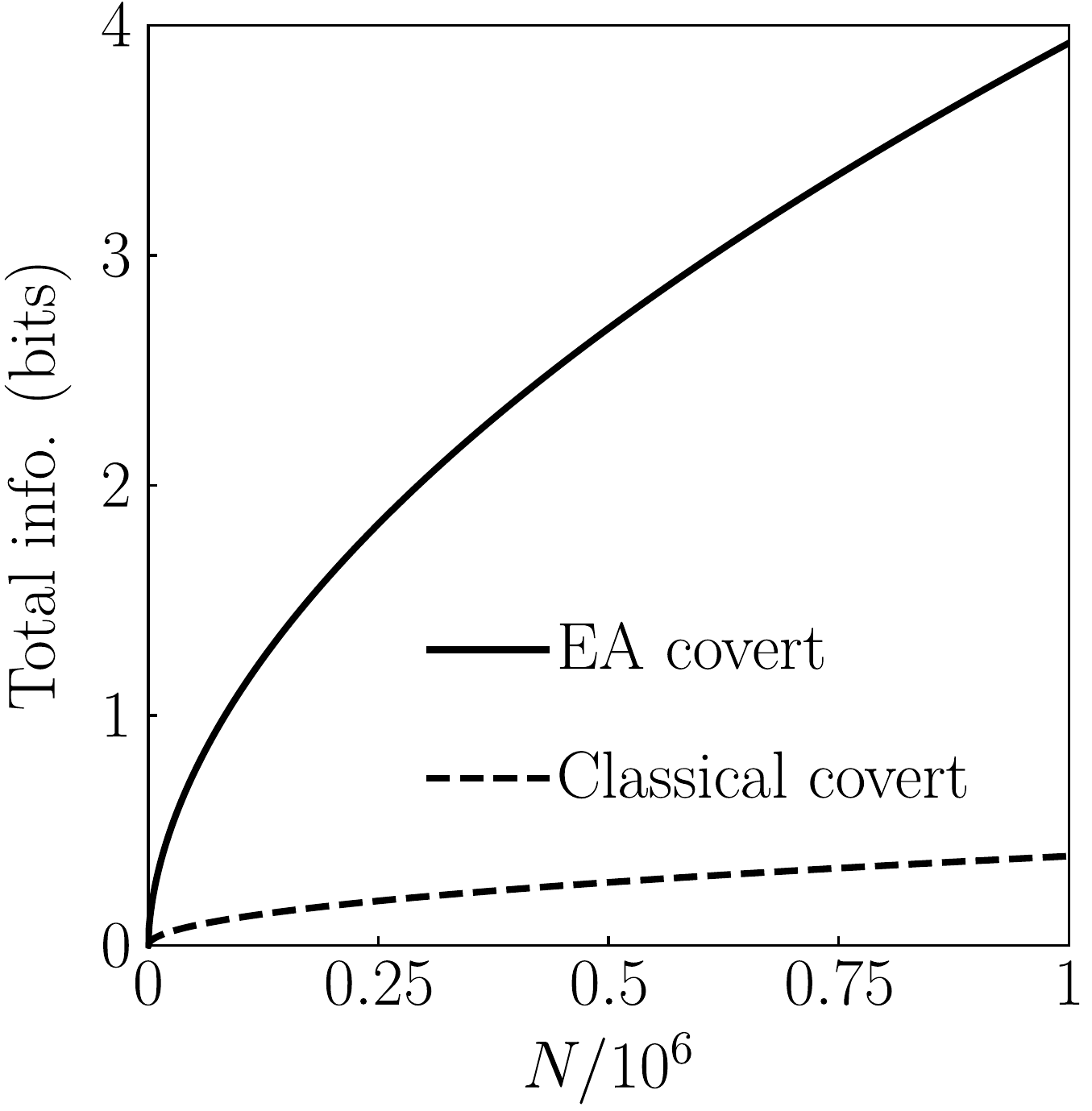}
\caption{Ultimate total covert information in bits transferred in $N$ modes by the entanglement-assisted (EA) covert communication and classical covert communication without EA. Channel transmissivity $\kappa=0.1$ and noise $N_B=10$, covertness $\delta=0.01$. 
\label{Ce_over_C_covert}
}
\end{figure}

An additional benefit of the EA communication protocol is its covertness and security~\cite{bash2015quantum,bullock2019fundamental}. 
Covert communication refers to the scenario that two parties are able to communicate, while the presence of the communication signals cannot be easily detected by any passive adversaries. It is possible when unavoidable environmental noise hides the weak signals. In the EA communication protocol, suppose that a passive adversary endeavors to detect Alice and Bob's communication attempt by monitoring the mode lost to the environment, but does not have access to the idler $\hat{a}_I$ since entanglement is pre-shared prior to communication. In the presence of EA communication with $N$ mode-pairs, the reduced state of the modes lost to the environment is a product of $N$ thermal states $\hat{\rho}_1$, each with mean photon number $n_1=\kappa N_B/(1-\kappa)+(1-\kappa)N_S\simeq \kappa N_B+N_S$, irrespective of the message being transmitted. In the absence of communication, the state $\hat{\rho}_0$ remains thermal with mean photon number $n_0=\kappa N_B/(1-\kappa)\simeq \kappa N_B$. With $\kappa N_B\gg1$, the difference between $\hat{\rho}_0$ and $\hat{\rho}_1$ is so small that communication covertness is warranted. The Helstrom bound of the error probability in distinguishing $\hat{\rho}_0^{\otimes N}$ from $\hat{\rho}_1^{\otimes N}$ can be numerically calculated as both states are diagonal in the number basis. Here, we use the quantum Chernoff bound~\cite{Pirandola2008,audenaert2007discriminating} to estimate the error probability of the adversary 
\be 
P_E\sim \exp\left[-NN_S^2/(8\kappa^2N_B^2)\right].
\ee
Under the requirement of $P_E\sim 1/2$, we can still communicate with $N\sim \kappa^2 N_B^2/N_S^2$ modes, which is large when $\kappa N_B\gg1$.

A more invovled calculation, similar to that in Ref.~\cite{bullock2019fundamental}, shows that under the requirement of $P_E\ge 1/2-\delta$, the relative entropy $D(\hat{\rho}_0^{\otimes N}\|\hat{\rho}_1^{\otimes N})\le 2\delta^2/\ln(2)$. Using the additivity of relative entropy and thermal state properties (or one can use Theorem 7 in Ref.~\cite{pirandola2017fundamental}), 
\begin{align}
&D(\hat{\rho}_0^{\otimes N}\|\hat{\rho}_1^{\otimes N})=N D(\hat{\rho}_0\|\hat{\rho}_1) \nonumber
\\
&=N \left\{\log_2 \left[\frac{n_1+1}{n_0+1}\right]+n_0\log_2 \left[\frac{n_0(n_1+1)}{n_1(n_0+1)}\right]\right\}.
\end{align}
Therefore, one has $N\le N_\delta\equiv 4\delta^2 \kappa N_B(\kappa N_B+1)/N_S^2+O(N_S^3)$. With a large $\kappa N_B$ and based on the capacity formula in Eq.~\ref{C_formula} and Eq.~\ref{CE_formula}, we expect the information transmitted in classical communication without EA to be $N_\delta C(\mathcal{L}^{\kappa,N_B})\simeq 4\kappa^3\delta^2 N_B/( N_S\ln(2))\sim \sqrt{N_\delta}\delta/\kappa^2$, which is often regarded as the square-root law for covert communication~\cite{bullock2019fundamental}. The EA communication, however, allows for transmitting a factor of $\ln(1/N_S)\sim \ln (N_\delta)$ more bits of information while maintaining the same level of covertness. The $\sqrt{N_\delta}\ln(N_\delta)$ scaling for EA covert communication thus breaks the square-root law for classical covert communication by a logarithmic factor, as illustrated by the example in Fig.~\ref{Ce_over_C_covert}.

Moreover, because the quantum states accessible to the adversary is identical for any encoded message, the adversary is unable to learn any information about the message. As such, the protocol is secure, as long as the pre-shared entanglement is perfect and the idler is retained securely in the Alice's laboratory. Note that the security is only conditioned on the entanglement being perfect, it does not rely on specific noise model for the communication channel. Pre-sharing perfect entanglement, however, is challenging in practice. Nonetheless, quantum internet is a technology under actively developed, and the challenges associated with entanglement distribution can be in principle overcome by quantum repeaters.

\section{Practical receiver structures}
\begin{figure}
\centering\includegraphics[width=0.34\textwidth]{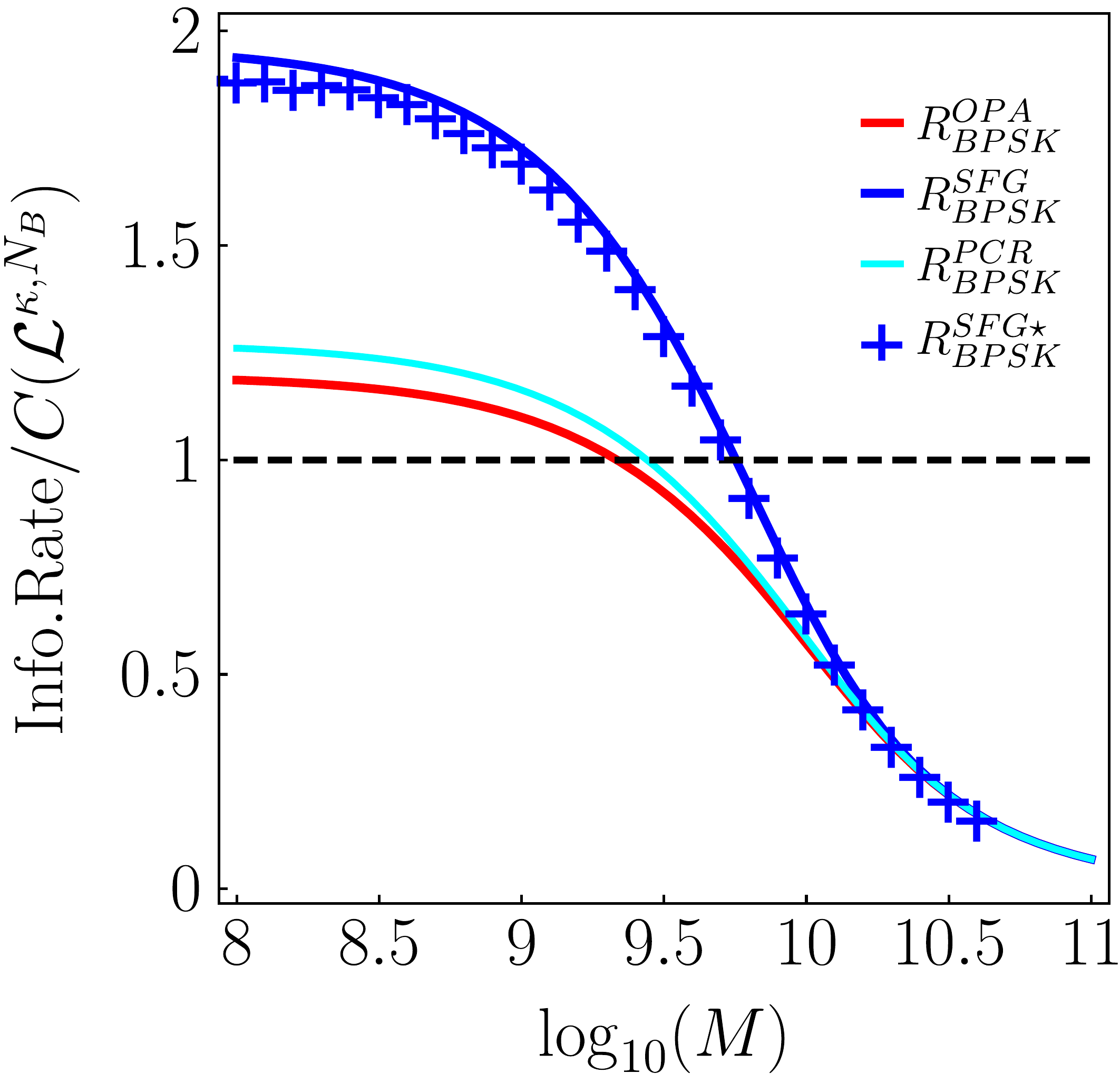}
\caption{Entanglement-assisted communication in comparison with classical communication. Information rates per mode normalized by the unassisted classical capacity $C(\mathcal{L}^{\kappa,N_B})$. The unassisted information rate $C(\mathcal{L}^{\kappa,N_B})$ is plotted in black dashed line as a benchmark. The plots are marked blue for sum-frequency generation (SFG) based receivers, cyan for phase conjugate receiver (PCR) and red for optical parametric amplifier (OPA) receiver. $\star$: The blue cross markers are the numerical result of Monte Carlo simulation of FF-SFG receiver with 50 adaptive measurement cycles and $8\times 10^5$ samples. 
Parameter: $N_s=10^{-3},N_B=10^4,\kappa=10^{-3}$. 
}
\label{fig:6curves}
\end{figure}

\subsection{Quantum receivers for discrete modulation and optimum hypothesis testing}
\label{sec:BPSK}
Section~\ref{sec:encoding_capacity} demonstrates the optimality of phase encoding in EA communication without specifying a structured receiver that approaches the channel capacity. In this section, we focus on practical receiver designs. To allow efficient error correction codes, we consider the repetition coding with the BPSK modulated state ensemble $\Sigma_{\rm BPSK}^M=\{\otimes_{k=1}^M \hat{\rho}_{R_kI_k}^\theta,\theta=0,\pi\}$. 
Formally, the decoding of BPSK may be viewed as a binary hypothesis testing task that discriminates two modulated phases $\theta=0,\pi$. Such a hypothesis testing task is similar to that of quantum illumination. It is known that quantum illumination's optical parametric amplifier (OPA) receiver and phase conjugate receiver (PCR)~\cite{Guha2009} both offer a 3-dB advantage in the error-probability exponent over that of the classical illumination, while the optimum quantum receiver offers a 6-dB error-probability exponent advantage. The advantage enabled by the OPA receiver has been demonstrated in a quantum-illumination experiment~\cite{Zheshen_15}. A more recent work discovered the optimum receiver~\cite{zhuang2017}, based on sum-frequency generation (SFG) and feed-forward (FF), to unleash quantum illumination's full advantage over the optimum classical scheme. Let the error probability of the symmetric hypothesis testing be $P_E$, the per-mode communication rate is given by 
\be
R_{P_E}=\frac{1}{M}\left(1+P_E\log_2 P_E+\left(1-P_E\right)\log_2\left(1-P_E\right)\right).
\label{I2}
\ee

The per-mode communication rates for the OPA receiver, the PCR, and the FF-SFG receiver in EA communication are evaluated and plotted in Fig.~\ref{fig:6curves}. 

\subsubsection{OPA receiver}
\label{sec:OPA}
\begin{figure}
\centering
\includegraphics[width=0.45\textwidth]{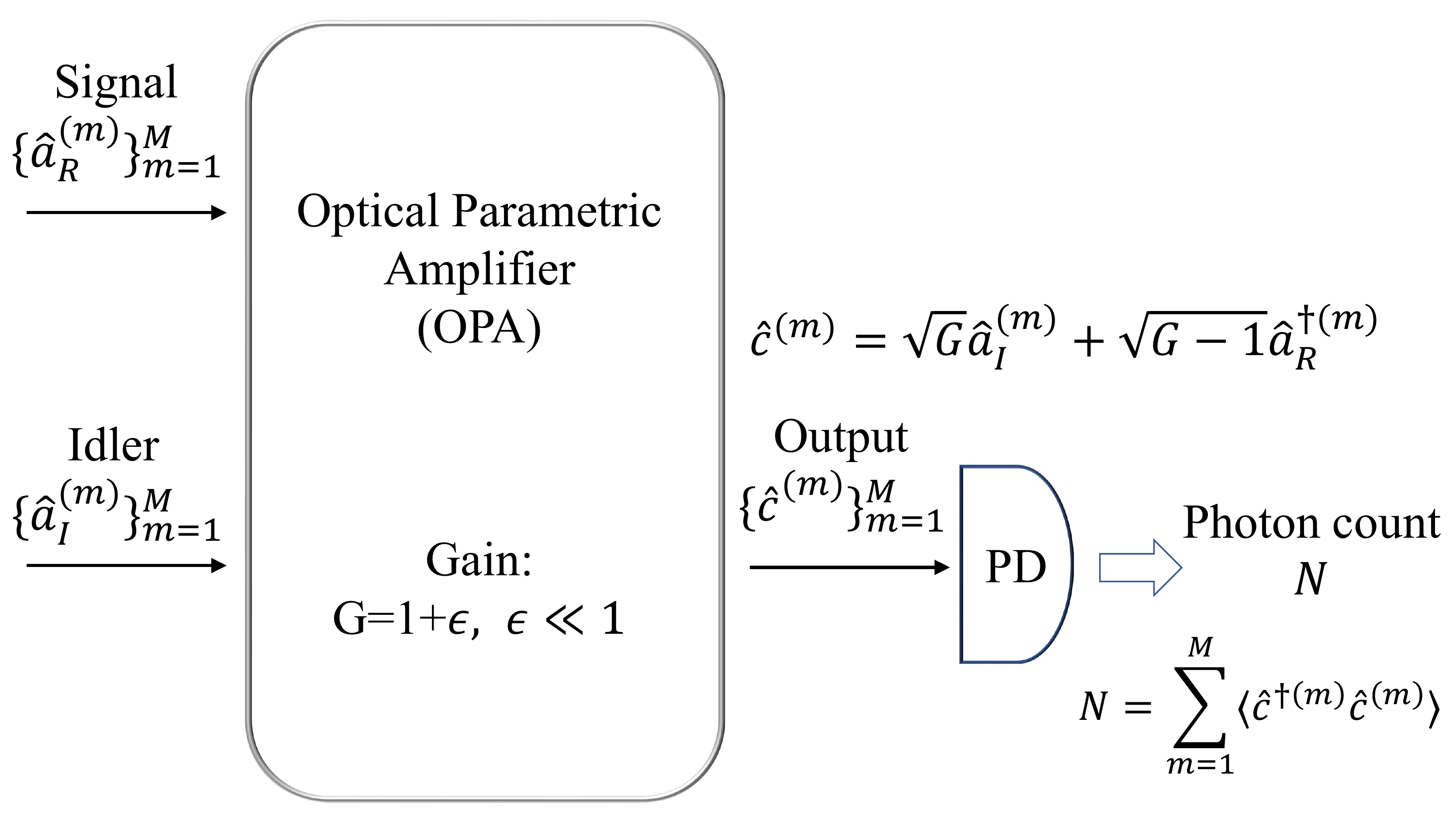}
\caption{The setup of optical parametric amplifier (OPA) receiver. PD: photo-detector. The returned signal $\hat a_R$ and idler $\hat a_I$ travel through the OPA ejecting amplified beams at two output ports, with the amplifier gain $G\simeq 1$. We collect photons at the port where the idler is amplified. 
}
\label{fig:OPAsetup}
\end{figure}

We first elaborate the OPA receiver (Fig.~\ref{fig:OPAsetup}). The OPA receiver applies parametric amplification across all returned-signal and retained-idler mode pairs $\{\hat a_{R}^{(m)},\hat a_{I}^{(m)}\}$ to transform the cross correlations between the input modes to photon-number differences. The two-mode squeezing in amplification produces the modes $\hat c^{(m)}=\sqrt{G} \hat a_{I}^{(m)}+\sqrt{G-1}\hat a_{R}^{\dagger(m)}$, with mean photon number 
\be
\bal
&\overline N(\theta)\equiv \expval{\hat c^{\dagger(m)} \hat c^{(m)}}\\
&\!=\!G N_S\!+\!(G\!-\!1)(\kappa N_S\!+\!N_B\!+\!1)\!+\!2\sqrt{G(G-1)}\cos\theta C_p
\label{eq:OPAmean}
\,,\eal\ee
for encoded phase $\theta$. The distribution of the total photon number across the $M$ modes can be obtained as
\be
P_{\rm OPA}(n|\theta; M)=\binom {n\!\!+\!\!M\!\!-\!\!1}{n}\!\left(\frac{\overline N(\theta)}{1+\overline N(\theta)}\right)^{\!n}\!\!\left(\frac{1}{1+\overline N(\theta)}\right)^{\!M}.
\label{Pn_OPA}
\ee

Given the conditional probability distribution, we evaluate the performance of maximum-likelihood decision in Appendix.~\ref{sec:OPAappendix}, with the error probability shown in Fig.~\ref{fig:DolinarPe} and the communication rate plotted in Fig.~\ref{fig:6curves}. An ideal OPA receiver applied on BPSK-encoded TMSV source (red line) beats the classical capacity by $\sim 18.6\%$ at $M=10^8$ and $\sim 10.0\%$ at $M=10^9$. As the number of modes $M$ in the repetition block increases, the rate per-mode decreases as expected. Note that the normalization $C(\mathcal{L}^{\kappa,N_B})$ does not change with $M$.

\subsubsection{Phase-conjugate receiver}
\begin{figure}
\centering
\includegraphics[width=0.51\textwidth]{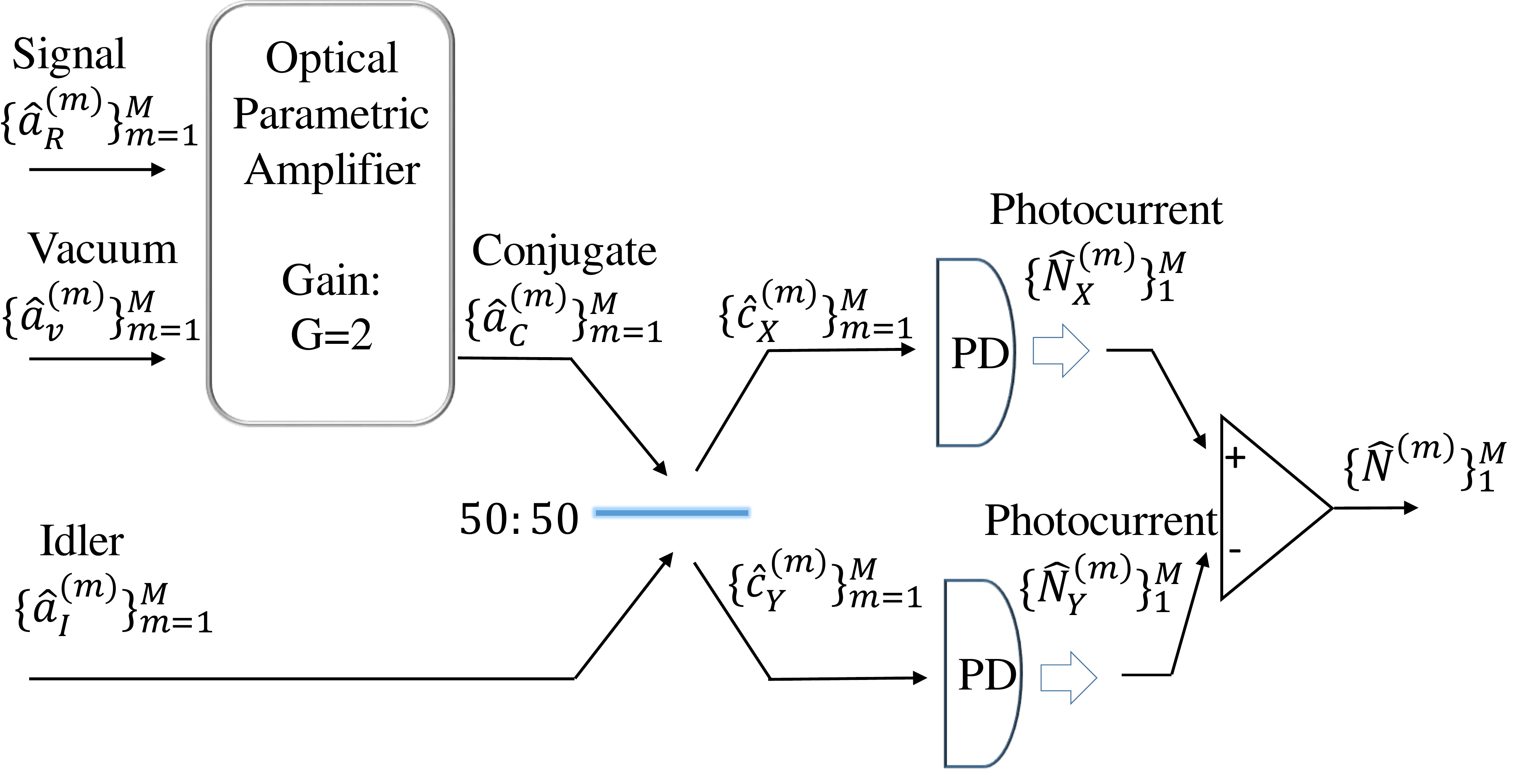}
\caption{The setup of a phase conjugate receiver (PCR). The returned signal $\hat a_R$ travels through an OPA with the amplifier gain $G=2$. The phase-conjugated field appears at the empty port, interfering with the idler $\hat a_I$ through a 50:50 beam splitter. The photon counts of the two arms of the interferometer are collected to derive the differential photon counts, based on which the message is decoded.}
\label{fig:PCRsetup}
\end{figure}

The PCR (Fig.~\ref{fig:PCRsetup}), a variant of the OPA receiver, reaches the same asymptotic error exponent with $N_S\ll1, N_B\gg 1$ but yields a slight advantage with nonzero $N_S$'s (see Fig.~\ref{fig:6curves}). The PCR conjugates the $M$ input modes $\hat a_R^{(m)}$ while amplifying the vacuum $\hat a_v^{(m)}$ at the empty port, i.e.,
\be
\hat a_C^{(m)}=\sqrt 2 \hat a_v^{(m)}+\hat a_R^{\dagger (m)}
\,.\ee
Then, the conjugated signal along with the idler is detected by a balanced difference detector with the photon count $\hat N^{(m)}=\hat N_X^{(m)}-\hat N_Y^{(m)}$, where $\hat N_X^{(m)}, \hat N_Y^{(m)}$ are the photon count of the two outputs of a 50-50 beam splitter: $\hat c_X^{(m)}=(\hat a_C^{(m)}+\hat a_I^{(m)})/\sqrt 2$, $\hat c_Y^{(m)}=(\hat a_C^{(m)}-\hat a_I^{(m)})/\sqrt 2$. In analogy to the OPA receiver, the decision is made according to the total photon count across the $M$ modes. 

A detailed analysis on the communication performance in Appendix~\ref{sec:PCRappendix} shows that PCR has a slight edge on the OPA receiver for that its signal-to-noise ratio is better in higher order terms. Illustrated by the cyan line in Fig.~\ref{fig:6curves}, an ideal PCR with BPSK encoded TMSV source overcomes the classical capacity by $\sim 26.0\%$ at $M=10^8$ and $\sim 16.3\%$ at $M=10^9$.

\subsubsection{FF-SFG receiver}
\label{sec:FFSFG}

\begin{figure}
\centering\includegraphics[width=0.49\textwidth]{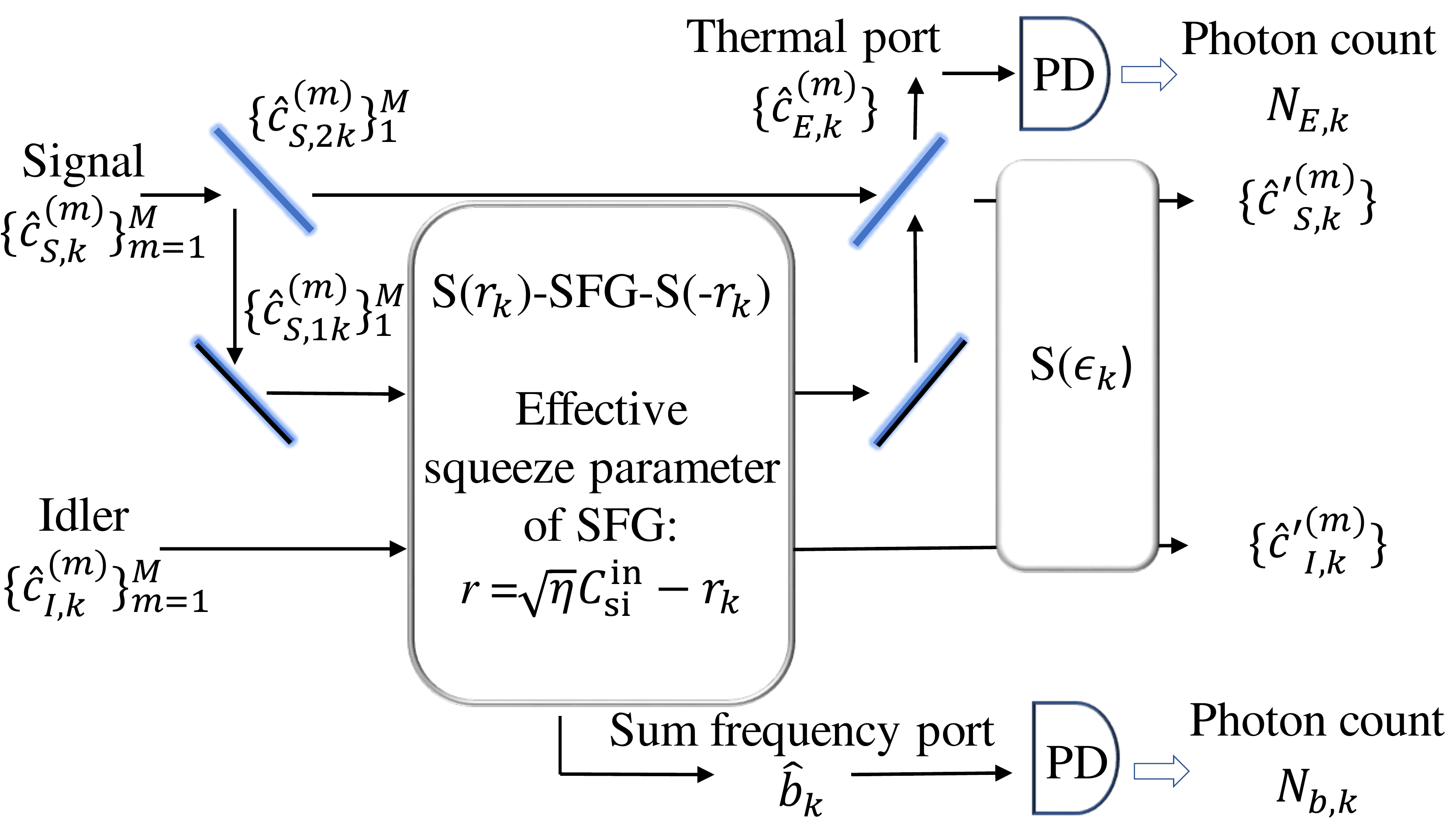}
\caption{The setup of a single cycle of the feed-forward (FF) sum-frequency generation (SFG) receiver. The signal in $k$th cycle $\hat c_{S,k}$ is first divided into a bright mainstream $\hat c_{S,2k}$ and a weak slice $\hat c_{S,1k}$ by a highly transmissive beam splitter with reflectivity $\sqrt\eta\ll 1$. The weak slice goes through a FF-SFG module containing three processes in sequence: $S(r_k), {\rm SFG}, S(-r_k)$, with the phase of squeeze parameter $r_k$ adaptively tuned. Eventually the processed weak slice $\hat c'_{S,1k}$ is merged back to the mainstream by a second highly transmissive beam splitter. We collect the photon counts at the sum frequency port of SFG and the thermal port of the second beam splitter.
}
\label{fig:FFSFGsetup}
\end{figure}

The FF-SFG receiver (Fig.~\ref{fig:FFSFGsetup}) improves the performance of the OPA receiver and is the optimum for quantum illumination in the strong noise and weak signal limit. Through an SFG process, 
the FF-SFG receiver converts the cross correlations between the signal-idler pairs and produces quantum states with the photon number statistics approximating a coherent state. Thus, by analogy with the Dolinar receiver, the optimum receiver for binary coherent-state discrimination, the FF-SFG receiver asymptotically achieves the quantum Chernoff bound for quantum illumination. The principle of the FF-SFG receiver is briefly introduced below (more details in supplemental materials of Ref.~\cite{zhuang2017}).

The FF-SFG receiver consists of a sequence of multiple cycles of adaptive detection. The measurement results of all previous cycles are combined through a Bayesian strategy that produces a posterior distribution of different hypotheses. In the $k$-th cycle, the prior probabilities $P^{(k)}_0,P^{(k)}_1$ for the hypotheses $\theta_0=0,\theta_1=\pi$ are used to design the measurements, whose results are used to obtain the posteriors (also the priors of $k+1$-th cycle) $P^{(k+1)}_0,P^{(k+1)}_1$ through a Bayesian formula. Denote the maximum-likelihood decision before the cycle as $\tilde{h}=\arg\max_\ell P^{(k)}_\ell$, while the true hypothesis is $h$. As shown in Fig.~\ref{fig:FFSFGsetup}, the FF-SFG slices a $\eta\ll1$ portion of the strong returned-signal modes $\hat{c}_{S,k}^{(m)}$'s to interact with the weak idler modes $\hat{c}_{I,k}^{(m)}$'s through a SFG process to produce a sum mode $\hat{b}_k$ for detection. Denote the cross correlation between $\hat{c}_{S,k}^{(m)}$ and $\hat{c}_{I,k}^{(m)}$ as $C_{si,k}^{\rm in}$. The interaction consists of: (1) two adaptively tuned two-mode squeezing modules $\hat{S}(r_k)$ and $\hat{S}(-r_k)$ that change the cross correlation, adopting the same feed-forward strategy as in the Dolinar receiver; (2) an SFG process that converts the cross correlation into a sum-frequency mode $\hat{b}_k$. The sum-frequency mode is approximately in a coherent state $\ket{e^{i\theta_h}\sqrt{M}r}$ with $r=\sqrt\eta C_{si,k}^{\rm in}-r_k$ plus thermal noise $\eta N_SN_B$, which is to be measured by photon counting. This presents an analogy to the Dolinar receiver~\cite{dolinar_processing_1973}, which, based on the maximum-likelihood decision $\tilde{h}$, chooses a $r_k$ to displace the coherent state to a near-vacuum state, i.e., $r \sim 0$, for optimum state discrimination at the Helstrom limit (more details in Appendix.~\ref{sec:SFGappendix}).

Subsequently, the sliced signal modes are recombined with the other part of the signal modes, forming an interferometer structure. An $M$-mode thermal state of $\hat{c}_{E,k}^{(m)}$'s is generated with the same mean photon number $M|r|^2$ at the dim port, which is thereby denoted as the \emph{thermal port}. The total number of photons at the thermal port is measured as well. 

Finally, the bright output goes through an additional two-mode squeezing $\hat{S}(\epsilon_k)$ that wipes out the $r_k$ dependence in the evolution of the cross correlation. The evolution is terminated when the cross correlation has been almost used up, i.e., when the residual cross-correlation is only a $\epsilon\ll1$ portion of the initial cross correlation.

\begin{figure}
\centering
\includegraphics[width=0.31\textwidth]{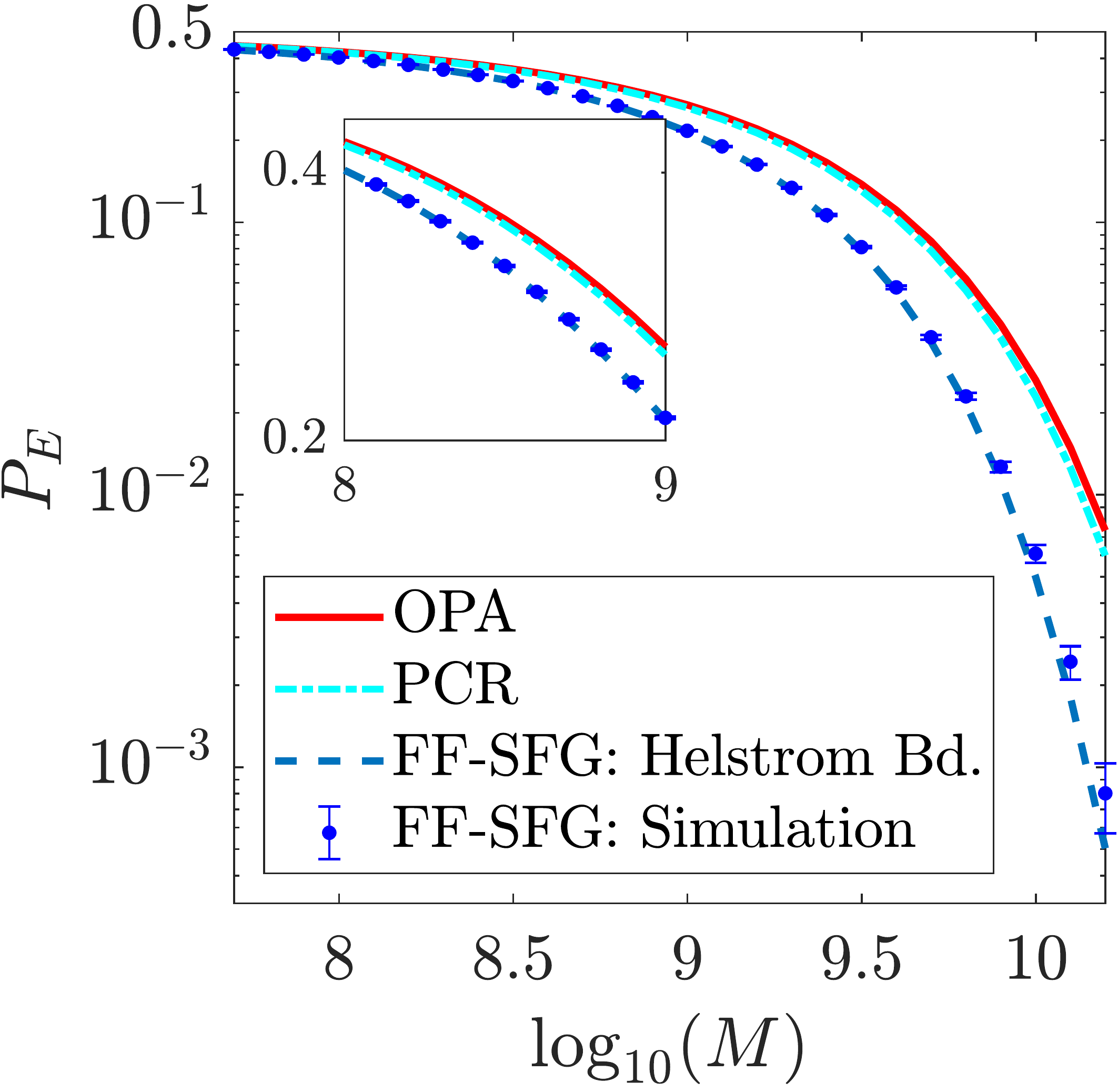}
\caption{The error probability of hypothesis testing between two encoded phases for optical parametric amplifier (OPA) receiver (red), phase conjugate receiver (PCR) (cyan), and feed-forward sum-frequency generation (FF-SFG) receiver (dashed line: theoretical bound; dots: numerical simulation with error bars). The dots are from Monte Carlo simulation with $8\times 10^5$ samples, which saturates the Helstrom bound. The inset is a zoom-in around error probability half, with error probability in linear scale, showing the synchronicity better.
Parameter: $N_S=10^{-3}, N_B=10^4, \kappa=10^{-3}, \eta =4\times 10^{-6}$. 
\label{fig:DolinarPe}
}
\end{figure}

Similar to the results for target detection~\cite{zhuang2017}, the FF-SFG receiver also demonstrates its optimality for phase discrimination. Monte-Carlo simulations on the FF-SFG receiver are performed under various parameters for EA communication as shown in Fig.~\ref{fig:DolinarPe}, in which the Helstrom limit under a uniform prior is estimated. From the error probabilities, the communication rate can be evaluated by Eq.~\ref{I2}. As is indicated by the blue stars in Fig.~\ref{fig:6curves}, an FF-SFG receiver with BPSK encoded TMSV source overwhelms the classical capacity by an advantage of $\sim 90\%$ for $M=10^8$ and $\sim 71\%$ for $M=10^9$. 
\subsection{Quantum receivers for continuous encoding and noisy phase estimation}
\label{sec:phase_estimation}

Although the BPSK encoding is handy for practical communications, its capacity is intrinsically bounded by one bit per symbol. This rapidly undermines the EA communication advantage as number of modes $M$ in a repetition block increases, as shown in Fig.~\ref{fig:6curves}. An immediate solution is increasing the alphabet size in the phase modulation. Continuous phase encoding is the limiting case when the alphabet size approaches infinity. With continuous phase encoding, decoding becomes a parameter estimation problem, in which one endeavors to acquire an estimation $\tilde{\theta}$ of the encoded phase $\theta$ based on the received state in the ensemble $\Sigma_{\theta}^M$. The conditional distribution $P(\tilde{\theta}|\theta)$ describes the measurement statistics. Since the encoding $\theta$ is uniformly distributed in $[0,2\pi)$, the per-mode communication rate reads
\begin{align}
&R_{P(\cdot|\cdot)}\!=\!
\nonumber
\\
&\frac{1}{M}\left(\log_2(2\pi)+\int_{0}^{2\pi}\!\!\frac{d\theta}{2\pi}\!\int_{0}^{2\pi}\!d\tilde\theta\! P(\tilde{\theta}|\theta) \log_2 P(\tilde{\theta}|\theta)\right).
\label{para_est_rate}
\end{align}

\begin{figure}[tbp]
\centering
\includegraphics[width=0.31\textwidth]{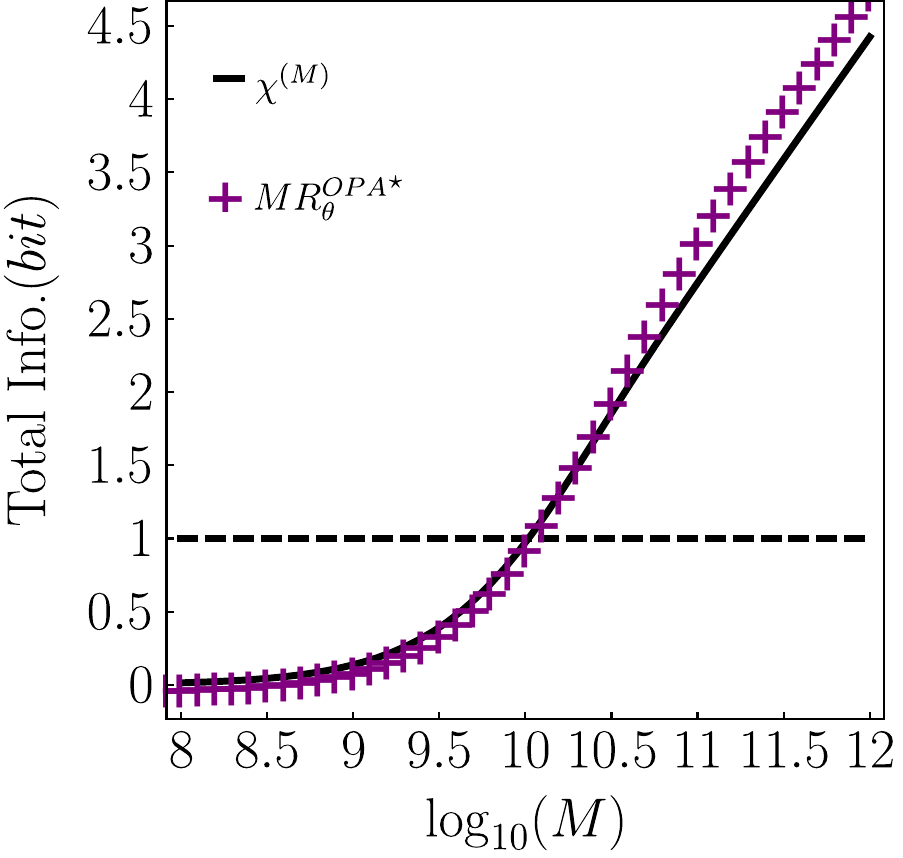}
\caption{The performance of continuous phase encoding with adaptive optical parametric amplifier (OPA) receiver. Purple crosses: $M$-mode information rate $MR_\theta^{\rm OPA}$ of OPA; black line: M-mode unassisted phase-encoding information rate $\chi^{(M)}$, repetition-coded with phase encoding. The 1 bit information upper bound of BPSK codes is plotted in dashed line as a benchmark. $\star$: Based on Monte Carlo simulation with 50 adaptive measurement cycles and $8\times 10^5$ samples. Parameter: $N_s=10^{-3},N_B=10^4,\kappa=10^{-3}$.}
\label{fig:2curves}
\end{figure}

The decoding of continuous phase requires a design of phase estimation in presence of large noise $N_B\gg1$. To this end, we first show that the TMSV is the asymptotic optimal input state for noisy phase estimation in the limit of strong noise and weak signal, as it maximizes the QFI among all states (see details in Appendix~\ref{App:adaptive_QFI}). Moreover, we derive an adaptive version of the OPA receiver which is the asymptotic optimal receiver for phase estimation with the TMSV state, as it saturates the QFI in the limit of strong noise and weak signal. Combining these results, a noisy phase-estimation protocol that is asymptotically optimal at the large number of copies limit is devised.

Unfortunately, noisy phase estimation operating in the large number of copies limit ($M\to\infty$) cannot be used as a decoding strategy, because it leads to a zero per-mode rate. Therefore, it is crucial to optimize the phase-estimation performance with finite or even a small number of modes. In Appendix~\ref{App:adaptive_Rx}, two adaptive receiver designs are presented. The basic idea is to introduce a sequence of measurements, each performed on a subset of $M_l$ modes. The setting for each measurement is determined by a prior probability distribution updated based on previous measurement results through the maximum Fisher information or maximum Van Trees information approach~\cite{van2004detection,martinez2017VanTrees,paris2009VanTrees}. We find the Van Trees approach gives much better performance.

Using the Van Trees approach, the total information rate is calculated using Eq.~\ref{para_est_rate}, and the results are depicted in Fig.~\ref{fig:2curves}. 
In the context of repetition coding, we take the $M$-mode unassisted phase-encoded Holevo information $\chi^{(M)}$ as a benchmark, assuming repetition coding of identical phase-encoded coherent states in $M$-mode blocks and no entanglement assistance (see Appendix~\ref{sec:stat} for details). Overall, in the region where BPSK saturates the one-bit bound, an extended practical EA advantage enabled by continuous encoding is observed over the repetition-phase-encoded classical communication performance $\chi^{(M)}$. Although the current numerical-simulation result for the adaptive OPA receiver shows no EA advantage over the unassisted classical capacity without repetition coding, a systematic optimization on the finite-copy phase-estimation protocol may further improve the EA performance.

\section{Experimental design}
\label{sec:exp}

\begin{figure}
\centering
\includegraphics[width=0.46\textwidth]{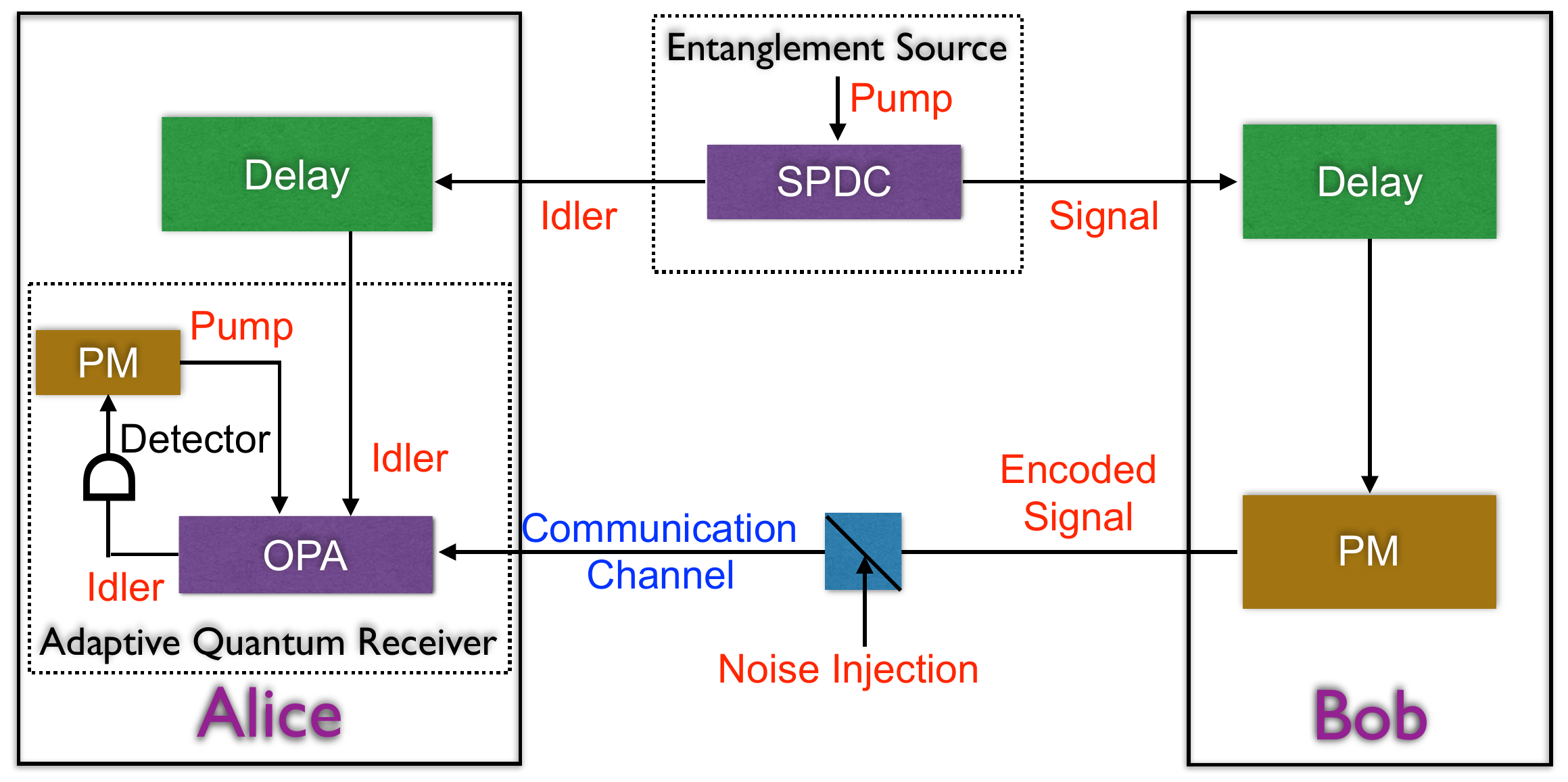}
\caption{The experimental setup of EA communication with an adaptive optical parametric amplifier (OPA) receiver. Broadband entangled signal and idler pairs are generated via spontaneous parametric down conversion (SPDC) and distributed to Bob and Alice. Bob employs phase modulation (PM) to encode on the signal and then sends the encoded photons to Alice. Alice applies an adaptive OPA receiver on the her retained idler and received signal to decode Bob's message. In the adaptive OPA receiver, the phase of the pump is adjusted based on prior measurement outcomes to achieve the optimal performance.}
\label{fig:experiment}
\end{figure}

A proof-of-concept experiment using the adaptive OPA receiver to beat the Holevo classical capacity can be readily built with off-the-shelf components, as conceptually illustrated in Fig.~\ref{fig:experiment}. Similar to the quantum illumination experiment~\cite{zhang2015}, broadband entanglement from spontaneous parametric down conversion (SPDC) can be generated and employed as the signal and the idler. A loosely focused pump is needed to achieve a > $99\%$ collection efficiency for the entanglement source. The idler photons can be stored in a spool of optical fibers with an efficiency in excess of $95\%$. Other experimental imperfections include free-space-to-fiber coupling loss (< $5\%$), detector loss ($1-\eta_D<2\%$), and filter losses (< $10\%$), which contributes to an overall exess loss on the signal $1-\kappa_S\sim 15\%$ and idler $1-\kappa_I\sim 15\%$ (combining the storage loss and filter loss). The noisy and lossy channel is usually induced by an adversary in a contested environment, which can be emulated by a beam splitter and a power-tunable amplified spontaneous emission source, e.g., an erbium-doped fiber amplifier, to deliver a $N_B$ up to $500 \times 10^3$. 

The adaptive OPA receiver can be realized by a field-programmable gate array (FPGA) that processes real-time detector output with > $100$-MHz bandwidth, capable of generating a feed-forward signal within $\sim$ $100$ ns. In conjunction with a $20$-GHz electro-optic phase modulator that controls the pump phase, the response time of the adaptive OPA receiver is sufficient to cope with 1 kbit/s communication rate, corresponding to $M = 2\times 10^9$. This experimental platform also allows for the demonstration of the optimal noisy phase estimation protocol described in Appendix~\ref{App:adaptive_Rx}.

To analyze the communication key rate, we include the extra losses $1-\kappa_I$, $1-\kappa_S$ and detector inefficiency $1-\eta_D$ in the theory analysis. We will focus on BPSK, which is easier to implement. The analysis is in parallel to section~\ref{sec:OPA}; With the imperfections, the mean photon count in Eq.~\ref{eq:OPAmean} changes to
$
\o N'(\theta)\!\!=\eta_D\ [\ G \kappa_I N_S+(G-1)(\kappa_S \kappa N_S+\kappa_S N_B+1)+2\sqrt{G(G-\!1)}\cos\theta\! \sqrt{\kappa_I\kappa_S} C_p\ ]
$.
As a result the optimum gain shifts to $G'=1+{\sqrt{\kappa_I N_S}}/{\kappa_S N_B}$. The distribution of the total photon number across $M$ modes is still given by Eq.~\ref{Pn_OPA}, with the new mean $\o N'(\theta)$.
With some algebra, we find that the variable inside the error function in Eq.~\ref{PE_OPA} is a factor of $\sqrt{\kappa_I\eta_D}$ smaller than the ideal case. It is independent of the excess signal loss because the large noise background $N_B$. 
As an example, at $M=10^9$ and using the same parameters as Fig.~\ref{fig:6curves}, $P_{E}^{\rm OPA \prime}={\rm erfc}(0.43\sqrt{\kappa_I\eta_D})/2$, where erfc denotes the complementary error function. To beat the classical capacity $C(\mathcal{L}^{\kappa,N_B})$, the efficiencies need to satisfy $\kappa_I\eta_D\gtrsim 90\%$. To reach this threshold, the efficiencies need to be improved upon the ones in the previous experiment~\cite{zhang2015}. In particular, if we replace the filter with a free space filter, the filter loss can be reduced to $<1\%$, thus leading to $1-\kappa_I\sim 5\%$ and $1-\eta_D\sim2\%$. In this case, the communication rate can have an advantage of $3\%$ over the ultimate unassisted classical capacity. When it comes to $M=10^8$, $P_{E}^{\rm OPA \prime}={\rm erfc}(0.14\sqrt{\kappa_I\eta_D})/2$. Under these parameters, the required efficiencies are subject to $\kappa_I\eta_D\gtrsim 84\%$. With the same loss $1-\kappa_I\sim 5\%$, $1-\eta_D\sim2\%$, the remaining advantage rises to $\sim10\%$.

\section{Blueprints for joint receivers}
\label{sec:SFG_est}
Before concluding, we point out some future directions of joint receiver design, via combining FF-SFG receiver and other receivers, for EA communication.
As shown in Ref.~\cite{zhuang2017}, conditioned on the encoded phase $\theta$ and at the $N_S\to 0$ limit, one effectively deals with a displaced thermal state $\hat{\rho}_{\lambda,n_e}^\theta$ with mean $\lambda=e^{i\theta} \sqrt{\kappa\left(1-\epsilon\right)MN_S(N_S+1)/(N_B+1)}$ and thermal noise $n_e\simeq N_S\ln\left(1/\epsilon\right)/2$ at the two output ports of the FF-SFG receiver. As explained in Section~\ref{sec:encoding}, the overall Holevo information for the repetition encoded ensemble $\Sigma_{\theta}^M$ is difficult to calculate. As an estimation, the Holevo information of the ensemble $\Sigma^M_{\rm SFG}= \{\hat{\rho}_{\lambda,n_e}^\theta, \theta\sim U[0,2\pi) \}$ is calculated. Although this is not the exact Holevo information of $\Sigma_{\theta}^M$, since the equivalence of the quantum states in the ensemble is only effective for SFG receiver's performance evaluation, one can still obtain interesting observations from this estimation.

The Holevo information of $\Sigma^M_{\rm SFG}$ can be efficiently calculated (details in Appendix.~\ref{sec:stat}).
The result for $\epsilon=0.05$ is in Fig.~\ref{Ce_over_C}. At $M=1$, this estimation well agrees with the exact result $\chi(\Sigma_\theta^1)$ and also reaches the EA capacity $C_E(\mathcal{L}^{\kappa,N_B})$. As $M$ increases, the per-mode Holevo information decreases, as expected. Nonetheless, the advantage over the classical capacity survives even at $M>10^5$.

This analogy inspires us to consider a concatenation of the FF-SFG receiver with Holevo-capacity-achieving receivers for classical communication, like the joint-detection receivers designed in Refs.~\cite{guha2011structured,wilde2012explicit}. While the FF-SFG receiver transforms the EA communication detection into a coherent state detection problem, the joint receiver optimally extracts information from the coherent states. A complete design for such a receiver is subject to future work.

\section{Conclusion}

In conclusion, we proposed a structured encoding and decoding devices to achieve EA advantages in communication over noisy bosonic channels. We showed that phase encoding on TMSV is asymptotically optimal as the noise increases. In particular, a simple BPSK encoding approaches the optimum Holevo information. In addition to offering higher-than-classical communication rates, the EA communication protocol is secure when the pre-shared entanglement is perfect and it beats the fundamental limit of covert communication without EA. 

We also showed that the practical repetition coding, e.g., on frequency modes, maintains a $\ln(1/N_S)$ rate advantage, even though the per-mode communication rate decreases. Moreover, with only lossy pre-shared entanglement available at the current stage of technology with no feasible quantum network, we showed that a slightly smaller advantage remains. For repetitive BPSK encoding, we analyzed practical receivers that offer a constant advantage over the classical capacity in the low signal power regime. 
For continuous phase encoding, we showed that TMSV with the practical receivers is asymptotically optimum for noisy phase estimation, in the high noise and large copies region. To optimize its parameter estimation performance with a finite copies of states, we developed adaptive Bayesian Van Trees phase estimation schemes, with fast convergence to the quantum Cram\'{e}r-Rao bound. However, the finite number of states effect prevents any quantum advantage, of which the optimization is still an open question. Nevertheless, the results on repetition coding provides a straightforward way to implement communications with a practically correctable error rate.

\begin{acknowledgements}
This research was sponsored by the Army Research Office and was accomplished under Grant Number W911NF-19-1-0418. The views and conclusions contained in this document are those of the authors and should not be interpreted as representing the official policies, either expressed or implied, of the Army Research Office or the U.S. Government. The U.S. Government is authorized to reproduce and distribute reprints for Government purposes notwithstanding any copyright notation herein. Z.Z. acknowledges National Science Foundation CCF-1907918. Q.Z. and Z.Z. also acknowledge the University of Arizona for support. Q.Z. acknowledges Boulat Bash, Saikat Guha, Jeffrey Shapiro and Nicolas Cerf for discussions. The authors also thank the comments from the anonymous referee which greatly improved the paper.
\end{acknowledgements}

\appendix

\section{Classical communication with coherent states}
\label{App:classical}

Classical communication protocols transmit coherent states $\ket\alpha$, encoding information on their real and imaginary quadratures with bounded average photon number $\o{|\alpha|^2}=N_S$. Then the transmitted states suffer thermal noises modeled by the thermal-loss channel $\mathcal L^{\kappa, N_B}$ channel with transmissivity $\kappa$ and noise $N_B$. Finally at the receiver the quadratures of the noisy coherent states are measured by homodyne or heterodyne, producing Gaussian measurement statistics. The information rate can be obtained through the Shannon capacity. For homodyne, the information is encoded on a single quadrature, thereby the average signal power is $4\kappa N_S$ and the white noise is $2N_B+1$; for heterodyne, both two quadratures are encoded so there are effectively two white-noise channels, each with average signal power $\kappa N_S$ (a factor of half from dividing the encoding power to two quadratures and a factor of two from heterodyne splitting) and noise $(2N_B+1)/2+1/2=N_B+1$. Then the information rate is given by (also can be found in Ref.~\cite{banaszek2019approaching})
\be
\bal
C_{\rm hom}&=\frac{1}{2}{\rm log}_2\left(1+\frac{4\kappa N_S}{1+2N_B}\right) \\
C_{\rm het}&={\rm log}_2\left(1+\frac{\kappa N_S}{1+N_B}\right) 
\eal
\,.\ee
In the asymptotical limit of $\kappa N_S\gg1$, we have
\be
C_{\rm het}/C(\mathcal L^{\kappa, N_B})=1+O(1/\ln(\kappa N_S),N_B).
\ee
In the asymptotic limit $\kappa N_S\ll N_B$, we have 
\be
C_{\rm hom}\simeq C_{\rm het}\simeq C(\mathcal L^{\kappa, N_B})\simeq \frac{\kappa N_S}{N_B{\rm ln}2 }.
\label{eq:Casym}
\ee

\section{Accessible information of TMSV with continuous phase encoding }
\label{sec:uncond_entropy}

The accessible (Holevo) information after the channel can be obtained by 
\be 
\chi\left(\Sigma_{\theta}^1\right)=S(\int_0^{2\pi}d\theta \hat{\rho}^\theta_{RI}/2\pi)-\int_0^{2\pi} d\theta S( \hat{\rho}^\theta_{RI})/2\pi
\label{chi_phase}
\ee 

The conditional entropy $S(\hat{\rho}^\theta_{RI})$ can be straightforwardly calculated because the state is Gaussian~\cite{Weedbrook_2012}. And note that $\hat{\rho}^\theta_{RI}=(\hat{U}_\theta \otimes \hat{I})\hat{\rho}_{RI}(\hat{U}_\theta^\dagger \otimes \hat{I})$, where $\hat{\rho}_{RI}\equiv \mathcal{L}^{\kappa,N_B}\left[ \hat{\psi}^{N_S}_{SI} \right]$ has covariance matrix ${\mathbf{\Lambda}}_{\theta=0}$ in Eq.~\ref{hk}.

Thus the unconditional term 
\ba 
&&\int_0^{2\pi} d\theta S( \hat{\rho}^\theta_{RI})/2\pi
=S(\hat{\rho}_{RI})
\nonumber
\\
&=&g((\mu_+-1)/2)+g((\mu_--1)/2).
\ea 
Here the symplectic eigenvalues of the covariance matrix ${\mathbf{\Lambda}}_{\theta=0}$ in Eq.~\ref{hk} are
$
\mu_\pm=\frac{1}{2}\left[\pm(S-A)+\sqrt{(A+S)^2-4C^2}\right],
$ 
with $A=2\left(N_B+\kappa N_S\right)+1, C=2C_p, S=2N_S+1$. In the limit of $N_B\gg1, N_S\ll1$, one can obtain
\begin{align}
&S(\hat{\rho}^{\theta=0}_{RI})=\log_2(N_B)-N_S \log_2(N_S)+\frac{1+N_S}{\ln(2)}+
\nonumber
\\
&\frac{\kappa N_S(1/\ln(2)+\log_2 (N_S))+1/2\ln(2)}{N_B}+O(N_S^2,1/N_B^2)
\label{Scond}
\end{align}

The number-basis matrix element of the unconditional state $\int_0^{2\pi} d\theta \hat{\rho}^\theta_{RI}/2\pi$ can be obtained analytically.
We first obtain the number bases density matrix , and then integrate over the unitary $\hat{U}_\theta \otimes \hat{I}$.

From the covariance matrix ${\mathbf{\Lambda}}_{\theta=0}$ of the Gaussian state $\hat{\rho}_{RI}$, we can obtain the density matrix in number bases $\matrixel{n_1,n_2}{\hat{\rho}_{RI}}{n_1^\prime, n_2\prime}$ is only non-zero when $n_1-n_1^\prime=n_2-n_2^\prime$, and the non-zero terms equals
\begin{align}
&\sqrt{\frac{n_1!n_2!}{n_1^\prime ! n_2^\prime !}}(-1)^{1+n_2+n_2^\prime}2^{2+n_2-n_2^\prime} C^{n_2-n_2^\prime}\times
\nonumber
\\
&\frac{(-1+C^2+E+S-E S)^{1+n_1^\prime +n_2}}{X^{1+n_1}Y^{1+n_2}}\times
\nonumber
\\
&F_R(1+n_1,1+n_2,1+n_2-n_2^\prime ,\frac{4C^2}{XY}),
\end{align}
where $F_R(a,b,c,z)$ is the regularized hypergeometric function and
$
X=(1+C^2+E-(1+E)S),
Y=(C^2-(E-1)(S+1))
$, with $C=2C_p, E=1+2(N_B+\kappa N_S), S=(1+2N_S)$.

Because $\hat{U}_\theta \otimes \hat{I} \ket{n_1}_R \ket{n_2}_I \bra{n_1^\prime}_R\bra{n_2^\prime}_I \hat{U}_\theta^\dagger \otimes \hat{I}=e^{i\theta(n_1-n_1^\prime)}\ket{n_1}_R \ket{n_2}_I \bra{n_1^\prime}_R\bra{n_2^\prime}_I$, the integration will lead to $n_1=n_1^\prime$. Combined with the fact that $n_1-n_1^\prime=n_2-n_2^\prime$, we see that the density matrix of $\int_0^{2\pi}d\theta \hat{\rho}^\theta_{RI}/2\pi$ is diagonal in number bases with $p(n_1,n_2)\equiv \matrixel{n_1,n_2}{\hat{\rho}_{RI}}{n_1, n_2}$ given by,
\begin{align}
&p(n_1,n_2)= -4 F_R(1+n_1,1+n_2,1,\frac{4C^2}{XY})
\nonumber
\\
&\frac{(-1+C^2+E+S-E S)^{1+n_1 +n_2}}{X^{1+n_1}Y^{1+n_2}}.
\end{align}
Thus the unconditional entropy
\be 
S(\int_0^{2\pi}d\theta \hat{\rho}^\theta_{RI}/2\pi)=-\sum_{n_1,n_2=0}^\infty p(n_1,n_2) \log_2 [p(n_1,n_2)].
\ee 
The rest of the analysis is to asymptotically expand the result. In the limit of $N_B\gg1$, we have $\frac{4C^2}{XY}=\kappa/(N_B(1-\kappa)+N_B^2)\ll1$, thus we can expand using $F_R(1+n_1,1+n_2,1,x)=1+(1+n_1+n_2+n_1n_2)x+O(x^2)$. With the above expansion, denote the first order result of $p(n_1,n_2)$ as $p_1(n_1,n_2)$, which is too long to display here. This expansion can be justified by checking the normalization
$
\sum_{n_1,n_2=0}^\infty p_1(n_1,n_2)=1-{\kappa^2(1+N_S)^2}/{N_B^2}+O(1/N_B^3),
$ 
which is accurate to high orders. 
Further expansion and summation leads to
\begin{align}
&S\left(\int_0^{2\pi}d\theta \hat{\rho}^\theta_{RI}/2\pi\right)=\log_2(N_B)-N_S \log_2(N_S)+\frac{1+N_S}{\ln(2)}+
\nonumber
\\
&\frac{2\kappa N_S/2\ln(2)+1/2\ln(2)}{N_B}+O(N_S^2,1/N_B^2).
\label{Suncond}
\end{align}
Overall, combing Eqs.~\ref{chi_phase}, \ref{Scond} and \ref{Suncond}, and noticing that all higher order terms in $N_S$ cancels, we have
\be 
\chi\left(\Sigma_{\theta}^1\right)=\frac{\kappa N_S (1+N_S)\log_2(1+1/N_S)}{N_B}+O(1/N_B^2)
\ee

In the limit of $N_B\gg1$, by comapring with EA classical capacity (Eq.~\ref{CE_formula}), we have Eq.~\ref{holevo_phase}.

\section{Information rates of receivers with binary phase shift keying}

\subsection{Optical parametric receiver}
\label{sec:OPAappendix}
We are interested in the limit $M\gg1$, while $M\kappa N_S/N_B\ll 1$ still holds, such that $P_{\rm OPA}(n|\theta=0;M)$ and $P_{\rm OPA}(n|\theta=\pi;M)$ are approximately the same Gaussian distribution. In this regime, the optimum binary encoding yields an approximately symmetric Gaussian channel. For equal priors, the maximum-likelihood decision rule gives the threshold $N_{th}=M[\sigma(\pi) \o N(0)+\sigma(0) \o N(\pi)]/[\sigma(0)+\sigma(\pi)]$, and the error probability 
\be
P_{E}^{\rm OPA}=\frac{1}{2}{\rm erfc}\left(\sqrt\frac{ M\mu_{\rm OPA}^2}{2\sigma_{\rm OPA}^2}\right)
\label{PE_OPA}
\ee
where $\mu_{\rm OPA}=|\o N(0)-\o N(\pi)|,\,\sigma_{\rm OPA}^2=[\sigma(0)+\sigma(\pi)]^2\simeq 4\o N(\frac{\pi}{2})[1+\o N(\frac{\pi}{2})]$.
with $N_B\gg1$ and the optimal gain $G=1+\sqrt{N_S}/{N_B}$, giving the leading order signal-to-noise ratio 
\be 
{\mu_{\rm OPA}^2}/{\sigma_{\rm OPA}^2}={4\kappa N_S(1+N_S)}/\left[{N_B(1+2\sqrt N_S+2N_S)}\right].
\ee 
Here, ${\rm erfc}(x)=1-2\int_0^x d t\ e^{-t^2}/\sqrt{\pi}$ is the complementary error function. With Eq.~\ref{I2}, we obtain the information rate by inputting the error probability in Eq.~\ref{PE_OPA}.

In the limit $N_S\ll 1, N_B\gg 1$, with Eq.~\ref{eq:OPAmean} we may simplify the variable in the error function as $\sqrt{M\mu_{\rm OPA}^2/2\sigma_{\rm OPA}^2}\simeq \sqrt{2M\kappa N_S/N_B}$. Expanding around $P_E=1/2$ we obtain 
\be
R_{BPSK}^{\rm OPA}\simeq 1.27\frac{\kappa N_S}{N_B {\rm ln}2}.
\,\ee
Compared with Eq.~\ref{eq:Casym},
the OPA receiver with BPSK theoretically offers an EA advantage of $\sim 27\%$ at best over the unassisted case. However, in the region that error rates are small enough for practical error correction, the advantage is smaller.

\subsection{Phase conjugate receiver}
\label{sec:PCRappendix}
With large $M$, the photon statistics of PCR is also approximately Gaussian and symmetric. With maximum-likelihood decision rule we have
\be
P_{E}^{\rm PCR}=\frac{1}{2}{\rm erfc}\left(\sqrt\frac{ M\mu_{\rm PCR}^2}{2\sigma_{\rm PCR}^2}\right),
\label{PE_PCR}
\ee
where $\mu_{\rm PCR}=| N_+- N_-|$ and $\sigma_{\rm PCR}^2=(\sigma_++\sigma_-)^2$. Here the means and variances, depending on the phase encoding $\theta\in \{0,\pi\}$, are given by $ N_\pm=\pm C_p$ and $\sigma^2_\pm=(1+ N_{X,\pm}) N_{X,\pm}+(1+ N_{Y,\pm}) N_{Y,\pm}-( N_{C}- N_{I})^2/2$, where the X arm contributes $ N_{X,\pm}=( N_{C}+ N_I)/2\pm C_p$ and the Y arm yields $ N_{Y,\pm}=( N_{C}+ N_I)/2\mp C_p$. Note that the photon number of the idler $N_I=N_S$ and that of the conjugated signal $N_{C}=\kappa N_S+N_B+1$ are independent with the phase, and the variances are symmetric, $\sigma_+^2=\sigma_-^2$. Finally, we have the signal-to-noise ratio 
\be 
\mu_{\rm PCR}^2/\sigma_{\rm PCR}^2= {4\kappa N_S(1+N_S)}/\left[{N_B(1+2N_S)}\right],
\ee 
in the limit of $N_B\gg 1$.

Note that PCR has the same signal-to-noise ratio to the leading order as the OPA and thus the same asymptotic advantage. However, at the practical error correctable region, we see the higher order term in the denominator for PCR is smaller than the OPA case, which enhances the performance especially when the influence of the higher terms compares with the EA advantage.

\subsection{Sum-frequency generation receiver}
\label{sec:SFGappendix}
Based on an analogy to the Dolinar receiver, the choice of $r_k$ is 
\be
\label{rk}
r_{k,\tilde{h}_k}=\sqrt{\eta}|C_{si,k}^{\rm in}|\!\left(\frac{(-1)^{\tilde{h}_k}}{\sqrt{1-\exp\left[-2 M(\sum_{\ell=0}^{k}\lambda_\ell^2-\lambda_k^2/2)\right]}}\right),
\ee
where $\lambda_k^2=4\eta |C_{si,k}^{in}|^2$. The intuition behind is that, when one guesses correctly $\tilde{h}_k=h$, with the information sufficiently extracted, i.e. $M\sum_{\ell=0}^k \lambda_\ell^2\gg1$, the condition reduces to $r_{k,\tilde{h}_k}\simeq \sqrt{\eta}C_{si,k}^{\rm in}$, leaving the sum-frequency mode $\hat{b}_k$ close to vacuum. In this case, any click of the photon detector implies, with a high likelihood, that a wrong hypothesis has been made. In doing so, nearly unambiguous information is obtained to improve the performance.

Akin to the Dolinar receiver, the minimum error probability of discriminating $\Sigma_{\rm BPSK}^M$ on the SFG receiver, determined by the Helstrom bound, can be estimated based on the discrimination between noisy coherent states with mean $e^{i\theta_h} \sqrt{(1-\epsilon)M\kappa N_S/N_B}$ and noise $-N_S\ln(\epsilon)/2$ with the residual correlation $\epsilon\ll 1$. The numerical results are plotted in Fig.~\ref{fig:DolinarPe}.

In the limit that $N_S\ll 1$, the noisy coherent state approximates to a pure coherent state. Its Helstrom bound yields
$
P_H=\frac{1}{2} \left[1-\sqrt{1-\exp\left(-4 M\kappa N_S/N_B\right)}\right].
$
With $M\kappa N_S/N_B\ll 1$, we have $P_H\simeq 1/2-\sqrt{M\kappa N_S/N_B}$. The Taylor expansion of Eq.~\ref{I2} around $P_E=1/2$ yields
\be
R_{\rm BPSK}^{\rm SFG}=\frac{2}{M{\rm ln}2}\left(P_H-\frac{1}{2}\right)^2=2 \frac{\kappa N_S}{N_B  {\rm ln}2},
\ee
which produces an EA advantage of 3dB.

\section{Adaptive noisy phase estimation}

\subsection{Precision limit of noisy phase estimation}
\label{App:adaptive_QFI}
The precision limit for the root-mean-square (rms) error in estimating a parameter $\theta$ on $M\gg1$ input states $\hat{\rho}_\theta$ is given by the quantum Cramm\'{e}r-Rao lower bound (CRLB): $\delta \theta\ge 1/\sqrt{M \mathcal{J}_\theta}$~\cite{Helstrom_1976,Holevo_1982,Yuen_1973}, where the single-parameter QFI~\cite{braun2018quantum,braunstein1994statistical,jarzyna2015true}
\be
\mathcal{J}_\theta=\lim_{d\theta\to0}8 \frac{1-\sqrt{\mathcal{F}\left(\hat{\rho}_\theta,\hat{\rho}_{\theta+d\theta}\right)}}{d\theta^2}
\ee
is obtained from the Uhlmann fidelity $\mathcal{F}\left(\hat{\rho},\hat{\sigma}\right)={\rm tr}\left(\sqrt{\sqrt{\hat{\rho}}\hat{\sigma}\sqrt{\hat{\rho}}}\right)^2$.

Although the well-known NOON state~\cite{bollinger1996optimal,dorner2009optimal} is the optimum for phase estimation in the absence of noise for fixed photon number, it quickly becomes impotent as noise and loss arise. While the optimum quantum state for noisy phase estimation remains unknown, an upper bound on the QFI has been found~\cite{gagatsos2017bounding}. It is straightforward to show that the maximum of the upper bound is achieved in a large photon number variance limit, i.e., $\Delta_{N_S}^2\to\infty$~\footnote{The photon number variance can be unbounded, e.g. $\left(1-p\right)\ket{0}\bra{0}+p\ket{N}\bra{N}$, with mean $pN=N_S$, has variance diverging as $\propto N$.} and 
\begin{align}
&\mathcal{J}^{\rm UB}_\theta=
\nonumber
\\
&\frac{4\kappa N_S\left(\kappa N_S+\left(1-\kappa\right)N_B+1\right)}{\left(1-\kappa\right)\left[\kappa N_S \left(2N_B+1\right)-\kappa N_B\left(N_B+1\right)+\left(N_B+1\right)^2\right]}.
\label{UB}
\end{align}
In the limit of $\kappa\ll 1, \kappa N_S\ll N_B,N_B\gg 1$, one has
$
\mathcal{J}^{\rm UB}_\theta\simeq  4\kappa N_S/N_B.
$
Since the rms error of phase estimation is bounded by the period $2\pi$, this QFI only holds in an asymptotic limit, at which the $1/\sqrt{M}$ factor decreases the rms error to $\delta \theta \ll 2\pi$.

With a TMSV source (TMSS), the joint state $\hat{\rho}^\theta_{RI}$ at the receiver in the EA communication protocol is Gaussian, thus the fidelity and the QFI can be analytically obtained~\cite{marian2016quantum,banchi2015quantum}:
\be 
\mathcal{J}^{\rm TMSS}_\theta=\frac{4\kappa N_S\left(N_S+1\right)}{1+N_B\left(1+2N_S\right)+N_S\left(1-\kappa\right)}.
\label{TMSV_opt}
\ee 
As a comparison, suppose one uses the coherent state $\ket{\sqrt{N_S}}$, in lieu of the TMSV, the returned state $\mathcal{L}^{\kappa,N_B}_\theta\left(\ket{\sqrt{N_S}}\bra{\sqrt{N_S}}\right)$ is a displaced thermal state with mean $e^{i\theta}\sqrt{\kappa N_S}$ and thermal noise $N_B$. It is straightforward to derive the fidelity~\cite{scutaru1998fidelity}, and
thus the QFI under this circumstance:
$
\mathcal{J}^{\rm coh}_\theta={4\kappa N_S}/{(1+2N_B)}.
$ 
In the limit of $N_B\gg1, \kappa\ll1,$ and $N_S\ll1$, one has 
$
\mathcal{J}^{\rm UB}_\theta\simeq \mathcal{J}^{\rm TMSS}_\theta\simeq 2\mathcal{J}^{\rm coh}_\theta.
$
Note that the QFI, in this limit, is only related to the mean of the displacement. As such, the coherent state is anticipated to also be the optimum state in the absence of EA. With EA, a 3-dB advantage can be achieved. In fact, the presented EA protocol based on the TMSS is asymptotically optimal in the limit of strong noise and weak signal. In the following, we describe the optimum receiver that saturates the maximum QFI. 

\subsection{Optimum receiver for noisy phase estimation---adaptive OPA receiver}
\label{App:adaptive_Rx}

Elaborated in Eq.~\ref{Pn_OPA}, the OPA receiver's photon number counting statistics are $P_{\rm OPA}(n|\theta; M)$, conditioned on the encoded phase $\theta$.
The corresponding classical Fisher information
$
\mathcal{J}^{\rm OPA}_\theta=\sum_{n=0}^\infty\left(\partial_\theta\mathrm{log}P_{\rm OPA}(n|\theta; M)\right)^2P_{\rm OPA}(n|\theta; M)
$ can be analytically solved:
\be 
\mathcal{J}^{\rm OPA}_\theta=\frac{4(G-1)GM\kappa N_S(1+N_S)\mathrm{sin}^2\theta}{\overline N(1+\overline N)}.
\label{eq:OPAFishersin}
\ee
For $N_B\gg1$ and $G=1+{\sqrt{N_S}}/{N_B}$, it becomes
$ 
\mathcal{J}^{\rm OPA}_\theta\simeq  M\mathrm{sin}^2\theta  \mathcal{J}^{\rm TMSS}_\theta
$.

The factor $\sin^2 \theta$ indicates that the QFI $\mathcal{J}^{\rm OPA}_\theta$ is phase dependent and is only maximized at $\theta=\pi/2$. Thus, a single-shot phase estimation of a random phase does not usually achieve the maximum QFI. However, with multiple copies of the joint signal-idler state available, viz., $M\gg1$, this phase-dependent factor can be asymptotically eliminated through an FF mechanism, as utilized in the achievability proof of single-parameter CRLB~\cite{fujiwara2006strong,gill2005state,hayashi2017quantum}. A simple FF approach involves first performing an OPA operation on $\sqrt{M}$ modes to obtain an initial estimation $\tilde{\theta}=\theta^\star+O(1/M^{1/4})$ of the true value $\theta^\star$, followed by a phase shift of $\Delta\theta=\pi/2-\tilde{\theta}$ to set the phases to $\theta^\star+\Delta\theta=\pi/2+O(1/M^{1/4})$ so that near-maximum QFI can be attained. A subsequent OPA operation on $M-\sqrt{M}$ modes gives a QFI of $\left(M-\sqrt{M}\right)\left(1-O(1/\sqrt{M})\right) \mathcal{J}^{\rm TMSS}_\theta$, which, to the first order, achieves $ M \mathcal{J}^{\rm TMSS}_\theta$. 

\begin{figure}
\centering
\includegraphics[width=0.45\textwidth]{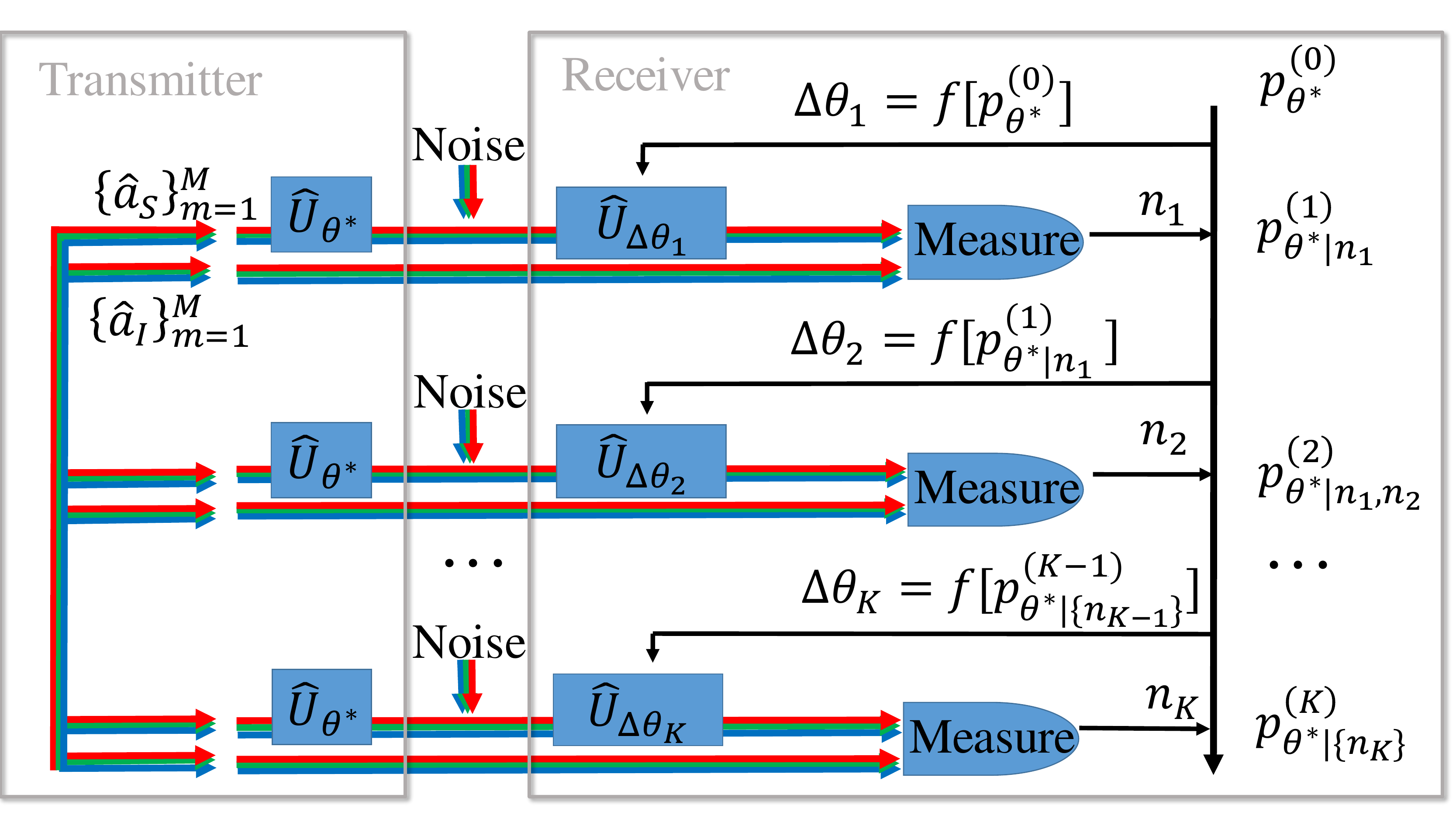}
\caption{Feed-forward setup of the adaptive schemes. On the transmitter side, the phase encoding unitary $\hat{U}_{\theta^\star}$ encodes identical information among multiple signal modes. On the receiver side, a phase compensation $\hat{U}_{\Delta \theta_k}$ is applied on the signal before the measurement. The compensation angle $\Delta\theta_k$ is determined from the posterior distribution $p_{\theta^\star|\{n_{k-1}\}}^{(k-1)}$.}
\label{fig:AdaptiveSetup}
\end{figure}

In EA communication, however, the rate of the convergence to the maximum QFI is important. Thus, a systematic Bayesian FF approach is adopted (schematic in Fig.~\ref{fig:AdaptiveSetup}). The entire $M$ mode pairs are measured in $K$ cycles, with each cycle consuming $M_k$ modes such that $\sum_{k=1}^K M_k=M$. By doing so, an adaptive strategy $\mathbb{S}_\bold{M}$ specified by the parameters
$\bold{M}=\{M_k, 1\le k \le K\}$ is executed as the following. Initially, the prior probability $p_{\theta^\star}^{(0)}(\theta)$ is set uniformly distributed in $[0,2\pi)$, because the phase encoding is uniform. In the $2\le k \le K$-th cycle, the prior-probability distribution $p_{\theta^\star|\{n_{k-1}\}}^{(k-1)}(\cdot|\{n_{k-1}\})$ equals the posterior in the $(k-1)$-th cycle, based on all previous measurement results $\{n_{k-1}\}\equiv \{n_1,\cdots, n_{k-1}\}$. Prior to the measurement, a phase shift $\hat{U}_{\Delta \theta_k}$ with $\Delta \theta_k=f\left[p_{\theta^\star|\{n_{k-1}\}}^{(k-1)}\right]$ is applied. The phase shift is a functional of the Bayesian posterior probability of the last cycle, which will be specified later. 

After the measurement, the posterior probability is updated, based on the measured photon number $n_k$ and the prior probability using the Bayesian formula
\be
p_{\theta^\star|\{n_k\}}^{(k)}(\theta|\{n_k\})\propto P_{\rm OPA}(n_k|\theta; M_k)p_{\theta^\star|\{n_{k-1}\}}^{(k-1)}(\theta|\{n_{k-1}\}).
\ee
From this, one can construct the estimator $\tilde{\theta}_k=\arg\max p_{\theta^\star|\{n_{k}\}}^{(k)}(\theta|\{n_{k}\})$. After all cycles are executed, the output from the last cycle is chosen as the final estimate.

The maximum Fisher information approach and the maximum Van Trees information~\cite{van2004detection,martinez2017VanTrees,paris2009VanTrees} approach are taken to determine the phase shift $\Delta \theta_k=f\left[p_{\theta^\star|\{n_{k-1}\}}^{(k-1)}\right]$. The Fisher information approach simply maximizes the Fisher information by taking $\Delta \theta_k=\arg\max_{\Delta\theta_k'}\mathcal{J}^{\rm OPA}_{\tilde\theta_{k-1}+\Delta\theta_k'}=\arg\max_{\Delta\theta_k'} \sin^2(\tilde\theta_{k-1}+\Delta\theta_k')$ based on the current estimator, giving $\Delta\theta_k=\pi/2-\tilde\theta_{k-1}$. The Van Trees approach maximizes the average Fisher information, also known as the Van Trees information:
\be
\Delta \theta_k=\arg\max_{\Delta\theta_k'}\int d\theta_0 p_{\theta^\star|\{n_{k-1}\}}^{(k-1)}(\theta_0|\{n_{k-1}\}) \mathcal{J}^{\rm OPA}_{\theta_0+\Delta\theta_k'}.
\ee
Because the Van Trees approach makes use of the entire posterior distribution, it yields a performance superior to that of the maximum Fisher information approach when the posterior probability has multiple peaks with similar heights.

\begin{figure}
    \subfigure{
    \centering
    \includegraphics[width=0.24\textwidth]{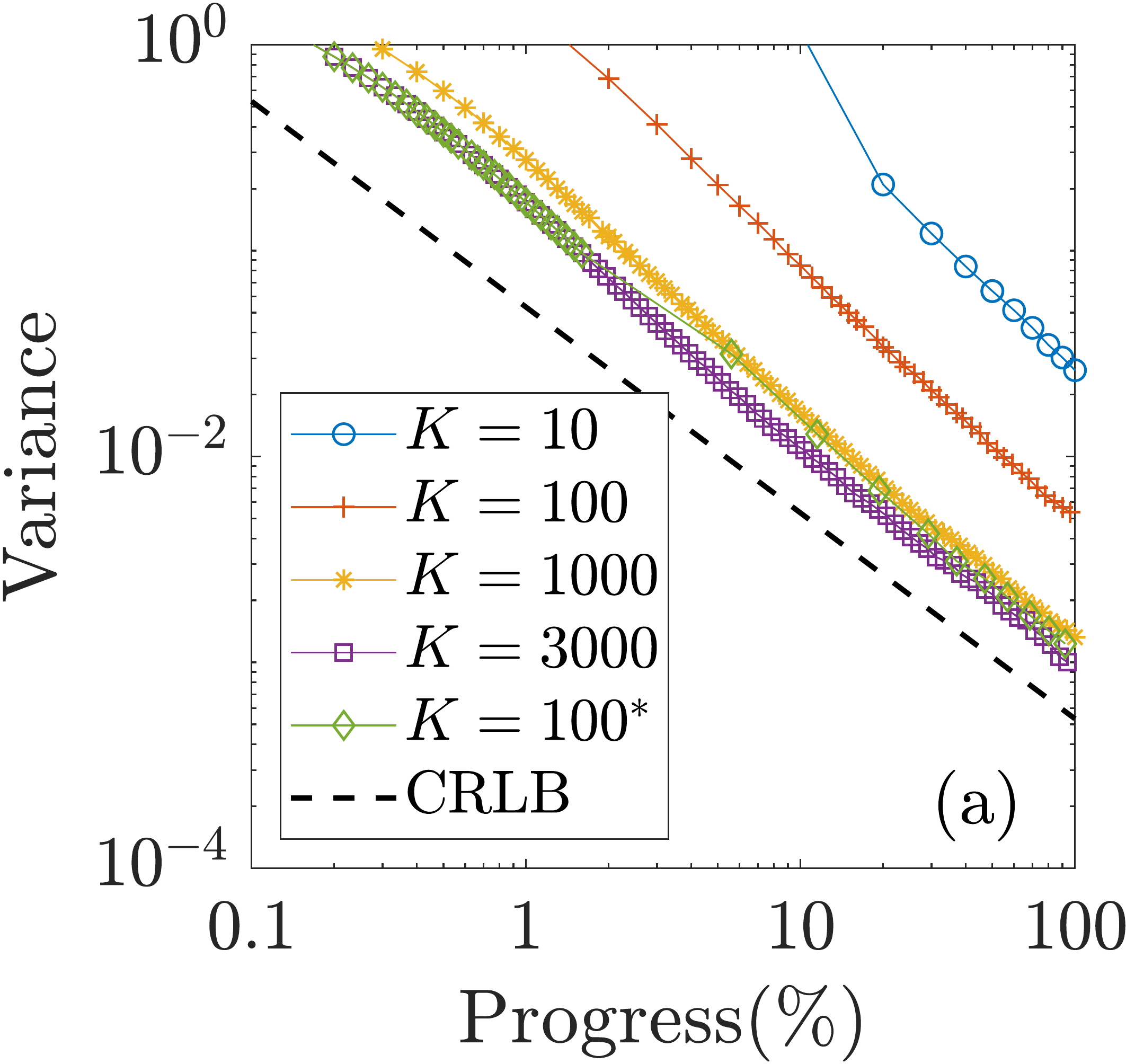}
    \label{fig:Bayes_fisher}
    }
    \subfigure{
    \centering
    \includegraphics[width=0.195\textwidth]{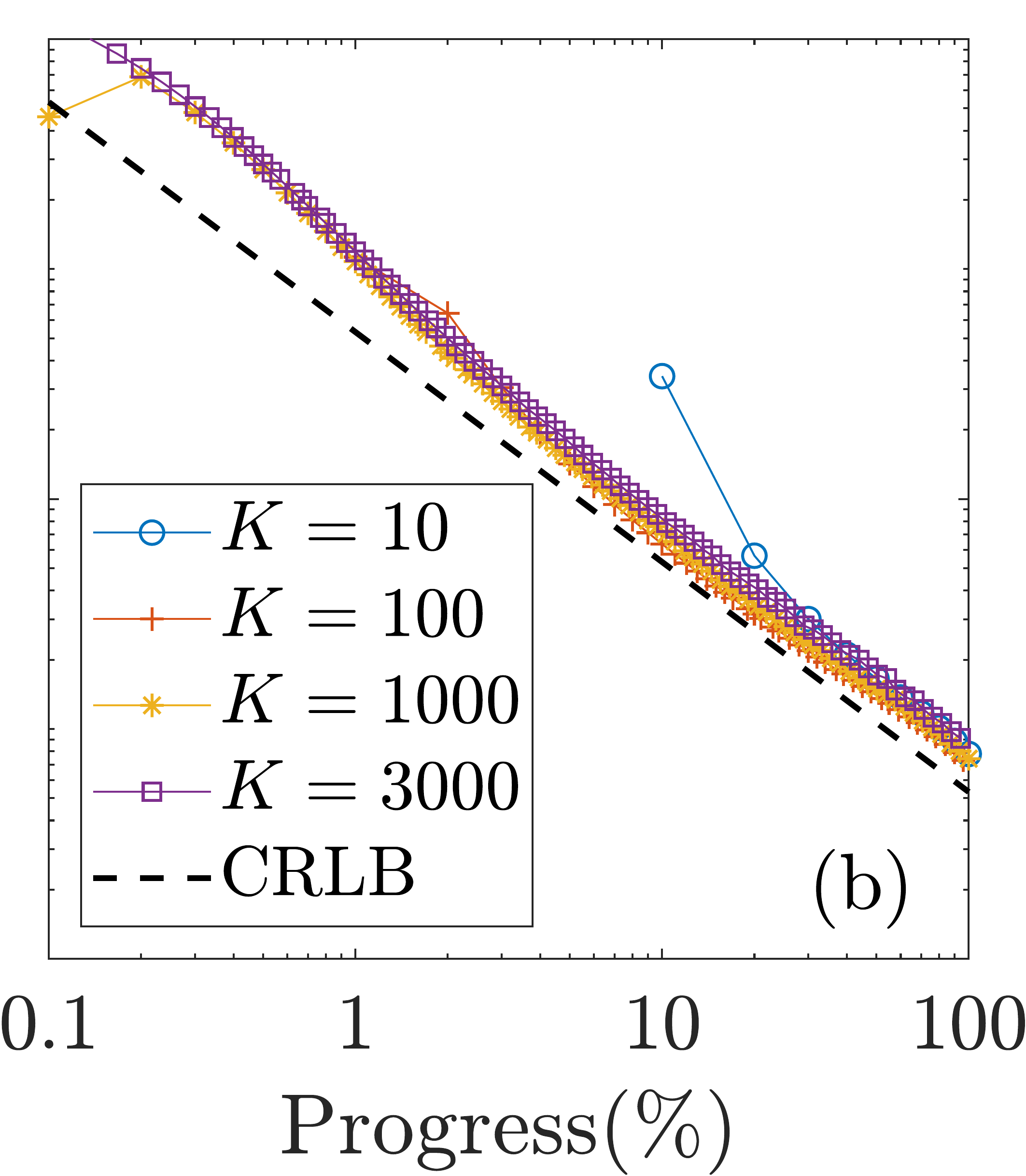}
    \label{fig:Bayes_van_trees}
    }
    \caption{Variance evolution of the Bayesian phase estimation using OPA receiver. (a) Maximum Fisher information approach. (b) Maximum Van Trees information approach. Parameter: $M=5\times 10^{12}, N_S=10^{-3}, N_B=10^{4}, \kappa=10^{-3}$. *The diamond marked line distributes resource heterogeneously to optimize the performance.}
    \label{fig:FisherVsVanTrees}
\end{figure}

Seeking an analytical solution for the ultimate posterior probability is challenging. We thus resort to a Monte Carlo simulation to evaluate the performance. We simulate the parameter estimation process with $8\times 10^5$ samples and record the evolution of the variance evaluated from the posterior probability $p_{\theta^\star|\{n_k\}}^{(k)}(\theta|\{n_k\})$ of each estimation cycle. In Fig.~\ref{fig:FisherVsVanTrees}, the variance at $k$-th cycle are plotted with the progress, i.e., the portion of the modes that have been utilized up to the current cycle $\sum_{\ell=1}^k M_\ell/M$. To benchmark the convergence, the CRLB in Eq.~\ref{TMSV_opt} for each $\sum_{\ell=1}^k M_\ell$ number of modes is shown. First, an equal slicing of $M_k=M/K$ is considered. In this case, the Fisher information approach has a variance converging to the CRLB as the number of cycles $K$ increases (Fig.~\ref{fig:Bayes_fisher}). Nevertheless, the Van Trees approach converges to the CRLB much faster. With $K=10$ slices, the variance is already close to the CRLB (Fig.~\ref{fig:Bayes_van_trees}). 

In practice, the implementation of the FF process can be challenging, so the number of cycles $K$ need be minimized. Hence, the Van Trees approach is favorable. One can reduce the number of cycles in the maximum Fisher approach by heterogeneously slicing $M$ into larger segments $M_k$ as we progress to a small variance region. As an example,the diamond marked line uses $K=100$ estimation cycles with heterogeneously distributed resource. The first 50 cycles are assigned with small $M_k$ equivalent to those of $K=3000$, whereas the latter 50 cycles are sliced wider with $M_k$ comparable to the uniform slices with $K=100$ (red crosses). A large advantage from the optimization of $\bold{M}=\{M_k, 1\le k \le K\}$ is observed. The systematic optimization of the parameter $\bm{M}$ is in general a dynamical programming problem subject to future work.

\section{Photon statistics of the displaced thermal state}
\label{sec:stat}

A displaced thermal state (DTS) $\hat{\rho}_{\lambda,n_e}^\theta$ with mean $\lambda=e^{i\theta}|\lambda|$ and thermal noise $n_e$ has the Glauber-Sudarshan P function
$
P(\alpha)=\exp\left[-|\alpha-\lambda|^2/(2\sigma_P^2)\right]/(2\pi\sigma_P^2),
$
where $\sigma_P^2=n_e/2$.
We immediately obtain the density matrix $\hat{\rho}_{\lambda,n_e}^\theta$ in the Fock basis
\be
\bal
&\matrixel{n}{\hat{\rho}_{\lambda,n_e}^\theta}{m}=\bra n \int d\alpha P(\alpha)\ket\alpha\bra\alpha\ket{m}=\\
&\frac{e^{-\frac{|\lambda|^2}{n_e}}\!e^{i(m\!-n)\theta}n_e^n\!|\lambda|^{m-n}\!\sqrt {m!}\,}{(1+n_e)^{m+1}\sqrt{n!}}\\
&\times _1\tilde F_1\left[m\!+\!1,m\!-\!n\!+\!1,\frac{|\lambda|^2}{n_e(1+n_e)}\right],
\eal
\ee
where $_1\tilde F_1$ is the regularized confluent hypergeometric function~\cite{Lachs_1965}. 
Also we can obtain the photon number distribution $P_{\rm DTS}(n;\lambda,n_e)$ by letting $n=m$, which leads to Laguerre statistics.

Now we calculate the Holevo information of an ensemble of uniformly phase encoded displaced thermal states 
i.e. $\Sigma_\theta^{M,C}=\{(\hat{\rho}_{\lambda,n_e}^\theta)^{\otimes M}, \theta\sim U[0,2\pi) \}$. Here $M$ is the number of repetition encoding, C implies classical states (cf. the TMSV ensemble $\Sigma_\theta^{M}$). First, we can use a balanced beam-splitter array to transform each state $\hat{\rho}_{\lambda,n_e}^\theta)^{\otimes M}$ to $\hat{\rho}^\theta_{\sqrt{M}\lambda,n_e}\otimes (\hat{\rho}_{0,n_e})^{\otimes M}$. Because Holevo information is unchanged under unitary and appending constant states, effectively we can consider the ensemble $\{\hat{\rho}^\theta_{\sqrt{M}\lambda,n_e}, \theta\sim U[0,2\pi) \}$.

Now notice that the conditional entropy $S(\hat{\rho}^\theta_{\sqrt{M}\lambda,n_e})$ is simply $g(n_e)$ due to the invariance of entropy under unitary transform. Furthermore, the unconditional single-mode state is diagonal in the photon number basis due to an average over uniform phase modulation $\theta\sim U[0,2\pi)$. As such, one only needs the Shannon entropy of the photon number distribution $P_{\rm DTS}(\cdot;\sqrt{M}\lambda, n_e)$ of displaced thermal state. The final result is
\be
\chi(\Sigma_\theta^{M,C})=H\left[P_{\rm DTS}(\cdot;\sqrt{M}\lambda, n_e)\right]-g(n_e),
\,,\ee 
which can be efficiently evaluated.


\begin{thebibliography}{80}%
\makeatletter
\providecommand \@ifxundefined [1]{%
 \@ifx{#1\undefined}
}%
\providecommand \@ifnum [1]{%
 \ifnum #1\expandafter \@firstoftwo
 \else \expandafter \@secondoftwo
 \fi
}%
\providecommand \@ifx [1]{%
 \ifx #1\expandafter \@firstoftwo
 \else \expandafter \@secondoftwo
 \fi
}%
\providecommand \natexlab [1]{#1}%
\providecommand \enquote  [1]{``#1''}%
\providecommand \bibnamefont  [1]{#1}%
\providecommand \bibfnamefont [1]{#1}%
\providecommand \citenamefont [1]{#1}%
\providecommand \href@noop [0]{\@secondoftwo}%
\providecommand \href [0]{\begingroup \@sanitize@url \@href}%
\providecommand \@href[1]{\@@startlink{#1}\@@href}%
\providecommand \@@href[1]{\endgroup#1\@@endlink}%
\providecommand \@sanitize@url [0]{\catcode `\\12\catcode `\$12\catcode
  `\&12\catcode `\#12\catcode `\^12\catcode `\_12\catcode `\%12\relax}%
\providecommand \@@startlink[1]{}%
\providecommand \@@endlink[0]{}%
\providecommand \url  [0]{\begingroup\@sanitize@url \@url }%
\providecommand \@url [1]{\endgroup\@href {#1}{\urlprefix }}%
\providecommand \urlprefix  [0]{URL }%
\providecommand \Eprint [0]{\href }%
\providecommand \doibase [0]{http://dx.doi.org/}%
\providecommand \selectlanguage [0]{\@gobble}%
\providecommand \bibinfo  [0]{\@secondoftwo}%
\providecommand \bibfield  [0]{\@secondoftwo}%
\providecommand \translation [1]{[#1]}%
\providecommand \BibitemOpen [0]{}%
\providecommand \bibitemStop [0]{}%
\providecommand \bibitemNoStop [0]{.\EOS\space}%
\providecommand \EOS [0]{\spacefactor3000\relax}%
\providecommand \BibitemShut  [1]{\csname bibitem#1\endcsname}%
\let\auto@bib@innerbib\@empty
\bibitem [{\citenamefont {Bennett}\ \emph {et~al.}(2002)\citenamefont
  {Bennett}, \citenamefont {Shor}, \citenamefont {Smolin},\ and\ \citenamefont
  {Thapliyal}}]{bennett2002entanglement}%
  \BibitemOpen
  \bibfield  {author} {\bibinfo {author} {\bibfnamefont {C.~H.}\ \bibnamefont
  {Bennett}}, \bibinfo {author} {\bibfnamefont {P.~W.}\ \bibnamefont {Shor}},
  \bibinfo {author} {\bibfnamefont {J.~A.}\ \bibnamefont {Smolin}}, \ and\
  \bibinfo {author} {\bibfnamefont {A.~V.}\ \bibnamefont {Thapliyal}},\
  }\href@noop {} {\bibfield  {journal} {\bibinfo  {journal} {IEEE Trans. Inf.
  Theory}\ }\textbf {\bibinfo {volume} {48}},\ \bibinfo {pages} {2637}
  (\bibinfo {year} {2002})}\BibitemShut {NoStop}%
\bibitem [{\citenamefont {Pirandola}\ \emph {et~al.}(2018)\citenamefont
  {Pirandola}, \citenamefont {Bardhan}, \citenamefont {Gehring}, \citenamefont
  {Weedbrook},\ and\ \citenamefont {Lloyd}}]{pirandola2018advances}%
  \BibitemOpen
  \bibfield  {author} {\bibinfo {author} {\bibfnamefont {S.}~\bibnamefont
  {Pirandola}}, \bibinfo {author} {\bibfnamefont {B.~R.}\ \bibnamefont
  {Bardhan}}, \bibinfo {author} {\bibfnamefont {T.}~\bibnamefont {Gehring}},
  \bibinfo {author} {\bibfnamefont {C.}~\bibnamefont {Weedbrook}}, \ and\
  \bibinfo {author} {\bibfnamefont {S.}~\bibnamefont {Lloyd}},\ }\href@noop {}
  {\bibfield  {journal} {\bibinfo  {journal} {Nat. Photonics}\ }\textbf
  {\bibinfo {volume} {12}},\ \bibinfo {pages} {724} (\bibinfo {year}
  {2018})}\BibitemShut {NoStop}%
\bibitem [{\citenamefont {Giovannetti}\ \emph {et~al.}(2004)\citenamefont
  {Giovannetti}, \citenamefont {Lloyd},\ and\ \citenamefont
  {Maccone}}]{giovannetti2004quantum}%
  \BibitemOpen
  \bibfield  {author} {\bibinfo {author} {\bibfnamefont {V.}~\bibnamefont
  {Giovannetti}}, \bibinfo {author} {\bibfnamefont {S.}~\bibnamefont {Lloyd}},
  \ and\ \bibinfo {author} {\bibfnamefont {L.}~\bibnamefont {Maccone}},\
  }\href@noop {} {\bibfield  {journal} {\bibinfo  {journal} {Science}\ }\textbf
  {\bibinfo {volume} {306}},\ \bibinfo {pages} {1330} (\bibinfo {year}
  {2004})}\BibitemShut {NoStop}%
\bibitem [{\citenamefont {Giovannetti}\ \emph {et~al.}(2011)\citenamefont
  {Giovannetti}, \citenamefont {Lloyd},\ and\ \citenamefont
  {Maccone}}]{giovannetti2011advances}%
  \BibitemOpen
  \bibfield  {author} {\bibinfo {author} {\bibfnamefont {V.}~\bibnamefont
  {Giovannetti}}, \bibinfo {author} {\bibfnamefont {S.}~\bibnamefont {Lloyd}},
  \ and\ \bibinfo {author} {\bibfnamefont {L.}~\bibnamefont {Maccone}},\
  }\href@noop {} {\bibfield  {journal} {\bibinfo  {journal} {Nat. Photonics}\
  }\textbf {\bibinfo {volume} {5}},\ \bibinfo {pages} {222} (\bibinfo {year}
  {2011})}\BibitemShut {NoStop}%
\bibitem [{\citenamefont {Shor}(1999)}]{shor1999polynomial}%
  \BibitemOpen
  \bibfield  {author} {\bibinfo {author} {\bibfnamefont {P.~W.}\ \bibnamefont
  {Shor}},\ }\href@noop {} {\bibfield  {journal} {\bibinfo  {journal} {SIAM
  Rev.}\ }\textbf {\bibinfo {volume} {41}},\ \bibinfo {pages} {303} (\bibinfo
  {year} {1999})}\BibitemShut {NoStop}%
\bibitem [{\citenamefont {Lloyd}(2008)}]{Lloyd2008}%
  \BibitemOpen
  \bibfield  {author} {\bibinfo {author} {\bibfnamefont {S.}~\bibnamefont
  {Lloyd}},\ }\href {\doibase 10.1126/science.1160627} {\bibfield  {journal}
  {\bibinfo  {journal} {Science}\ }\textbf {\bibinfo {volume} {321}},\ \bibinfo
  {pages} {1463} (\bibinfo {year} {2008})}\BibitemShut {NoStop}%
\bibitem [{\citenamefont {Tan}\ \emph {et~al.}(2008)\citenamefont {Tan},
  \citenamefont {Erkmen}, \citenamefont {Giovannetti}, \citenamefont {Guha},
  \citenamefont {Lloyd}, \citenamefont {Maccone}, \citenamefont {Pirandola},\
  and\ \citenamefont {Shapiro}}]{tan2008quantum}%
  \BibitemOpen
  \bibfield  {author} {\bibinfo {author} {\bibfnamefont {S.-H.}\ \bibnamefont
  {Tan}}, \bibinfo {author} {\bibfnamefont {B.~I.}\ \bibnamefont {Erkmen}},
  \bibinfo {author} {\bibfnamefont {V.}~\bibnamefont {Giovannetti}}, \bibinfo
  {author} {\bibfnamefont {S.}~\bibnamefont {Guha}}, \bibinfo {author}
  {\bibfnamefont {S.}~\bibnamefont {Lloyd}}, \bibinfo {author} {\bibfnamefont
  {L.}~\bibnamefont {Maccone}}, \bibinfo {author} {\bibfnamefont
  {S.}~\bibnamefont {Pirandola}}, \ and\ \bibinfo {author} {\bibfnamefont
  {J.~H.}\ \bibnamefont {Shapiro}},\ }\href@noop {} {\bibfield  {journal}
  {\bibinfo  {journal} {Phys. Rev. Lett.}\ }\textbf {\bibinfo {volume} {101}},\
  \bibinfo {pages} {253601} (\bibinfo {year} {2008})}\BibitemShut {NoStop}%
\bibitem [{\citenamefont {Zhuang}\ \emph
  {et~al.}(2017{\natexlab{a}})\citenamefont {Zhuang}, \citenamefont {Zhang},\
  and\ \citenamefont {Shapiro}}]{zhuang2017}%
  \BibitemOpen
  \bibfield  {author} {\bibinfo {author} {\bibfnamefont {Q.}~\bibnamefont
  {Zhuang}}, \bibinfo {author} {\bibfnamefont {Z.}~\bibnamefont {Zhang}}, \
  and\ \bibinfo {author} {\bibfnamefont {J.~H.}\ \bibnamefont {Shapiro}},\
  }\href {\doibase 10.1103/PhysRevLett.118.040801} {\bibfield  {journal}
  {\bibinfo  {journal} {Phys. Rev. Lett.}\ }\textbf {\bibinfo {volume} {118}},\
  \bibinfo {pages} {040801} (\bibinfo {year} {2017}{\natexlab{a}})}\BibitemShut
  {NoStop}%
\bibitem [{\citenamefont {Barzanjeh}\ \emph {et~al.}(2015)\citenamefont
  {Barzanjeh}, \citenamefont {Guha}, \citenamefont {Weedbrook}, \citenamefont
  {Vitali}, \citenamefont {Shapiro},\ and\ \citenamefont
  {Pirandola}}]{barzanjeh2015microwave}%
  \BibitemOpen
  \bibfield  {author} {\bibinfo {author} {\bibfnamefont {S.}~\bibnamefont
  {Barzanjeh}}, \bibinfo {author} {\bibfnamefont {S.}~\bibnamefont {Guha}},
  \bibinfo {author} {\bibfnamefont {C.}~\bibnamefont {Weedbrook}}, \bibinfo
  {author} {\bibfnamefont {D.}~\bibnamefont {Vitali}}, \bibinfo {author}
  {\bibfnamefont {J.~H.}\ \bibnamefont {Shapiro}}, \ and\ \bibinfo {author}
  {\bibfnamefont {S.}~\bibnamefont {Pirandola}},\ }\href@noop {} {\bibfield
  {journal} {\bibinfo  {journal} {Phys. Rev. Lett.}\ }\textbf {\bibinfo
  {volume} {114}},\ \bibinfo {pages} {080503} (\bibinfo {year}
  {2015})}\BibitemShut {NoStop}%
\bibitem [{\citenamefont {Zhang}\ \emph {et~al.}(2013)\citenamefont {Zhang},
  \citenamefont {Tengner}, \citenamefont {Zhong}, \citenamefont {Wong},\ and\
  \citenamefont {Shapiro}}]{zhang2013}%
  \BibitemOpen
  \bibfield  {author} {\bibinfo {author} {\bibfnamefont {Z.}~\bibnamefont
  {Zhang}}, \bibinfo {author} {\bibfnamefont {M.}~\bibnamefont {Tengner}},
  \bibinfo {author} {\bibfnamefont {T.}~\bibnamefont {Zhong}}, \bibinfo
  {author} {\bibfnamefont {F.~N.~C.}\ \bibnamefont {Wong}}, \ and\ \bibinfo
  {author} {\bibfnamefont {J.~H.}\ \bibnamefont {Shapiro}},\ }\href {\doibase
  10.1103/PhysRevLett.111.010501} {\bibfield  {journal} {\bibinfo  {journal}
  {Phys. Rev. Lett.}\ }\textbf {\bibinfo {volume} {111}},\ \bibinfo {pages}
  {010501} (\bibinfo {year} {2013})}\BibitemShut {NoStop}%
\bibitem [{\citenamefont {Zhang}\ \emph
  {et~al.}(2015{\natexlab{a}})\citenamefont {Zhang}, \citenamefont {Mouradian},
  \citenamefont {Wong},\ and\ \citenamefont {Shapiro}}]{zhang2015}%
  \BibitemOpen
  \bibfield  {author} {\bibinfo {author} {\bibfnamefont {Z.}~\bibnamefont
  {Zhang}}, \bibinfo {author} {\bibfnamefont {S.}~\bibnamefont {Mouradian}},
  \bibinfo {author} {\bibfnamefont {F.~N.~C.}\ \bibnamefont {Wong}}, \ and\
  \bibinfo {author} {\bibfnamefont {J.~H.}\ \bibnamefont {Shapiro}},\ }\href
  {\doibase 10.1103/PhysRevLett.114.110506} {\bibfield  {journal} {\bibinfo
  {journal} {Phys. Rev. Lett.}\ }\textbf {\bibinfo {volume} {114}},\ \bibinfo
  {pages} {110506} (\bibinfo {year} {2015}{\natexlab{a}})}\BibitemShut
  {NoStop}%
\bibitem [{\citenamefont {Barzanjeh}\ \emph {et~al.}(2019)\citenamefont
  {Barzanjeh}, \citenamefont {Pirandola}, \citenamefont {Vitali},\ and\
  \citenamefont {Fink}}]{barzanjeh2019experimental}%
  \BibitemOpen
  \bibfield  {author} {\bibinfo {author} {\bibfnamefont {S.}~\bibnamefont
  {Barzanjeh}}, \bibinfo {author} {\bibfnamefont {S.}~\bibnamefont
  {Pirandola}}, \bibinfo {author} {\bibfnamefont {D.}~\bibnamefont {Vitali}}, \
  and\ \bibinfo {author} {\bibfnamefont {J.}~\bibnamefont {Fink}},\ }\href@noop
  {} {\bibfield  {journal} {\bibinfo  {journal} {arXiv:1908.03058}\ } (\bibinfo
  {year} {2019})}\BibitemShut {NoStop}%
\bibitem [{\citenamefont {Bennett}\ and\ \citenamefont
  {Wiesner}(1992)}]{bennett1992}%
  \BibitemOpen
  \bibfield  {author} {\bibinfo {author} {\bibfnamefont {C.~H.}\ \bibnamefont
  {Bennett}}\ and\ \bibinfo {author} {\bibfnamefont {S.~J.}\ \bibnamefont
  {Wiesner}},\ }\href@noop {} {\bibfield  {journal} {\bibinfo  {journal} {Phys.
  Rev. Lett.}\ }\textbf {\bibinfo {volume} {69}},\ \bibinfo {pages} {2881}
  (\bibinfo {year} {1992})}\BibitemShut {NoStop}%
\bibitem [{\citenamefont {Bennett}\ \emph {et~al.}(1999)\citenamefont
  {Bennett}, \citenamefont {Shor}, \citenamefont {Smolin},\ and\ \citenamefont
  {Thapliyal}}]{bennett1999entanglement}%
  \BibitemOpen
  \bibfield  {author} {\bibinfo {author} {\bibfnamefont {C.~H.}\ \bibnamefont
  {Bennett}}, \bibinfo {author} {\bibfnamefont {P.~W.}\ \bibnamefont {Shor}},
  \bibinfo {author} {\bibfnamefont {J.~A.}\ \bibnamefont {Smolin}}, \ and\
  \bibinfo {author} {\bibfnamefont {A.~V.}\ \bibnamefont {Thapliyal}},\
  }\href@noop {} {\bibfield  {journal} {\bibinfo  {journal} {Phys. Rev. Lett.}\
  }\textbf {\bibinfo {volume} {83}},\ \bibinfo {pages} {3081} (\bibinfo {year}
  {1999})}\BibitemShut {NoStop}%
\bibitem [{\citenamefont {Holevo}(2002)}]{holevo02}%
  \BibitemOpen
  \bibfield  {author} {\bibinfo {author} {\bibfnamefont {A.~S.}\ \bibnamefont
  {Holevo}},\ }\href {\doibase 10.1063/1.1495877} {\bibfield  {journal}
  {\bibinfo  {journal} {J. Math. Phys.}\ }\textbf {\bibinfo {volume} {43}},\
  \bibinfo {pages} {4326} (\bibinfo {year} {2002})}\BibitemShut {NoStop}%
\bibitem [{\citenamefont {Hsieh}\ \emph {et~al.}(2008)\citenamefont {Hsieh},
  \citenamefont {Devetak},\ and\ \citenamefont
  {Winter}}]{hsieh2008entanglement}%
  \BibitemOpen
  \bibfield  {author} {\bibinfo {author} {\bibfnamefont {M.-H.}\ \bibnamefont
  {Hsieh}}, \bibinfo {author} {\bibfnamefont {I.}~\bibnamefont {Devetak}}, \
  and\ \bibinfo {author} {\bibfnamefont {A.}~\bibnamefont {Winter}},\
  }\href@noop {} {\bibfield  {journal} {\bibinfo  {journal} {IEEE Trans. Inf.
  Theory}\ }\textbf {\bibinfo {volume} {54}},\ \bibinfo {pages} {3078}
  (\bibinfo {year} {2008})}\BibitemShut {NoStop}%
\bibitem [{Note1()}]{Note1}%
  \BibitemOpen
  \bibinfo {note} {The initial proof is for finite-dimensional systems, and
  Refs.~\cite {holevo2003entanglement,holevo2013classical} proved the EA
  capacity formula for infinite dimensional channel with more
  rigor.}\BibitemShut {Stop}%
\bibitem [{\citenamefont {Hausladen}\ \emph {et~al.}(1996)\citenamefont
  {Hausladen}, \citenamefont {Jozsa}, \citenamefont {Schumacher}, \citenamefont
  {Westmoreland},\ and\ \citenamefont {Wootters}}]{hausladen1996classical}%
  \BibitemOpen
  \bibfield  {author} {\bibinfo {author} {\bibfnamefont {P.}~\bibnamefont
  {Hausladen}}, \bibinfo {author} {\bibfnamefont {R.}~\bibnamefont {Jozsa}},
  \bibinfo {author} {\bibfnamefont {B.}~\bibnamefont {Schumacher}}, \bibinfo
  {author} {\bibfnamefont {M.}~\bibnamefont {Westmoreland}}, \ and\ \bibinfo
  {author} {\bibfnamefont {W.~K.}\ \bibnamefont {Wootters}},\ }\href@noop {}
  {\bibfield  {journal} {\bibinfo  {journal} {Phys. Rev. A}\ }\textbf {\bibinfo
  {volume} {54}},\ \bibinfo {pages} {1869} (\bibinfo {year}
  {1996})}\BibitemShut {NoStop}%
\bibitem [{\citenamefont {Schumacher}\ and\ \citenamefont
  {Westmoreland}(1997)}]{schumacher1997sending}%
  \BibitemOpen
  \bibfield  {author} {\bibinfo {author} {\bibfnamefont {B.}~\bibnamefont
  {Schumacher}}\ and\ \bibinfo {author} {\bibfnamefont {M.~D.}\ \bibnamefont
  {Westmoreland}},\ }\href@noop {} {\bibfield  {journal} {\bibinfo  {journal}
  {Phys. Rev. A}\ }\textbf {\bibinfo {volume} {56}},\ \bibinfo {pages} {131}
  (\bibinfo {year} {1997})}\BibitemShut {NoStop}%
\bibitem [{\citenamefont {Holevo}(1998)}]{holevo1998capacity}%
  \BibitemOpen
  \bibfield  {author} {\bibinfo {author} {\bibfnamefont {A.~S.}\ \bibnamefont
  {Holevo}},\ }\href@noop {} {\bibfield  {journal} {\bibinfo  {journal} {IEEE
  Trans. Inf. Theory}\ }\textbf {\bibinfo {volume} {44}},\ \bibinfo {pages}
  {269} (\bibinfo {year} {1998})}\BibitemShut {NoStop}%
\bibitem [{Note2()}]{Note2}%
  \BibitemOpen
  \bibinfo {note} {Similar large improvement can also happen in large-dimension
  depolarizing channels~\cite {holevo02,holevo2013information}}\BibitemShut
  {NoStop}%
\bibitem [{\citenamefont {Banaszek}\ \emph {et~al.}(2019)\citenamefont
  {Banaszek}, \citenamefont {Kunz}, \citenamefont {Jarzyna},\ and\
  \citenamefont {Jachura}}]{banaszek2019approaching}%
  \BibitemOpen
  \bibfield  {author} {\bibinfo {author} {\bibfnamefont {K.}~\bibnamefont
  {Banaszek}}, \bibinfo {author} {\bibfnamefont {L.}~\bibnamefont {Kunz}},
  \bibinfo {author} {\bibfnamefont {M.}~\bibnamefont {Jarzyna}}, \ and\
  \bibinfo {author} {\bibfnamefont {M.}~\bibnamefont {Jachura}},\ }\href@noop
  {} {\bibfield  {journal} {\bibinfo  {journal} {Proc. SPIE 10910, Free-Space
  Laser Communications XXXI}\ ,\ \bibinfo {pages} {109100A}} (\bibinfo {year}
  {2019})}\BibitemShut {NoStop}%
\bibitem [{\citenamefont {Bash}\ \emph {et~al.}(2015)\citenamefont {Bash},
  \citenamefont {Gheorghe}, \citenamefont {Patel}, \citenamefont {Habif},
  \citenamefont {Goeckel}, \citenamefont {Towsley},\ and\ \citenamefont
  {Guha}}]{bash2015quantum}%
  \BibitemOpen
  \bibfield  {author} {\bibinfo {author} {\bibfnamefont {B.~A.}\ \bibnamefont
  {Bash}}, \bibinfo {author} {\bibfnamefont {A.~H.}\ \bibnamefont {Gheorghe}},
  \bibinfo {author} {\bibfnamefont {M.}~\bibnamefont {Patel}}, \bibinfo
  {author} {\bibfnamefont {J.~L.}\ \bibnamefont {Habif}}, \bibinfo {author}
  {\bibfnamefont {D.}~\bibnamefont {Goeckel}}, \bibinfo {author} {\bibfnamefont
  {D.}~\bibnamefont {Towsley}}, \ and\ \bibinfo {author} {\bibfnamefont
  {S.}~\bibnamefont {Guha}},\ }\href@noop {} {\bibfield  {journal} {\bibinfo
  {journal} {Nat. Commun.}\ }\textbf {\bibinfo {volume} {6}},\ \bibinfo {pages}
  {8626} (\bibinfo {year} {2015})}\BibitemShut {NoStop}%
\bibitem [{\citenamefont {Bullock}\ \emph {et~al.}(2019)\citenamefont
  {Bullock}, \citenamefont {Gagatsos}, \citenamefont {Guha},\ and\
  \citenamefont {Bash}}]{bullock2019fundamental}%
  \BibitemOpen
  \bibfield  {author} {\bibinfo {author} {\bibfnamefont {M.~S.}\ \bibnamefont
  {Bullock}}, \bibinfo {author} {\bibfnamefont {C.~N.}\ \bibnamefont
  {Gagatsos}}, \bibinfo {author} {\bibfnamefont {S.}~\bibnamefont {Guha}}, \
  and\ \bibinfo {author} {\bibfnamefont {B.~A.}\ \bibnamefont {Bash}},\
  }\href@noop {} {\bibfield  {journal} {\bibinfo  {journal} {arXiv:1907.04228}\
  } (\bibinfo {year} {2019})}\BibitemShut {NoStop}%
\bibitem [{\citenamefont {Prevedel}\ \emph {et~al.}(2011)\citenamefont
  {Prevedel}, \citenamefont {Lu}, \citenamefont {Matthews}, \citenamefont
  {Kaltenbaek},\ and\ \citenamefont {Resch}}]{prevedel_2011}%
  \BibitemOpen
  \bibfield  {author} {\bibinfo {author} {\bibfnamefont {R.}~\bibnamefont
  {Prevedel}}, \bibinfo {author} {\bibfnamefont {Y.}~\bibnamefont {Lu}},
  \bibinfo {author} {\bibfnamefont {W.}~\bibnamefont {Matthews}}, \bibinfo
  {author} {\bibfnamefont {R.}~\bibnamefont {Kaltenbaek}}, \ and\ \bibinfo
  {author} {\bibfnamefont {K.~J.}\ \bibnamefont {Resch}},\ }\href {\doibase
  10.1103/PhysRevLett.106.110505} {\bibfield  {journal} {\bibinfo  {journal}
  {Phys. Rev. Lett.}\ }\textbf {\bibinfo {volume} {106}},\ \bibinfo {pages}
  {110505} (\bibinfo {year} {2011})}\BibitemShut {NoStop}%
\bibitem [{\citenamefont {Chiuri}\ \emph {et~al.}(2013)\citenamefont {Chiuri},
  \citenamefont {Giacomini}, \citenamefont {Macchiavello},\ and\ \citenamefont
  {Mataloni}}]{chiuri_2013}%
  \BibitemOpen
  \bibfield  {author} {\bibinfo {author} {\bibfnamefont {A.}~\bibnamefont
  {Chiuri}}, \bibinfo {author} {\bibfnamefont {S.}~\bibnamefont {Giacomini}},
  \bibinfo {author} {\bibfnamefont {C.}~\bibnamefont {Macchiavello}}, \ and\
  \bibinfo {author} {\bibfnamefont {P.}~\bibnamefont {Mataloni}},\ }\href
  {\doibase 10.1103/PhysRevA.87.022333} {\bibfield  {journal} {\bibinfo
  {journal} {Phys. Rev. A}\ }\textbf {\bibinfo {volume} {87}},\ \bibinfo
  {pages} {022333} (\bibinfo {year} {2013})}\BibitemShut {NoStop}%
\bibitem [{\citenamefont {Holevo}\ and\ \citenamefont
  {Werner}(2001)}]{holevo01}%
  \BibitemOpen
  \bibfield  {author} {\bibinfo {author} {\bibfnamefont {A.~S.}\ \bibnamefont
  {Holevo}}\ and\ \bibinfo {author} {\bibfnamefont {R.~F.}\ \bibnamefont
  {Werner}},\ }\href {\doibase 10.1103/PhysRevA.63.032312} {\bibfield
  {journal} {\bibinfo  {journal} {Phys. Rev. A}\ }\textbf {\bibinfo {volume}
  {63}},\ \bibinfo {pages} {032312} (\bibinfo {year} {2001})}\BibitemShut
  {NoStop}%
\bibitem [{\citenamefont {De~Palma}\ \emph {et~al.}(2017)\citenamefont
  {De~Palma}, \citenamefont {Trevisan},\ and\ \citenamefont
  {Giovannetti}}]{de2017gaussian}%
  \BibitemOpen
  \bibfield  {author} {\bibinfo {author} {\bibfnamefont {G.}~\bibnamefont
  {De~Palma}}, \bibinfo {author} {\bibfnamefont {D.}~\bibnamefont {Trevisan}},
  \ and\ \bibinfo {author} {\bibfnamefont {V.}~\bibnamefont {Giovannetti}},\
  }\href@noop {} {\bibfield  {journal} {\bibinfo  {journal} {Phys. Rev. Lett.}\
  }\textbf {\bibinfo {volume} {118}},\ \bibinfo {pages} {160503} (\bibinfo
  {year} {2017})}\BibitemShut {NoStop}%
\bibitem [{\citenamefont {Ban}(1999)}]{ban1999quantum}%
  \BibitemOpen
  \bibfield  {author} {\bibinfo {author} {\bibfnamefont {M.}~\bibnamefont
  {Ban}},\ }\href@noop {} {\bibfield  {journal} {\bibinfo  {journal} {J. Opt.
  B: Quantum Semiclassical Opt.}\ }\textbf {\bibinfo {volume} {1}},\ \bibinfo
  {pages} {L9} (\bibinfo {year} {1999})}\BibitemShut {NoStop}%
\bibitem [{\citenamefont {Braunstein}\ and\ \citenamefont
  {Kimble}(2000)}]{braunstein2000}%
  \BibitemOpen
  \bibfield  {author} {\bibinfo {author} {\bibfnamefont {S.~L.}\ \bibnamefont
  {Braunstein}}\ and\ \bibinfo {author} {\bibfnamefont {H.~J.}\ \bibnamefont
  {Kimble}},\ }\href {\doibase 10.1103/PhysRevA.61.042302} {\bibfield
  {journal} {\bibinfo  {journal} {Phys. Rev. A}\ }\textbf {\bibinfo {volume}
  {61}},\ \bibinfo {pages} {042302} (\bibinfo {year} {2000})}\BibitemShut
  {NoStop}%
\bibitem [{\citenamefont {Ban}(2000)}]{ban2000quantum}%
  \BibitemOpen
  \bibfield  {author} {\bibinfo {author} {\bibfnamefont {M.}~\bibnamefont
  {Ban}},\ }\href@noop {} {\bibfield  {journal} {\bibinfo  {journal} {J. Opt.
  B: Quantum Semiclassical Opt.}\ }\textbf {\bibinfo {volume} {2}},\ \bibinfo
  {pages} {786} (\bibinfo {year} {2000})}\BibitemShut {NoStop}%
\bibitem [{\citenamefont {Sohma}\ and\ \citenamefont
  {Hirota}(2003)}]{sohma2003}%
  \BibitemOpen
  \bibfield  {author} {\bibinfo {author} {\bibfnamefont {M.}~\bibnamefont
  {Sohma}}\ and\ \bibinfo {author} {\bibfnamefont {O.}~\bibnamefont {Hirota}},\
  }\href {\doibase 10.1103/PhysRevA.68.022303} {\bibfield  {journal} {\bibinfo
  {journal} {Phys. Rev. A}\ }\textbf {\bibinfo {volume} {68}},\ \bibinfo
  {pages} {022303} (\bibinfo {year} {2003})}\BibitemShut {NoStop}%
\bibitem [{\citenamefont {Mizuno}\ \emph {et~al.}(2005)\citenamefont {Mizuno},
  \citenamefont {Wakui}, \citenamefont {Furusawa},\ and\ \citenamefont
  {Sasaki}}]{mizuno2005experimental}%
  \BibitemOpen
  \bibfield  {author} {\bibinfo {author} {\bibfnamefont {J.}~\bibnamefont
  {Mizuno}}, \bibinfo {author} {\bibfnamefont {K.}~\bibnamefont {Wakui}},
  \bibinfo {author} {\bibfnamefont {A.}~\bibnamefont {Furusawa}}, \ and\
  \bibinfo {author} {\bibfnamefont {M.}~\bibnamefont {Sasaki}},\ }\href@noop {}
  {\bibfield  {journal} {\bibinfo  {journal} {Phys. Rev. A}\ }\textbf {\bibinfo
  {volume} {71}},\ \bibinfo {pages} {012304} (\bibinfo {year}
  {2005})}\BibitemShut {NoStop}%
\bibitem [{\citenamefont {Barzanjeh}\ \emph {et~al.}(2013)\citenamefont
  {Barzanjeh}, \citenamefont {Pirandola},\ and\ \citenamefont
  {Weedbrook}}]{barzanjeh2013}%
  \BibitemOpen
  \bibfield  {author} {\bibinfo {author} {\bibfnamefont {S.}~\bibnamefont
  {Barzanjeh}}, \bibinfo {author} {\bibfnamefont {S.}~\bibnamefont
  {Pirandola}}, \ and\ \bibinfo {author} {\bibfnamefont {C.}~\bibnamefont
  {Weedbrook}},\ }\href {\doibase 10.1103/PhysRevA.88.042331} {\bibfield
  {journal} {\bibinfo  {journal} {Phys. Rev. A}\ }\textbf {\bibinfo {volume}
  {88}},\ \bibinfo {pages} {042331} (\bibinfo {year} {2013})}\BibitemShut
  {NoStop}%
\bibitem [{\citenamefont {Li}\ \emph {et~al.}(2002)\citenamefont {Li},
  \citenamefont {Pan}, \citenamefont {Jing}, \citenamefont {Zhang},
  \citenamefont {Xie},\ and\ \citenamefont {Peng}}]{li2002quantum}%
  \BibitemOpen
  \bibfield  {author} {\bibinfo {author} {\bibfnamefont {X.}~\bibnamefont
  {Li}}, \bibinfo {author} {\bibfnamefont {Q.}~\bibnamefont {Pan}}, \bibinfo
  {author} {\bibfnamefont {J.}~\bibnamefont {Jing}}, \bibinfo {author}
  {\bibfnamefont {J.}~\bibnamefont {Zhang}}, \bibinfo {author} {\bibfnamefont
  {C.}~\bibnamefont {Xie}}, \ and\ \bibinfo {author} {\bibfnamefont
  {K.}~\bibnamefont {Peng}},\ }\href@noop {} {\bibfield  {journal} {\bibinfo
  {journal} {Phys. Rev. Lett.}\ }\textbf {\bibinfo {volume} {88}},\ \bibinfo
  {pages} {047904} (\bibinfo {year} {2002})}\BibitemShut {NoStop}%
\bibitem [{\citenamefont {Wilde}\ \emph
  {et~al.}(2012{\natexlab{a}})\citenamefont {Wilde}, \citenamefont {Hayden},\
  and\ \citenamefont {Guha}}]{wilde2012information}%
  \BibitemOpen
  \bibfield  {author} {\bibinfo {author} {\bibfnamefont {M.~M.}\ \bibnamefont
  {Wilde}}, \bibinfo {author} {\bibfnamefont {P.}~\bibnamefont {Hayden}}, \
  and\ \bibinfo {author} {\bibfnamefont {S.}~\bibnamefont {Guha}},\ }\href@noop
  {} {\bibfield  {journal} {\bibinfo  {journal} {Phys. Rev. Lett.}\ }\textbf
  {\bibinfo {volume} {108}},\ \bibinfo {pages} {140501} (\bibinfo {year}
  {2012}{\natexlab{a}})}\BibitemShut {NoStop}%
\bibitem [{\citenamefont {Anshu}\ \emph {et~al.}(2019)\citenamefont {Anshu},
  \citenamefont {Jain},\ and\ \citenamefont {Warsi}}]{anshu2019building}%
  \BibitemOpen
  \bibfield  {author} {\bibinfo {author} {\bibfnamefont {A.}~\bibnamefont
  {Anshu}}, \bibinfo {author} {\bibfnamefont {R.}~\bibnamefont {Jain}}, \ and\
  \bibinfo {author} {\bibfnamefont {N.~A.}\ \bibnamefont {Warsi}},\ }\href@noop
  {} {\bibfield  {journal} {\bibinfo  {journal} {IEEE Trans. Inf. Theory}\
  }\textbf {\bibinfo {volume} {65}},\ \bibinfo {pages} {1287} (\bibinfo {year}
  {2019})}\BibitemShut {NoStop}%
\bibitem [{\citenamefont {Qi}\ \emph {et~al.}(2018)\citenamefont {Qi},
  \citenamefont {Wang},\ and\ \citenamefont {Wilde}}]{qi2018applications}%
  \BibitemOpen
  \bibfield  {author} {\bibinfo {author} {\bibfnamefont {H.}~\bibnamefont
  {Qi}}, \bibinfo {author} {\bibfnamefont {Q.}~\bibnamefont {Wang}}, \ and\
  \bibinfo {author} {\bibfnamefont {M.~M.}\ \bibnamefont {Wilde}},\ }\href@noop
  {} {\bibfield  {journal} {\bibinfo  {journal} {J. Phys. A: Math. Theor.}\
  }\textbf {\bibinfo {volume} {51}},\ \bibinfo {pages} {444002} (\bibinfo
  {year} {2018})}\BibitemShut {NoStop}%
\bibitem [{\citenamefont {Khabbazi~Oskouei}\ \emph {et~al.}(2019)\citenamefont
  {Khabbazi~Oskouei}, \citenamefont {Mancini},\ and\ \citenamefont
  {Wilde}}]{khabbazi2019union}%
  \BibitemOpen
  \bibfield  {author} {\bibinfo {author} {\bibfnamefont {S.}~\bibnamefont
  {Khabbazi~Oskouei}}, \bibinfo {author} {\bibfnamefont {S.}~\bibnamefont
  {Mancini}}, \ and\ \bibinfo {author} {\bibfnamefont {M.~M.}\ \bibnamefont
  {Wilde}},\ }\href@noop {} {\bibfield  {journal} {\bibinfo  {journal} {Proc.
  Royal Soc. Lond.}\ }\textbf {\bibinfo {volume} {475}},\ \bibinfo {pages}
  {20180612} (\bibinfo {year} {2019})}\BibitemShut {NoStop}%
\bibitem [{\citenamefont {Guha}\ and\ \citenamefont {Erkmen}(2009)}]{Guha2009}%
  \BibitemOpen
  \bibfield  {author} {\bibinfo {author} {\bibfnamefont {S.}~\bibnamefont
  {Guha}}\ and\ \bibinfo {author} {\bibfnamefont {B.~I.}\ \bibnamefont
  {Erkmen}},\ }\href {\doibase 10.1103/PhysRevA.80.052310} {\bibfield
  {journal} {\bibinfo  {journal} {Phys. Rev. A}\ }\textbf {\bibinfo {volume}
  {80}},\ \bibinfo {pages} {052310} (\bibinfo {year} {2009})}\BibitemShut
  {NoStop}%
\bibitem [{\citenamefont {Gagatsos}\ \emph {et~al.}(2017)\citenamefont
  {Gagatsos}, \citenamefont {Bash}, \citenamefont {Guha},\ and\ \citenamefont
  {Datta}}]{gagatsos2017bounding}%
  \BibitemOpen
  \bibfield  {author} {\bibinfo {author} {\bibfnamefont {C.~N.}\ \bibnamefont
  {Gagatsos}}, \bibinfo {author} {\bibfnamefont {B.~A.}\ \bibnamefont {Bash}},
  \bibinfo {author} {\bibfnamefont {S.}~\bibnamefont {Guha}}, \ and\ \bibinfo
  {author} {\bibfnamefont {A.}~\bibnamefont {Datta}},\ }\href@noop {}
  {\bibfield  {journal} {\bibinfo  {journal} {Phys. Rev. A}\ }\textbf {\bibinfo
  {volume} {96}},\ \bibinfo {pages} {062306} (\bibinfo {year}
  {2017})}\BibitemShut {NoStop}%
\bibitem [{\citenamefont {Giovannetti}\ \emph {et~al.}(2014)\citenamefont
  {Giovannetti}, \citenamefont {Garc{\'\i}a-Patr{\'o}n}, \citenamefont {Cerf},\
  and\ \citenamefont {Holevo}}]{GiovannettiV2014}%
  \BibitemOpen
  \bibfield  {author} {\bibinfo {author} {\bibfnamefont {V.}~\bibnamefont
  {Giovannetti}}, \bibinfo {author} {\bibfnamefont {R.}~\bibnamefont
  {Garc{\'\i}a-Patr{\'o}n}}, \bibinfo {author} {\bibfnamefont {N.~J.}\
  \bibnamefont {Cerf}}, \ and\ \bibinfo {author} {\bibfnamefont {A.~S.}\
  \bibnamefont {Holevo}},\ }\href@noop {} {\bibfield  {journal} {\bibinfo
  {journal} {Nat. Photon.}\ }\textbf {\bibinfo {volume} {8}},\ \bibinfo {pages}
  {796} (\bibinfo {year} {2014})}\BibitemShut {NoStop}%
\bibitem [{\citenamefont {Holevo}(1973)}]{holevo1973bounds}%
  \BibitemOpen
  \bibfield  {author} {\bibinfo {author} {\bibfnamefont {A.~S.}\ \bibnamefont
  {Holevo}},\ }\href@noop {} {\bibfield  {journal} {\bibinfo  {journal}
  {Problemy Peredachi Informatsii}\ }\textbf {\bibinfo {volume} {9}},\ \bibinfo
  {pages} {3} (\bibinfo {year} {1973})}\BibitemShut {NoStop}%
\bibitem [{\citenamefont {Shapiro}(2009)}]{Shapiro2009ieee}%
  \BibitemOpen
  \bibfield  {author} {\bibinfo {author} {\bibfnamefont {J.~H.}\ \bibnamefont
  {Shapiro}},\ }\href@noop {} {\bibfield  {journal} {\bibinfo  {journal} {IEEE
  J. Sel. Top. Quantum Electron}\ }\textbf {\bibinfo {volume} {15}},\ \bibinfo
  {pages} {1547} (\bibinfo {year} {2009})}\BibitemShut {NoStop}%
\bibitem [{\citenamefont {Shor}(2004)}]{shor2004classical}%
  \BibitemOpen
  \bibfield  {author} {\bibinfo {author} {\bibfnamefont {P.~W.}\ \bibnamefont
  {Shor}},\ }\href@noop {} {\bibfield  {journal} {\bibinfo  {journal} {arXiv
  quant-ph/0402129}\ } (\bibinfo {year} {2004})}\BibitemShut {NoStop}%
\bibitem [{\citenamefont {Zhuang}\ \emph
  {et~al.}(2017{\natexlab{b}})\citenamefont {Zhuang}, \citenamefont {Zhu},\
  and\ \citenamefont {Shor}}]{zhuang2017additive}%
  \BibitemOpen
  \bibfield  {author} {\bibinfo {author} {\bibfnamefont {Q.}~\bibnamefont
  {Zhuang}}, \bibinfo {author} {\bibfnamefont {E.~Y.}\ \bibnamefont {Zhu}}, \
  and\ \bibinfo {author} {\bibfnamefont {P.~W.}\ \bibnamefont {Shor}},\
  }\href@noop {} {\bibfield  {journal} {\bibinfo  {journal} {Phys. Rev. Lett.}\
  }\textbf {\bibinfo {volume} {118}},\ \bibinfo {pages} {200503} (\bibinfo
  {year} {2017}{\natexlab{b}})}\BibitemShut {NoStop}%
\bibitem [{\citenamefont {Wilde}\ and\ \citenamefont
  {Hsieh}(2012)}]{wilde2012quantum}%
  \BibitemOpen
  \bibfield  {author} {\bibinfo {author} {\bibfnamefont {M.~M.}\ \bibnamefont
  {Wilde}}\ and\ \bibinfo {author} {\bibfnamefont {M.-H.}\ \bibnamefont
  {Hsieh}},\ }\href@noop {} {\bibfield  {journal} {\bibinfo  {journal} {Quantum
  Inf. Process.}\ }\textbf {\bibinfo {volume} {11}},\ \bibinfo {pages} {1431}
  (\bibinfo {year} {2012})}\BibitemShut {NoStop}%
\bibitem [{\citenamefont {Zhu}\ \emph {et~al.}(2017)\citenamefont {Zhu},
  \citenamefont {Zhuang},\ and\ \citenamefont {Shor}}]{zhu2017}%
  \BibitemOpen
  \bibfield  {author} {\bibinfo {author} {\bibfnamefont {E.~Y.}\ \bibnamefont
  {Zhu}}, \bibinfo {author} {\bibfnamefont {Q.}~\bibnamefont {Zhuang}}, \ and\
  \bibinfo {author} {\bibfnamefont {P.~W.}\ \bibnamefont {Shor}},\ }\href
  {\doibase 10.1103/PhysRevLett.119.040503} {\bibfield  {journal} {\bibinfo
  {journal} {Phys. Rev. Lett.}\ }\textbf {\bibinfo {volume} {119}},\ \bibinfo
  {pages} {040503} (\bibinfo {year} {2017})}\BibitemShut {NoStop}%
\bibitem [{\citenamefont {Zhu}\ \emph {et~al.}(2018)\citenamefont {Zhu},
  \citenamefont {Zhuang}, \citenamefont {Hsieh},\ and\ \citenamefont
  {Shor}}]{zhu2018superadditivity}%
  \BibitemOpen
  \bibfield  {author} {\bibinfo {author} {\bibfnamefont {E.~Y.}\ \bibnamefont
  {Zhu}}, \bibinfo {author} {\bibfnamefont {Q.}~\bibnamefont {Zhuang}},
  \bibinfo {author} {\bibfnamefont {M.-H.}\ \bibnamefont {Hsieh}}, \ and\
  \bibinfo {author} {\bibfnamefont {P.~W.}\ \bibnamefont {Shor}},\ }\href@noop
  {} {\bibfield  {journal} {\bibinfo  {journal} {IEEE Trans. Inf. Theory}\ }
  (\bibinfo {year} {2018})}\BibitemShut {NoStop}%
\bibitem [{\citenamefont {Weedbrook}\ \emph {et~al.}(2012)\citenamefont
  {Weedbrook}, \citenamefont {Pirandola}, \citenamefont {Garc\'{\i}a-Patr\'on},
  \citenamefont {Cerf}, \citenamefont {Ralph}, \citenamefont {Shapiro},\ and\
  \citenamefont {Lloyd}}]{Weedbrook_2012}%
  \BibitemOpen
  \bibfield  {author} {\bibinfo {author} {\bibfnamefont {C.}~\bibnamefont
  {Weedbrook}}, \bibinfo {author} {\bibfnamefont {S.}~\bibnamefont
  {Pirandola}}, \bibinfo {author} {\bibfnamefont {R.}~\bibnamefont
  {Garc\'{\i}a-Patr\'on}}, \bibinfo {author} {\bibfnamefont {N.~J.}\
  \bibnamefont {Cerf}}, \bibinfo {author} {\bibfnamefont {T.~C.}\ \bibnamefont
  {Ralph}}, \bibinfo {author} {\bibfnamefont {J.~H.}\ \bibnamefont {Shapiro}},
  \ and\ \bibinfo {author} {\bibfnamefont {S.}~\bibnamefont {Lloyd}},\ }\href
  {\doibase 10.1103/RevModPhys.84.621} {\bibfield  {journal} {\bibinfo
  {journal} {Rev. Mod. Phys.}\ }\textbf {\bibinfo {volume} {84}},\ \bibinfo
  {pages} {621} (\bibinfo {year} {2012})}\BibitemShut {NoStop}%
\bibitem [{\citenamefont {Zhuang}\ \emph {et~al.}(2016)\citenamefont {Zhuang},
  \citenamefont {Zhang}, \citenamefont {Dove}, \citenamefont {Wong},\ and\
  \citenamefont {Shapiro}}]{zhuang2016floodlight}%
  \BibitemOpen
  \bibfield  {author} {\bibinfo {author} {\bibfnamefont {Q.}~\bibnamefont
  {Zhuang}}, \bibinfo {author} {\bibfnamefont {Z.}~\bibnamefont {Zhang}},
  \bibinfo {author} {\bibfnamefont {J.}~\bibnamefont {Dove}}, \bibinfo {author}
  {\bibfnamefont {F.~N.}\ \bibnamefont {Wong}}, \ and\ \bibinfo {author}
  {\bibfnamefont {J.~H.}\ \bibnamefont {Shapiro}},\ }\href@noop {} {\bibfield
  {journal} {\bibinfo  {journal} {Phys. Rev. A}\ }\textbf {\bibinfo {volume}
  {94}},\ \bibinfo {pages} {012322} (\bibinfo {year} {2016})}\BibitemShut
  {NoStop}%
\bibitem [{\citenamefont {Pirandola}\ and\ \citenamefont
  {Lloyd}(2008)}]{Pirandola2008}%
  \BibitemOpen
  \bibfield  {author} {\bibinfo {author} {\bibfnamefont {S.}~\bibnamefont
  {Pirandola}}\ and\ \bibinfo {author} {\bibfnamefont {S.}~\bibnamefont
  {Lloyd}},\ }\href {\doibase 10.1103/PhysRevA.78.012331} {\bibfield  {journal}
  {\bibinfo  {journal} {Phys. Rev. A}\ }\textbf {\bibinfo {volume} {78}},\
  \bibinfo {pages} {012331} (\bibinfo {year} {2008})}\BibitemShut {NoStop}%
\bibitem [{\citenamefont {Audenaert}\ \emph {et~al.}(2007)\citenamefont
  {Audenaert}, \citenamefont {Calsamiglia}, \citenamefont {Munoz-Tapia},
  \citenamefont {Bagan}, \citenamefont {Masanes}, \citenamefont {Acin},\ and\
  \citenamefont {Verstraete}}]{audenaert2007discriminating}%
  \BibitemOpen
  \bibfield  {author} {\bibinfo {author} {\bibfnamefont {K.~M.}\ \bibnamefont
  {Audenaert}}, \bibinfo {author} {\bibfnamefont {J.}~\bibnamefont
  {Calsamiglia}}, \bibinfo {author} {\bibfnamefont {R.}~\bibnamefont
  {Munoz-Tapia}}, \bibinfo {author} {\bibfnamefont {E.}~\bibnamefont {Bagan}},
  \bibinfo {author} {\bibfnamefont {L.}~\bibnamefont {Masanes}}, \bibinfo
  {author} {\bibfnamefont {A.}~\bibnamefont {Acin}}, \ and\ \bibinfo {author}
  {\bibfnamefont {F.}~\bibnamefont {Verstraete}},\ }\href@noop {} {\bibfield
  {journal} {\bibinfo  {journal} {Phys. Rev. Lett.}\ }\textbf {\bibinfo
  {volume} {98}},\ \bibinfo {pages} {160501} (\bibinfo {year}
  {2007})}\BibitemShut {NoStop}%
\bibitem [{\citenamefont {Pirandola}\ \emph {et~al.}(2017)\citenamefont
  {Pirandola}, \citenamefont {Laurenza}, \citenamefont {Ottaviani},\ and\
  \citenamefont {Banchi}}]{pirandola2017fundamental}%
  \BibitemOpen
  \bibfield  {author} {\bibinfo {author} {\bibfnamefont {S.}~\bibnamefont
  {Pirandola}}, \bibinfo {author} {\bibfnamefont {R.}~\bibnamefont {Laurenza}},
  \bibinfo {author} {\bibfnamefont {C.}~\bibnamefont {Ottaviani}}, \ and\
  \bibinfo {author} {\bibfnamefont {L.}~\bibnamefont {Banchi}},\ }\href@noop {}
  {\bibfield  {journal} {\bibinfo  {journal} {Nat. Commun.}\ }\textbf {\bibinfo
  {volume} {8}},\ \bibinfo {pages} {15043} (\bibinfo {year}
  {2017})}\BibitemShut {NoStop}%
\bibitem [{\citenamefont {Zhang}\ \emph
  {et~al.}(2015{\natexlab{b}})\citenamefont {Zhang}, \citenamefont {Mouradian},
  \citenamefont {Wong},\ and\ \citenamefont {Shapiro}}]{Zheshen_15}%
  \BibitemOpen
  \bibfield  {author} {\bibinfo {author} {\bibfnamefont {Z.}~\bibnamefont
  {Zhang}}, \bibinfo {author} {\bibfnamefont {S.}~\bibnamefont {Mouradian}},
  \bibinfo {author} {\bibfnamefont {F.~N.~C.}\ \bibnamefont {Wong}}, \ and\
  \bibinfo {author} {\bibfnamefont {J.~H.}\ \bibnamefont {Shapiro}},\ }\href
  {\doibase 10.1103/PhysRevLett.114.110506} {\bibfield  {journal} {\bibinfo
  {journal} {Phys. Rev. Lett.}\ }\textbf {\bibinfo {volume} {114}},\ \bibinfo
  {pages} {110506} (\bibinfo {year} {2015}{\natexlab{b}})}\BibitemShut
  {NoStop}%
\bibitem [{\citenamefont {Dolinar}(1973)}]{dolinar_processing_1973}%
  \BibitemOpen
  \bibfield  {author} {\bibinfo {author} {\bibfnamefont {S.~J.}\ \bibnamefont
  {Dolinar}},\ }\href@noop {} {\emph {\bibinfo {title} {Processing and
  {Transmission} of {Information}}}},\ \bibinfo {type} {Technical {Report}}\
  (\bibinfo  {institution} {Research Laboratory of Electronics (RLE) at the
  Massachusetts Institute of Technology (MIT)},\ \bibinfo {year}
  {1973})\BibitemShut {NoStop}%
\bibitem [{\citenamefont {Van~Trees}(2004)}]{van2004detection}%
  \BibitemOpen
  \bibfield  {author} {\bibinfo {author} {\bibfnamefont {H.~L.}\ \bibnamefont
  {Van~Trees}},\ }\href@noop {} {\emph {\bibinfo {title} {Detection,
  estimation, and modulation theory, part I: detection, estimation, and linear
  modulation theory}}}\ (\bibinfo  {publisher} {John Wiley \& Sons},\ \bibinfo
  {year} {2004})\BibitemShut {NoStop}%
\bibitem [{\citenamefont {Mart{\'\i}nez-Vargas}\ \emph
  {et~al.}(2017)\citenamefont {Mart{\'\i}nez-Vargas}, \citenamefont {Pineda},
  \citenamefont {Leyvraz},\ and\ \citenamefont
  {Barberis-Blostein}}]{martinez2017VanTrees}%
  \BibitemOpen
  \bibfield  {author} {\bibinfo {author} {\bibfnamefont {E.}~\bibnamefont
  {Mart{\'\i}nez-Vargas}}, \bibinfo {author} {\bibfnamefont {C.}~\bibnamefont
  {Pineda}}, \bibinfo {author} {\bibfnamefont {F.}~\bibnamefont {Leyvraz}}, \
  and\ \bibinfo {author} {\bibfnamefont {P.}~\bibnamefont
  {Barberis-Blostein}},\ }\href@noop {} {\bibfield  {journal} {\bibinfo
  {journal} {Phys. Rev. A}\ }\textbf {\bibinfo {volume} {95}},\ \bibinfo
  {pages} {012136} (\bibinfo {year} {2017})}\BibitemShut {NoStop}%
\bibitem [{\citenamefont {Paris}(2009)}]{paris2009VanTrees}%
  \BibitemOpen
  \bibfield  {author} {\bibinfo {author} {\bibfnamefont {M.~G.}\ \bibnamefont
  {Paris}},\ }\href@noop {} {\bibfield  {journal} {\bibinfo  {journal} {Int. J.
  Quant. Inf.}\ }\textbf {\bibinfo {volume} {7}},\ \bibinfo {pages} {125}
  (\bibinfo {year} {2009})}\BibitemShut {NoStop}%
\bibitem [{\citenamefont {Guha}(2011)}]{guha2011structured}%
  \BibitemOpen
  \bibfield  {author} {\bibinfo {author} {\bibfnamefont {S.}~\bibnamefont
  {Guha}},\ }\href@noop {} {\bibfield  {journal} {\bibinfo  {journal} {Phys.
  Rev. Lett.}\ }\textbf {\bibinfo {volume} {106}},\ \bibinfo {pages} {240502}
  (\bibinfo {year} {2011})}\BibitemShut {NoStop}%
\bibitem [{\citenamefont {Wilde}\ \emph
  {et~al.}(2012{\natexlab{b}})\citenamefont {Wilde}, \citenamefont {Guha},
  \citenamefont {Tan},\ and\ \citenamefont {Lloyd}}]{wilde2012explicit}%
  \BibitemOpen
  \bibfield  {author} {\bibinfo {author} {\bibfnamefont {M.~M.}\ \bibnamefont
  {Wilde}}, \bibinfo {author} {\bibfnamefont {S.}~\bibnamefont {Guha}},
  \bibinfo {author} {\bibfnamefont {S.-H.}\ \bibnamefont {Tan}}, \ and\
  \bibinfo {author} {\bibfnamefont {S.}~\bibnamefont {Lloyd}},\ }\href@noop {}
  {\bibfield  {journal} {\bibinfo  {journal} {IEEE ISIT}\ ,\ \bibinfo {pages}
  {551}} (\bibinfo {year} {2012}{\natexlab{b}})}\BibitemShut {NoStop}%
\bibitem [{\citenamefont {Holevo}(2003)}]{holevo2003entanglement}%
  \BibitemOpen
  \bibfield  {author} {\bibinfo {author} {\bibfnamefont {A.~S.}\ \bibnamefont
  {Holevo}},\ }in\ \href@noop {} {\emph {\bibinfo {booktitle} {First
  International Symposium on Quantum Informatics}}},\ Vol.\ \bibinfo {volume}
  {5128}\ (\bibinfo {organization} {International Society for Optics and
  Photonics},\ \bibinfo {year} {2003})\ pp.\ \bibinfo {pages}
  {62--70}\BibitemShut {NoStop}%
\bibitem [{\citenamefont {Holevo}\ and\ \citenamefont
  {Shirokov}(2013)}]{holevo2013classical}%
  \BibitemOpen
  \bibfield  {author} {\bibinfo {author} {\bibfnamefont {A.~S.}\ \bibnamefont
  {Holevo}}\ and\ \bibinfo {author} {\bibfnamefont {M.~E.}\ \bibnamefont
  {Shirokov}},\ }\href@noop {} {\bibfield  {journal} {\bibinfo  {journal}
  {Problems of Information Transmission}\ }\textbf {\bibinfo {volume} {49}},\
  \bibinfo {pages} {15} (\bibinfo {year} {2013})}\BibitemShut {NoStop}%
\bibitem [{\citenamefont {Holevo}(2013)}]{holevo2013information}%
  \BibitemOpen
  \bibfield  {author} {\bibinfo {author} {\bibfnamefont {A.}~\bibnamefont
  {Holevo}},\ }\href@noop {} {\bibfield  {journal} {\bibinfo  {journal} {Phys.
  Scr.}\ }\textbf {\bibinfo {volume} {2013}},\ \bibinfo {pages} {014034}
  (\bibinfo {year} {2013})}\BibitemShut {NoStop}%
\bibitem [{\citenamefont {Helstrom}(1976)}]{Helstrom_1976}%
  \BibitemOpen
  \bibfield  {author} {\bibinfo {author} {\bibfnamefont {C.}~\bibnamefont
  {Helstrom}},\ }\href {https://books.google.com/books?id=fv9SAAAAMAAJ} {\emph
  {\bibinfo {title} {Quantum Detection and Estimation Theory}}}\ (\bibinfo
  {publisher} {Academic Press},\ \bibinfo {year} {1976})\BibitemShut {NoStop}%
\bibitem [{\citenamefont {Holevo}(1982)}]{Holevo_1982}%
  \BibitemOpen
  \bibfield  {author} {\bibinfo {author} {\bibfnamefont {A.}~\bibnamefont
  {Holevo}},\ }\href@noop {} {\emph {\bibinfo {title} {Probabilistic and
  Statistical Aspects of Quantum Mechanics}}}\ (\bibinfo  {publisher}
  {North-Holland, Amsterdam},\ \bibinfo {year} {1982})\BibitemShut {NoStop}%
\bibitem [{\citenamefont {Yuen}\ and\ \citenamefont {Lax}(1973)}]{Yuen_1973}%
  \BibitemOpen
  \bibfield  {author} {\bibinfo {author} {\bibfnamefont {H.}~\bibnamefont
  {Yuen}}\ and\ \bibinfo {author} {\bibfnamefont {M.}~\bibnamefont {Lax}},\
  }\href@noop {} {\bibfield  {journal} {\bibinfo  {journal} {IEEE Trans. Inf.
  Theory}\ }\textbf {\bibinfo {volume} {19}},\ \bibinfo {pages} {740} (\bibinfo
  {year} {1973})}\BibitemShut {NoStop}%
\bibitem [{\citenamefont {Braun}\ \emph {et~al.}(2018)\citenamefont {Braun},
  \citenamefont {Adesso}, \citenamefont {Benatti}, \citenamefont {Floreanini},
  \citenamefont {Marzolino}, \citenamefont {Mitchell},\ and\ \citenamefont
  {Pirandola}}]{braun2018quantum}%
  \BibitemOpen
  \bibfield  {author} {\bibinfo {author} {\bibfnamefont {D.}~\bibnamefont
  {Braun}}, \bibinfo {author} {\bibfnamefont {G.}~\bibnamefont {Adesso}},
  \bibinfo {author} {\bibfnamefont {F.}~\bibnamefont {Benatti}}, \bibinfo
  {author} {\bibfnamefont {R.}~\bibnamefont {Floreanini}}, \bibinfo {author}
  {\bibfnamefont {U.}~\bibnamefont {Marzolino}}, \bibinfo {author}
  {\bibfnamefont {M.~W.}\ \bibnamefont {Mitchell}}, \ and\ \bibinfo {author}
  {\bibfnamefont {S.}~\bibnamefont {Pirandola}},\ }\href@noop {} {\bibfield
  {journal} {\bibinfo  {journal} {Rev. Mod. Phys.}\ }\textbf {\bibinfo {volume}
  {90}},\ \bibinfo {pages} {035006} (\bibinfo {year} {2018})}\BibitemShut
  {NoStop}%
\bibitem [{\citenamefont {Braunstein}\ and\ \citenamefont
  {Caves}(1994)}]{braunstein1994statistical}%
  \BibitemOpen
  \bibfield  {author} {\bibinfo {author} {\bibfnamefont {S.~L.}\ \bibnamefont
  {Braunstein}}\ and\ \bibinfo {author} {\bibfnamefont {C.~M.}\ \bibnamefont
  {Caves}},\ }\href@noop {} {\bibfield  {journal} {\bibinfo  {journal} {Phys.
  Rev. Lett.}\ }\textbf {\bibinfo {volume} {72}},\ \bibinfo {pages} {3439}
  (\bibinfo {year} {1994})}\BibitemShut {NoStop}%
\bibitem [{\citenamefont {Jarzyna}\ and\ \citenamefont
  {Demkowicz-Dobrza{\'n}ski}(2015)}]{jarzyna2015true}%
  \BibitemOpen
  \bibfield  {author} {\bibinfo {author} {\bibfnamefont {M.}~\bibnamefont
  {Jarzyna}}\ and\ \bibinfo {author} {\bibfnamefont {R.}~\bibnamefont
  {Demkowicz-Dobrza{\'n}ski}},\ }\href@noop {} {\bibfield  {journal} {\bibinfo
  {journal} {New J. Phys.}\ }\textbf {\bibinfo {volume} {17}},\ \bibinfo
  {pages} {013010} (\bibinfo {year} {2015})}\BibitemShut {NoStop}%
\bibitem [{\citenamefont {Bollinger}\ \emph {et~al.}(1996)\citenamefont
  {Bollinger}, \citenamefont {Itano}, \citenamefont {Wineland},\ and\
  \citenamefont {Heinzen}}]{bollinger1996optimal}%
  \BibitemOpen
  \bibfield  {author} {\bibinfo {author} {\bibfnamefont {J.~J.}\ \bibnamefont
  {Bollinger}}, \bibinfo {author} {\bibfnamefont {W.~M.}\ \bibnamefont
  {Itano}}, \bibinfo {author} {\bibfnamefont {D.~J.}\ \bibnamefont {Wineland}},
  \ and\ \bibinfo {author} {\bibfnamefont {D.}~\bibnamefont {Heinzen}},\
  }\href@noop {} {\bibfield  {journal} {\bibinfo  {journal} {Phys. Rev. A}\
  }\textbf {\bibinfo {volume} {54}},\ \bibinfo {pages} {R4649} (\bibinfo {year}
  {1996})}\BibitemShut {NoStop}%
\bibitem [{\citenamefont {Dorner}\ \emph {et~al.}(2009)\citenamefont {Dorner},
  \citenamefont {Demkowicz-Dobrzanski}, \citenamefont {Smith}, \citenamefont
  {Lundeen}, \citenamefont {Wasilewski}, \citenamefont {Banaszek},\ and\
  \citenamefont {Walmsley}}]{dorner2009optimal}%
  \BibitemOpen
  \bibfield  {author} {\bibinfo {author} {\bibfnamefont {U.}~\bibnamefont
  {Dorner}}, \bibinfo {author} {\bibfnamefont {R.}~\bibnamefont
  {Demkowicz-Dobrzanski}}, \bibinfo {author} {\bibfnamefont {B.}~\bibnamefont
  {Smith}}, \bibinfo {author} {\bibfnamefont {J.}~\bibnamefont {Lundeen}},
  \bibinfo {author} {\bibfnamefont {W.}~\bibnamefont {Wasilewski}}, \bibinfo
  {author} {\bibfnamefont {K.}~\bibnamefont {Banaszek}}, \ and\ \bibinfo
  {author} {\bibfnamefont {I.}~\bibnamefont {Walmsley}},\ }\href@noop {}
  {\bibfield  {journal} {\bibinfo  {journal} {Phys. Rev. Lett.}\ }\textbf
  {\bibinfo {volume} {102}},\ \bibinfo {pages} {040403} (\bibinfo {year}
  {2009})}\BibitemShut {NoStop}%
\bibitem [{Note3()}]{Note3}%
  \BibitemOpen
  \bibinfo {note} {The photon number variance can be unbounded, e.g. $\left
  (1-p\right )\ket {0}\bra {0}+p\ket {N}\bra {N}$, with mean $pN=N_S$, has
  variance diverging as $\propto N$.}\BibitemShut {Stop}%
\bibitem [{\citenamefont {Marian}\ and\ \citenamefont
  {Marian}(2016)}]{marian2016quantum}%
  \BibitemOpen
  \bibfield  {author} {\bibinfo {author} {\bibfnamefont {P.}~\bibnamefont
  {Marian}}\ and\ \bibinfo {author} {\bibfnamefont {T.~A.}\ \bibnamefont
  {Marian}},\ }\href@noop {} {\bibfield  {journal} {\bibinfo  {journal} {Phys.
  Rev. A}\ }\textbf {\bibinfo {volume} {93}},\ \bibinfo {pages} {052330}
  (\bibinfo {year} {2016})}\BibitemShut {NoStop}%
\bibitem [{\citenamefont {Banchi}\ \emph {et~al.}(2015)\citenamefont {Banchi},
  \citenamefont {Braunstein},\ and\ \citenamefont
  {Pirandola}}]{banchi2015quantum}%
  \BibitemOpen
  \bibfield  {author} {\bibinfo {author} {\bibfnamefont {L.}~\bibnamefont
  {Banchi}}, \bibinfo {author} {\bibfnamefont {S.~L.}\ \bibnamefont
  {Braunstein}}, \ and\ \bibinfo {author} {\bibfnamefont {S.}~\bibnamefont
  {Pirandola}},\ }\href@noop {} {\bibfield  {journal} {\bibinfo  {journal}
  {Phys. Rev. Lett.}\ }\textbf {\bibinfo {volume} {115}},\ \bibinfo {pages}
  {260501} (\bibinfo {year} {2015})}\BibitemShut {NoStop}%
\bibitem [{\citenamefont {Scutaru}(1998)}]{scutaru1998fidelity}%
  \BibitemOpen
  \bibfield  {author} {\bibinfo {author} {\bibfnamefont {H.}~\bibnamefont
  {Scutaru}},\ }\href@noop {} {\bibfield  {journal} {\bibinfo  {journal} {J.
  Phys. A: Math. Gen.}\ }\textbf {\bibinfo {volume} {31}},\ \bibinfo {pages}
  {3659} (\bibinfo {year} {1998})}\BibitemShut {NoStop}%
\bibitem [{\citenamefont {Fujiwara}(2006)}]{fujiwara2006strong}%
  \BibitemOpen
  \bibfield  {author} {\bibinfo {author} {\bibfnamefont {A.}~\bibnamefont
  {Fujiwara}},\ }\href@noop {} {\bibfield  {journal} {\bibinfo  {journal} {J.
  Phys. A: Math. Gen.}\ }\textbf {\bibinfo {volume} {39}},\ \bibinfo {pages}
  {12489} (\bibinfo {year} {2006})}\BibitemShut {NoStop}%
\bibitem [{\citenamefont {Gill}\ and\ \citenamefont
  {Massar}(2005)}]{gill2005state}%
  \BibitemOpen
  \bibfield  {author} {\bibinfo {author} {\bibfnamefont {R.~D.}\ \bibnamefont
  {Gill}}\ and\ \bibinfo {author} {\bibfnamefont {S.}~\bibnamefont {Massar}},\
  }in\ \href@noop {} {\emph {\bibinfo {booktitle} {Asymptotic Theory Of Quantum
  Statistical Inference: Selected Papers}}}\ (\bibinfo  {publisher} {World
  Scientific},\ \bibinfo {year} {2005})\ pp.\ \bibinfo {pages}
  {178--214}\BibitemShut {NoStop}%
\bibitem [{\citenamefont {Hayashi}(2017)}]{hayashi2017quantum}%
  \BibitemOpen
  \bibfield  {author} {\bibinfo {author} {\bibfnamefont {M.}~\bibnamefont
  {Hayashi}},\ }\href@noop {} {\emph {\bibinfo {title} {Quantum Information
  Theory}}}\ (\bibinfo  {publisher} {Springer},\ \bibinfo {year}
  {2017})\BibitemShut {NoStop}%
\bibitem [{\citenamefont {Lachs}(1965)}]{Lachs_1965}%
  \BibitemOpen
  \bibfield  {author} {\bibinfo {author} {\bibfnamefont {G.}~\bibnamefont
  {Lachs}},\ }\href {\doibase 10.1103/PhysRev.138.B1012} {\bibfield  {journal}
  {\bibinfo  {journal} {Phys. Rev.}\ }\textbf {\bibinfo {volume} {138}},\
  \bibinfo {pages} {B1012} (\bibinfo {year} {1965})}\BibitemShut {NoStop}%
\end{thebibliography}
\end{document}